\documentclass[twocolumn]{openjournal}
\usepackage{graphicx} 

\usepackage[titletoc]{appendix}
\usepackage[utf8]{inputenc}
\DeclareUnicodeCharacter{03BC}{\ensuremath{\mu}}
\usepackage{xspace}
\usepackage[colorlinks=true,urlcolor=blue,citecolor=NavyBlue]{hyperref}
\setcitestyle{authoryear,open={(},close={)}} 

\usepackage{tikz}

\newcommand{\tridownmarker}{%
  \tikz[scale=0.12,baseline=-0.4ex]{%
    \draw[line width=0.7pt] (0,0) -- (270:1);
    \draw[line width=0.7pt] (0,0) -- (30:1);
    \draw[line width=0.7pt] (0,0) -- (150:1);
  }%
}

\makeatletter
\renewcommand\appendix{\par
  \setcounter{section}{0}%
  \setcounter{subsection}{0}%
  \renewcommand\thesection{\@Alph\c@section}%
  \renewcommand\thesubsection{\thesection.\@arabic\c@subsection}}
\makeatother

\def\referee#1{{#1}}

\usepackage{placeins}
\usepackage[svgnames]{xcolor}

\providecommand{\apjl}{ApJL} 
\providecommand{\apjs}{ApJS} 
\providecommand{\aap}{A\&A} 

\newcommand{\um}{\ensuremath{\mu\mathrm{m}}\xspace}
\renewcommand{\micron}{\ensuremath{\mu\mathrm{m}}\xspace}

\newcommand{\persc}{\ensuremath{\mathrm{cm}^{-2}}\xspace}
\newcommand{\percc}{\ensuremath{\mathrm{cm}^{-3}}\xspace}
\newcommand{\water}{\ensuremath{\mathrm{H}_2\mathrm{O}}\xspace}
\newcommand{\methanol}{\ensuremath{\mathrm{CH}_3\mathrm{OH}}\xspace}
\newcommand{\ammonia}{\ensuremath{\mathrm{NH}_3}\xspace}
\newcommand{\ethanol}{\ensuremath{\mathrm{C}_2\mathrm{H}_5\mathrm{OH}}\xspace}
\def\software#1{\textbf{Software:} \texttt{#1}\xspace} 

\begin{document}

\title{The Colors of Ices: Measuring ice column density through photometry}


\author{Adam Ginsburg$^1$,
Savannah R. Gramze$^1$,
Matthew L. N. Ashby$^2$,
Brandt A. L. Gaches$^3$,
Nazar Budaiev$^1$,
Miriam G. Santa-Maria$^{1,4}$,
Alyssa Bulatek$^1$,
A.~T.~Barnes$^5$,
Desmond Jeff$^1$,
Neal J. Evans II$^6$,
Cara D. Battersby$^7$}

\affiliation{$^1$Department of Astronomy, University of Florida, Gainesville, FL 32611 USA\\
$^2$Center for Astrophysics $|$ Harvard \& Smithsonian, 60 Garden Street, Cambridge, MA 02138, USA\\
$^3$Faculty of Physics, University of Duisburg-Essen, Lotharstraße 1, 47057 Duisburg, Germany \\
$^4$Instituto de Física Fundamental (CSIC). Calle Serrano 121-123, 28006, Madrid, Spain \\
$^5$European Southern Observatory (ESO), Karl-Schwarzschild-Stra{\ss}e 2, 85748 Garching, Germany\\
$^6$Department of Astronomy, The University of Texas at Austin, 2515 Speedway, Stop C1400, Austin, Texas 78712-1205, USA\\
$^7$University of Connecticut, Department of Physics, 196A Auditorium Road, Unit 3046, Storrs, CT 06269}


\begin{abstract}
    Ices imprint strong absorption features in the near- and mid-infrared, but until recently they have been studied almost exclusively with spectroscopy toward small samples of bright sources. 
    We show that \textit{JWST} photometry alone can reveal and quantify interstellar ices, and we present a new open-source modeling tool, \texttt{icemodels}, to produce synthetic photometry of ices based on laboratory measurements. 
    We provide reference tables indicating which filters are likely to be observably affected by ice absorption. 
    Applying these models to NIRCam data of background stars behind \referee{several} Galactic Center (GC) clouds \referee{(dust ridge clouds A [the Brick], C, and D)}, and validating against NIRSpec spectra of Galactic disk sources, we find clear signatures of CO, H$_2$O, and CO$_2$ ices and evidence for excess absorption in the F356W filter likely caused by CH-bearing species such as \methanol. 
    The ice ratios differ between the Galactic disk and Center, with GC clouds showing a higher H$_2$O fraction. 
    \referee{A} large ice abundance \referee{is observed} in CO, \water, and possibly complex molecules, \referee{which implies that there is substantial freezeout and therefore potential for} ice-phase chemistry in non-star-forming gas. 
    Accounting for all likely ices, we infer that $>25\%$ of the total carbon is frozen into CO ice in the GC, which exceeds the entire solar-neighborhood carbon budget. 
    By assuming the freezeout fraction is the same in GC and disk clouds, we obtain a metallicity measurement indicating that $Z_{GC}\gtrsim2.5Z_\odot$. 
    These results demonstrate that photometric ice measurements are feasible with \textit{JWST} and capable of probing the metallicity structure of the cold interstellar medium.

\end{abstract}

\section{Introduction}
\label{sec:introduction}

The coldest part of the interstellar medium is molecular clouds, in which nearly all material except noble gases is in molecular form.
Since these regions are too cold to excite molecular hydrogen, we trace them with more easily excited molecules like CO.
However, in the densest, coldest parts of the cloud, even these tracers freeze out.
A total inventory of the tracer molecules is needed to accurately measure the total gas, but our inventories have historically been limited to gas-phase molecules that emit observable radiation.

Ices are cool, but they are difficult to observe.
The mid- to far-infrared absorption features they generate have until recently been observable only toward a handful of sources.
Prior to JWST, hundreds to thousands of spectra with ices had been observed by ISO \citep[e.g.,][]{Gibb2004}, Spitzer \citep[e.g.][]{Boogert2008,Oberg2008,Pontoppidan2008}, Akari \citep{Aikawa2012}, and a few ground-based observatories \citep[e.g.][]{Jang2022}.
JWST NIRSpec and NIRCam WFSS have added a few hundred more to that count, and SPHEREx \citep[]{Bock2026} will soon add thousands to millions more \referee{\citep[J. Hora et al.\ in prep]{2026Melnick,Ashby2023}}.
However, JWST NIRCam and MIRI have already detected many thousands, and possibly millions, of stars behind icy gas clouds, enabling photometric measurements of these ices.

There is a limited history of using photometric filters to study interstellar absorption features in the infrared.
\citet{Ginsburg2023} showed that the F466N filter, centered close to the CO ice absorption peak, is sensitive to ice and very weakly sensitive to gas.
In particular, they showed that at typical CMZ conditions (abundance of CO, gas temperature, and linewidths), CO gas is limited to an absorption of $\lesssim10\%$ (or magnitude difference $\Delta m < 0.1$), much less than is observed throughout the target cloud, the Brick.
\citet{Gunay2025} showed that NIRCam and MIRI photometry can be used to measure spectral features of dust grains, specifically the aliphatic \referee{C-H} feature around 3.4 \um.
\citet{Meingast2025} demonstrated that a combination of ground- and space-based filters can be used to measure the \water ice column density.
To our knowledge, no other programs have attempted to measure ice photometrically in the last two decades \citep{Harker1997}, and the only attempts to map ice distribution in clouds have used spectroscopic measurements \citep{Pontoppidan2006, Noble2017, Smith2025}.

Photometric ice mapping holds the potential to expand our knowledge of clouds on several otherwise inaccessible axes.
Photometry reaches deeper than spectroscopy, enabling relatively inexpensive ice column measurements at higher extinction.
It also \referee{typically} covers wider fields, including far more stars.
These larger, deeper samples allow exploration of cloud structure, comparison of populations of clouds, and examination of variation across many lines of sight.
Further, they provide optimal selections for spectroscopic followup.

In this work, we develop a technique to partially infer the composition of ices and measure column density using JWST photometric data.
Using column density measurements of CO ice, we then compare clouds in the Galactic disk, bar, and center, and demonstrate that they differ most likely because of changing metallicity with galactocentric radius.


Here, we adopt a few common assumptions about ice and dust abundance as a basis for analysis and re-examine each one after confronting it with data.
We adopt N(H)/A$_V$=2.21$\times10^{21}$ \persc mag$^{-1}$ \citep{Guver2009}, close to the N(H$_2$)/A$_V=1\times10^{21}$ found by \citet{Lacy2017}.
We adopt a CO abundance with respect to hydrogen, X(CO) = N(CO)/N(H$_2$) = $10^{-4}$ (N(CO)/N(H) = $0.5\times10^{-4}$), which assumes that $10\%$ of C is in CO for Solar neighborhood (HII region) carbon abundance X(C) = N(C)/N(H) = $10^{-3.3} = 5\times10^{-4}$ \citep{Asplund2009}.
\referee{The HII region abundance is the highest reported in \citet{Asplund2009}; solar photosphere abundances are lower, at $\sim3.0-3.6\times10^{-4}$.
In diffuse lines of sight, the gas-phase carbon abundance is poorly understood, but a value of $2.0\times10^{-4}$ is commonly adopted based on a handful of measurements \citep{Jenkins2009,Hensley2021}.
Jenkins' table 4 yields a logarithmic range of gas-phase carbon depletion $[X/H]_0 = -0.112 \pm 0.194$, i.e., the total carbon depletion is consistent with zero (all in gas phase) to 50\% depleted (half in dust) on diffuse lines of sight, providing a very weak upper limit to the fraction of gas available to form CO, while along fully-depleted lines of sight, the carbon depletion is $[X/H]_1 = -0.213\pm0.075$, i.e., 50-75\% is assumed to be locked into dust - but their work does not differentiate between CO on dust and carbon integrated into dust.
The assumption that $\sim50\%$ of carbon is incorporated into dust grains, unavailable to form CO, appears reasonable, but the evidence is weak.
}
The Galactic Center has higher-metallicity gas than the solar neighborhood, with $Z\approx2.2-2.5 Z_\odot$ reported \citep[e.g.,][]{Nandakumar2018,Do2015} based on stars and about 3$Z_\odot$ based on gas \citep{Simpson2018}, which hints that X(CO) might be 2-3$\times$ greater in the GC.
We will show that this is indeed the case for ice-phase CO.


This manuscript is organized as follows.
In Section \ref{sec:data}, we present the JWST data analyzed here.
Section \ref{sec:methods} describes the \texttt{icemodels} package used to analyze the data and infer column densities.
Section \ref{sec:icebands} examines each JWST filter involved in this study and shows which NIRCam filters are expected to exhibit ice absorption features.
Section \ref{sec:colors} then discusses color-color diagrams, showing how they can be used to measure specific ices.
Section \ref{sec:measuringicecolumn} derives column density of CO ice, then explores the implications of these measurements.
We conclude and summarize in Section \ref{sec:conclusion}, after which several appendices provide additional reference material and characterization of uncertainty.

\section{Data}
\label{sec:data}

\subsection{JWST NIRCam}
We discuss photometric data obtained with JWST NIRCam toward The Brick (G0.253+0.016; projects 1182 and 2221), dust ridge clouds C and D (project 2221), and Sgr B2 (project 5365).
The photometric measurement process is described in \citet{Ginsburg2023}, \citet{Gramze2025}, and \citet{Budaiev2025}.
In brief, \texttt{photutils} \citep{Bradley2025} was used with \texttt{webbpsf} and \texttt{stpsf} \citep{Perrin2012,Perrin2014} to perform PSF photometry on each individual frame, then the frames were cross-matched.
The cross-matching included all stars whose nearest neighbor is $>0.1$\arcsec\ away from any others in the catalog as a new entry, and merged with existing entries those that had a neighbor within $\leq0.1$\arcsec.
Individual star measurements were kept only if both their \texttt{qfit} and \texttt{cfit} values were $<0.4$.
The star measurements (of which there are up to 24) were then filtered with a sigma-clipping algorithm with $\sigma=3$ and 5 iterations.
Any stars with fewer than 4 independent measurements after this clipping were removed from the catalog.
The final combined measurement was the inverse-variance-weighted average of the individual frame measurements.

The individual filters' catalogs were then cross-matched into a band-merged catalog.
Stars were considered the same object if their separation is $<0.1$\arcsec.
For analysis in this work, we keep only measurements with detections in each of the 6 filters in program 2221 (F182M, F187N, F212N, F405N, F410M, and F466N) and uncertainty less than 0.1 mag, i.e., S/N$>10$.
Similarly, for project 1182 data, we keep only stars with uncertainty $<0.1$ mag in F200W, F356W, and F444W (we do not require a detection in F115W, since few stars in the Galactic Center are detected at such short wavelengths).
Finally, stars with $m_{F182M} < 15.5$ were excluded because they are saturated.
Stars with $m_{F410M}<13.7$ and F405N$-$F410M $<-0.2$ or $m_{F410M}>17$ and F405N$-$F410M$<-1$ were also excluded for saturation and an additional currently-unexplained effect on the faint end \referee{in which some faint sources have blue colors in these filters}.

We use \referee{Vega} magnitudes throughout.
We calculate the \referee{Vega} magnitude using zero points acquired from the SVO Filter Profile Service \citep{Rodrigo2024}.

There is an apparent systematic offset in colors that affects the F405N and F410M filters, the former more than the latter.
We do not know the origin of this offset, though it may be from atmospheric absorption from the Br$\alpha$ line in the stellar spectra.
It adds a 5\% (0.05 mag) systematic uncertainty to the measurements.

\subsection{JWST NIRSpec}
\label{sec:spectra}


We downloaded all spectra of star-forming regions and dark clouds from the JWST archive as of June 2025.
We include in our analysis only spectra that cover at least the range 1.6 to 4.8 \um, as we need both the short wavelengths for extinction measurements and the long wavelengths for ices.
These include those obtained with the PRISM grating, which is low-resolution and covers the full wavelength range from 1-5 $\um$, and sources observed with both the G235M and G395M gratings, allowing us to construct synthetic photometry in both the short (F182M and F212N) and long (F405N, F410M, and F466N) bands.

We include spectra from three programs that were immediately useful and comparable to our data: projects 1186, 3222, and 5804.
These spectra come from highly embedded protostars that have strong infrared excesses (program HEFE looking at protostars in Orion, project 5804) and from stars behind clouds (project 3222 observing stars behind IRAS 16293 and project 1186 targeting stars in Serpens).
The data from 3222 do not always fully cover the F182M band used for extinction measurement because of a chip gap \referee{(i.e., the full spectrum is not always measured with high-resolution gratings because part of the spectrum is dispersed off the detector)}; we interpolate across this gap using a Gaussian filter with width 10 pixels if fewer than 20\% of the pixels are blank.
We \referee{excluded} any spectra with a larger gap or bad data overlapping with the required filters, and we visually excluded those with apparently low signal-to-noise in the spectra (most of these spectra have many negative pixels, indicating no significant detection).
These exclusions are not intended to be systematic, as the spectra are being used to guide our interpretation of photometry, not as primary targets of study.
We also examined spectra from project 2770, which targeted many stars in and behind the Orion Nebula, but found that this sample was dominated by emission-line sources and spectra with background subtraction issues, and it is unclear which stars are behind or in the nebula, so we do not include these data in our analysis.

\begin{figure*}[htbp]
    \centering
    \includegraphics[width=0.48\linewidth]{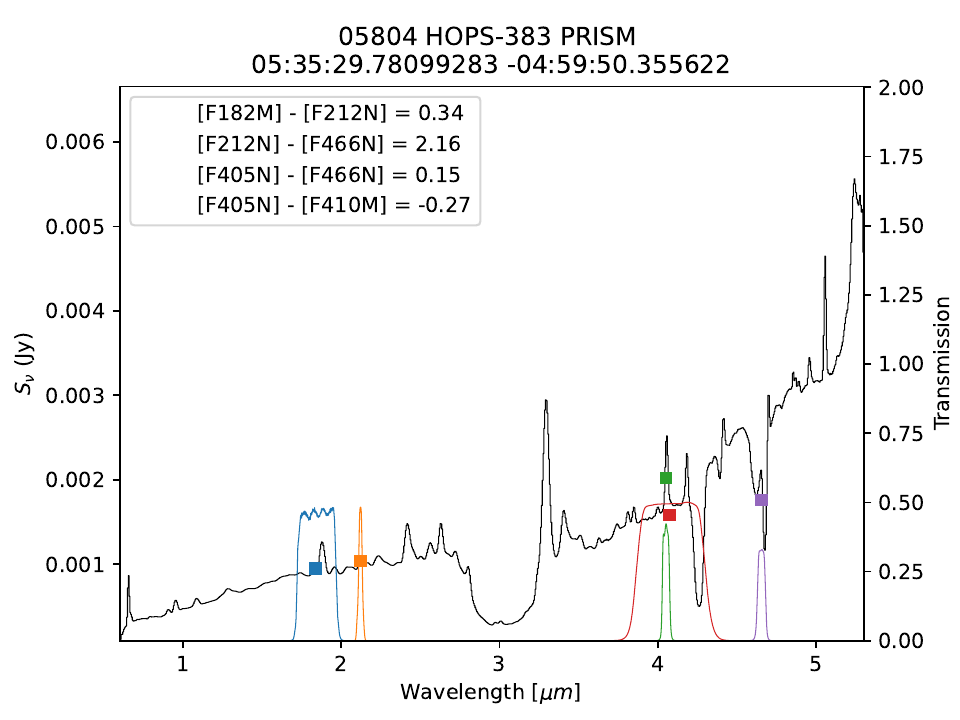}
    \includegraphics[width=0.48\linewidth]{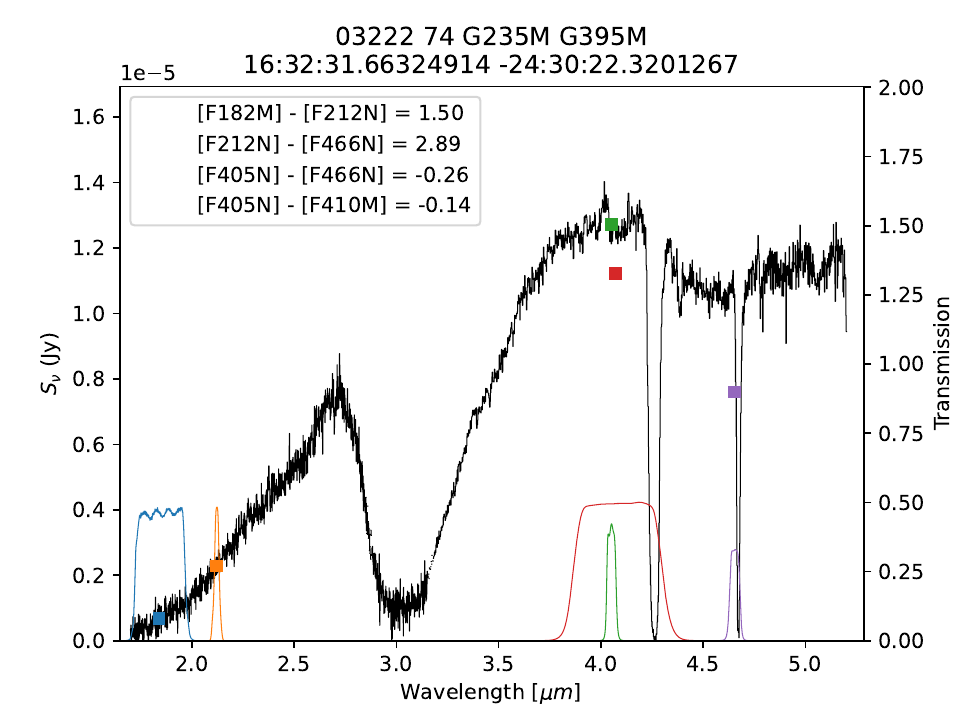}
    \caption{Example JWST NIRSpec archival spectra  with the filters used in project 2221 and emphasized throughout this work, F182M, F212N, F405N, F410M, and F466N, shown.
    HOPS-383  (left) from program 5804  is a Class 0 YSO \citep{Megeath2012}.
    IRAS16293 slit 74 from program 3222 is a star behind the IRAS16293 molecular cloud.
    The legend shows the synthetic magnitudes and colors computed from the spectrum.
    \referee{The title provides the source of the spectrum, including the source name from the project (HOPS-383, left, and spectrum ID 74, right), the program ID (5804, left, and 3222, right), and the grating used (PRISM, left, and G235M+G395M, right).
    The second line of the title provides the star's ICRS coordinates.
    The colored squares show the transmission-profile-weighted flux density, i.e., that which would be measured in the corresponding filter. 
    The right axis shows the relative transmission of each filter.
    }
    }
    \label{fig:sample_spectra_medbands}
\end{figure*}

The example spectra are shown in Figure \ref{fig:sample_spectra_medbands}.
To calculate the synthetic photometry from these spectra for plotting in color-color diagrams, we multiply the spectrum by the filter transmission profile and integrate over it following the same approach described below in \S \ref{sec:rteqns}.
The code for the synthetic photometry calculation is available on the github repository\footnote{\url{https://github.com/keflavich/brick-jwst-2221/blob/main/brick2221/analysis/JWST\_Archive\_Spectra.py}}.

\section{Methods}
\label{sec:methods}

\subsection{The \texttt{icemodels} package}
We present the \texttt{icemodels}\footnote{\url{https://github.com/keflavich/icemodels/}} package, a library to access and load laboratory measurements of ice \referee{absorption} opacity, perform simple radiative transfer modeling, and compute the expected flux in photometric filters.
It provides interfaces to the UNIVAP, LIDA \citep{Rocha2022}, and OCDB \footnote{\url{https://ocdb.smce.nasa.gov/}} databases, though because the databases are relatively small ($<2$ GB as of June 2025), the recommended approach is to download the full databases.

The main use of \texttt{icemodels} in this paper is to compute synthetic photometry of stars behind icy clouds.
The toolkit can be used for more general modeling, e.g., spectroscopic fitting as in ENIIGMA \citep[][]{Rocha2021}, but that use case is not explored here.
The intended use case is similar to \texttt{SynthIceSpec} \citep{Taillard2025}, but we use laboratory-measured opacities instead of Gaussians.
To obtain photometric predictions, it uses background models from the Kurucz stellar atmosphere library as provided by the \texttt{mysg}\footnote{\url{https://github.com/astrofrog/mysg}} package, which is the same as used in the \citet{Robitaille2017} and \citet{Richardson2024} model grids; \referee{\texttt{mysg} simply reads model spectra and metadata}.
The stellar atmospheres are then attenuated by opacities as a function of wavelength as described below in \S \ref{sec:rteqns} using the \texttt{absorbed\_spectrum} function.
The fluxes are then multiplied by filter transfer functions obtained from the Spanish Virtual Observatory Filter Profile Service \citep[SVO-FPS][]{Rodrigo2024} as provided by \texttt{astroquery} \citep{Ginsburg2019}, integrated, and compared to filter zero-points to obtain in-band magnitudes.

The script \texttt{absorbance\_in\_filters.py} loops over all OCDB, UNIVAP, and LIDA ice compounds and computes the colors for 50 column densities in the range N(ice)=$[10^{17}, 10^{21}]$ \persc using 4000 K Kurucz \citep{Kurucz1979} models via the \texttt{mysg} wrapper as the backlight\footnote{Variations in stellar atmospheres affect the observed filters by $<0.1$ magnitudes ($\lesssim10\%$) above 4000 K.  While intrinsic stellar colors do have some effect on the short wavelength bands ($<2$ \micron), they do not produce any systematic effects in the ice absorption bands.}.
\referee{\citet{Kurucz1979} 4000 K models are adopted as a simple reference for the cool stars that are expected to be the dominant population toward the Galactic Center, in which almost all stars have ages between 1-10 Gyr \citep{Nogueras-Lara2020}.
Cooler stars are expected to be a negligible fraction of the population: Main-sequence M-stars are below the detection limit in most filters (even at t=0, when they are brightest, M0V $M_{K,abs}\approx5$ \citep{Pecaut2013}, and for the GC's distance modulus of 14.5 mag plus $A_K\approx2.5$ mag, they would appear at $m_K\approx22$ mag), while the more luminous main-sequence stars and giant stars are all dominated by warmer stars, K-type and earlier; cooler red supergiants, while detectable, are intrinsically rare and comprise only a few percent of the total population \citep{Girardi2005}.
To validate these assumptions, we used the TRILEGAL model \citep{Girardi2012} modified with $A_V / \mathrm{kpc} = 2$ mag and $R_{gal,\odot}=8.1$ kpc to model a somewhat realistic stellar population; while TRILEGAL under-predicts the total number of stars because it does not explicitly model the nuclear stellar disk or cluster, it still gives a reasonable stellar synthesis model.
Selecting only those stars with extincted $14 < m_{F212N} < 20$ mag at distances $>7$ kpc, $>99.5\%$ of detectable stars have $T_{eff} > 4000$ K, confirming the assumed temperatures are reasonable.
}

We translated the lab-measured absorbances\footnote{Not to be confused with extinction $A_\mathrm{V}$.} $A = \log I/I_0$ into $k$ values adopting very simple assumptions, i.e., that $k=A \ln(10) \lambda / 4 \pi d$ \citep{Rocha2022}, where $d$ is the ice slab thickness and is computed as $d = N / n$, $N$ is the ice column density supplied in the LIDA database and $n=\rho / \mu$ is the ice number density; $\rho$ is assumed to be 1 g \percc.
Note that several entries in the LIDA database are not usable as they do not report the ice column density used in the experiment.

\subsection{Radiative Transfer Equations}
\label{sec:rteqns}
In the plots below, we define $\kappa_{eff}$ to be the constant by which one multiplies the ice molecule's number column density $N$, with units \persc, to obtain the optical depth $\tau$.
\referee{In so doing, we explicitly ignore scattering effects, which are expected to be important when the ice is circumstellar but negligible in the case we consider, in which the star is treated as a backlight behind a very distant icy molecular cloud.}

Adopting the equation
\begin{equation}
\tau_\nu = N \frac{\alpha_\nu}{\rho_N} = N \frac{4 \pi \tilde{\nu} k_\nu}{\rho_N}
\end{equation}
derived from \citet{Hudgins1993}, that means the effective opacity in units of \persc per molecule is
\begin{equation}
\label{eqn:kappaeff}
    \kappa_{eff}(\nu) =  \frac{4 \pi \tilde{\nu} k_\nu}{\rho_N}
\end{equation}
where $k_\nu$ is the absorption opacity constant at a given wavelength obtained from laboratory measurements \referee{(i.e., the imaginary part of the complex index of refraction)}, the number density $\rho_N$ is
$ \rho_N = \rho/ \mu$, $\rho$ is the  mass density of the ice (assumed to be $\approx1$ g \percc; see \S \ref{sec:icemixes}), $\mu$ is the mean molecular weight (see Eqn. \ref{eqn:mu}), and $\tilde{\nu}$ is the observed wavenumber (in cm$^{-1}$).

The input stellar spectrum is then absorbed by the above optical depth, i.e.,
\begin{equation}
    S_\nu = S_{\nu,0} e^{-\tau_\nu}
\end{equation}
where $S_\nu$ is the flux density, generally in Janskys.
Then, to get an integrated flux in a filter, the flux density is multiplied by the filter transfer function \referee{$T_\nu$}, integrated over, and normalized:
\begin{equation}
    F_\nu =\frac{\int S_\nu T_\nu d\nu}{\int T_\nu d\nu}.
\end{equation}
To compute the integral, the $S_\nu$ array is interpolated onto the transmission curve's grid, which is sampled at 10-20 \AA\ per pixel for the NIRCam filters.
The resulting flux density is converted to a magnitude with
\begin{equation}
    m = -2.5 \log \left( \frac{F_\nu}{F_{0}}\right)
\end{equation}
where $F_0$ is the magnitude zero-point in the specified filter.
We compute tables of delta-magnitude (flux ratio) by taking the difference between the stellar magnitude and the ice-absorbed stellar magnitude,
\begin{equation}
    dm = m_* - m_{*,iced}.
\end{equation}

The models are implemented in the \texttt{icemodels} package.

\subsection{Ice Mixtures}
\label{sec:icemixes}
Laboratory measurements frequently use mixtures of ices.
Additionally, \texttt{icemodels} supports linear combinations of ice models.
In this work, we use linear combinations of pure ice models in most of our modeling below, since that allows us to freely change the relative abundance of these molecules.
This approach does not account for interactions between the ices: the opacity of CO$_2$ embedded in \water ice is different from pure CO$_2$ ice, for example \citet{Bernstein2005,Oberg2007}.
Nevertheless, these linear combinations provide reasonable estimates and hint at which ice mixtures need to be measured in lab.

The composition of any given ice is converted into a molecular weight using the \texttt{molscomps} function.
In the code, and in laboratory reports, the ice ratios are given as, e.g., H$_2$O:CO$_2$ (100:14), which are number ratios, i.e., for every 14 CO$_2$ molecules, there are 100 H$_2$O in the ice.

All ices are assumed to have a mass density \mbox{$\rho = 1$ g \percc.}
This is a reasonable approximation that adds less than 30\% uncertainty in most measured cases \citep{Hudson2020,Satorre2008}.
This value is in between the value for low-density amorphous (LDA) water ice ($\rho = 0.94$ g \percc) and high-density amorphous (HDA) water ice ($\rho=1.1$ g \percc) \citep{Eltareb2024}.
Since astrophysical ice is mostly comprised of water \citep[e.g.,][]{Gibb2004}, and CO ice has roughly the same density \citep{Luna2022}, this assumption is physically warranted.
\referee{CO$_2$ ice has higher density, up to $\rho\lesssim1.53$ g \percc \citep{Satorre2008}, but it is a small fraction of the total ice mass.}

%
When performing a linear combination of multiple ice species, the $k_\nu$ values are computed by taking the number-weighted mean of the effective opacities $\kappa_{eff}$:
\begin{equation}
\kappa_{eff, \nu,i} = \frac{\sum_i \kappa_{eff, \nu,i}  N_i}{\sum_i N_i}
\end{equation}
where $N_i$ is the fraction (e.g., $N_{\mathrm{H_2O}}=100$ and $N_{\mathrm{CO_2}}=14$ in the example above).
This $\kappa_{eff}$ is then inverted back to a $k$ value and incorporated into an opacity table with the same structure as those retrieved from the online databases.
The mean molecular weight $\mu$ is then computed as
\begin{equation}
    \label{eqn:mu}
    \mu = \frac{\sum_i \mu_i  N_i}{\sum_i N_i}
\end{equation}
where $\mu_i$ is the molecular weight of the $i$th ice species (e.g., $\mu(\mathrm{CO}_2)=44$, and $\mu(\mathrm{H_2O:CO_2 (100:14)}) = 21.2$).


\section{JWST Filters with Ice}
\label{sec:icebands}
We discuss each of the JWST NIRCam filters that is potentially affected by known ices in this section.
We limit our discussion to NIRCam, excluding MIRI and NIRISS, because large data sets with $>10^4$ stars (possibly $>>$) already exist in the Galactic Center \citep[][and projects 1182, 2092,]{Ginsburg2023,Gramze2025,Budaiev2025,Crowe2025}.
The tools we present are equally applicable to MIRI, \referee{but because we lack MIRI photometry, we do not model ices in the MIRI bands here.}

\referee{A theme throughout this section is that water ice contributes to opacity in most bands, even when it is not the dominant absorber in a specific filter.}

\begin{figure}
    \centering
    \includegraphics[width=\linewidth]{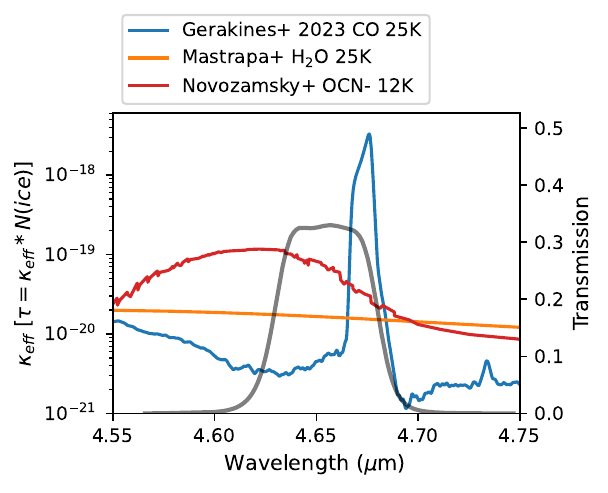}
    \caption{F466N transmission spectrum (grey; \referee{label on right axis}) and ice effective opacity $\kappa_{eff}$ (Eqn \ref{eqn:kappaeff}) as a function of wavelength across the F466N band.
    The colored curves show single-ice opacities for CO at 25 K \citep{Gerakines2023}, \water at 25 K \citep{Mastrapa2009}, and OCN$^-$ (with contaminants) from \citet{Novozamsky2001} as described in the legend.
    }
    \label{fig:F466Nplusopacities}
\end{figure}

\subsection{F466N is absorbed by CO, \water, \referee{and maybe OCN$^-$}}
By design, the F466N filter has the greatest overlap with the CO ice feature at 4.675 microns.
While the opacity in this line is very high and CO ice is expected to be highly abundant, it only covers about 20\% of the filter's width.
That implies that, even for high CO ice column densities N(CO) $> 10^{18}$ \persc, the maximum flux loss in-band is 20\%, which corresponds to 0.24 magnitudes \citep{Ginsburg2023}.
The observed F466N color differences reported in \citet{Ginsburg2023} vastly exceed this value.
This implies that there is some substantial opacity crossing the whole filter width, producing the observed flux losses of $\sim1.5$ magnitudes (roughly, the highest observed at high signal-to-noise ratio), i.e., 75\% flux loss.
Figure \ref{fig:F466Nplusopacities} shows how this might be possible: the broad water ice feature centered at $\sim4.5\um$ (the peak is not shown in the figure, only the wing that overlaps the relevant filter) has opacity about two orders of magnitude lower than the CO peak, but nevertheless high enough to matter at high column densities.
Given that water ice is expected to form everywhere CO ice is observed, since oxygen is more abundant than carbon and \water has a higher freezing temperature, we propose that this is the correct interpretation of the \citet{Ginsburg2023} results, and we use water plus CO ice opacity in discussion below.


The XCN feature, possibly assigned to OCN$^-$ \citep{Novozamsky2001}, is another possible significant contributor to the opacity in this band, with individual molecular opacity exceeding that of water but abundance expected to be much lower.
We do not model the XCN feature primarily because of a lack of usable laboratory spectra.
The only published opacity curves in OCDB \& LIDA come from photolysis-produced OCN$^-$ \citep{Novozamsky2001}, and the published measurements therefore include other molecular contaminants that may not be present in the ISM and that affect other bands' opacities; this means that including OCN$^-$ in the models cannot produce the observed colors in JWST bands.
Additionally, as discussed in \S \ref{sec:colors466410}, the observed spectra of background stars behind local clouds do not exhibit the XCN feature; protostars do show OCN$^-$, but protostars have extremely red colors that render them inconsistent with the observed colors of backlight stars, so we are confident we are not observing a large population of protostars.
There is some reason to expect the XCN feature to be a prominent component that may affect our measurements: the Galactic Center stars observed with ISO \& UKIRT have prominent absorption features with depth comparable to CO centered at 4.62\um \citep[see Appendix \ref{sec:isospectra};][]{Chiar2002}.
However, \citet{Onaka2022} suggested that formation of OCN$^-$ requires photolysis by UV photons, which may limit this line to be in the vicinity of massive stars or accreting low-mass stars; in the Galactic center, the higher background UV field may supply these photons, but still the extent should be more limited than CO.

\citet{Taillard2025} recently published laboratory measurements of CS$_2$ opacity.
It overlaps with the F466N filter, but we have not included it in our modeling because no opacity tables were available.
We expect CS$_2$ ice to be a minor constituent since sulfur is more than an order of magnitude less abundant than CO, but we cannot exclude it definitively.

\subsection{F410M is absorbed by CO$_2$ and \water}
\begin{figure*}
    \centering
    \includegraphics[width=\linewidth]{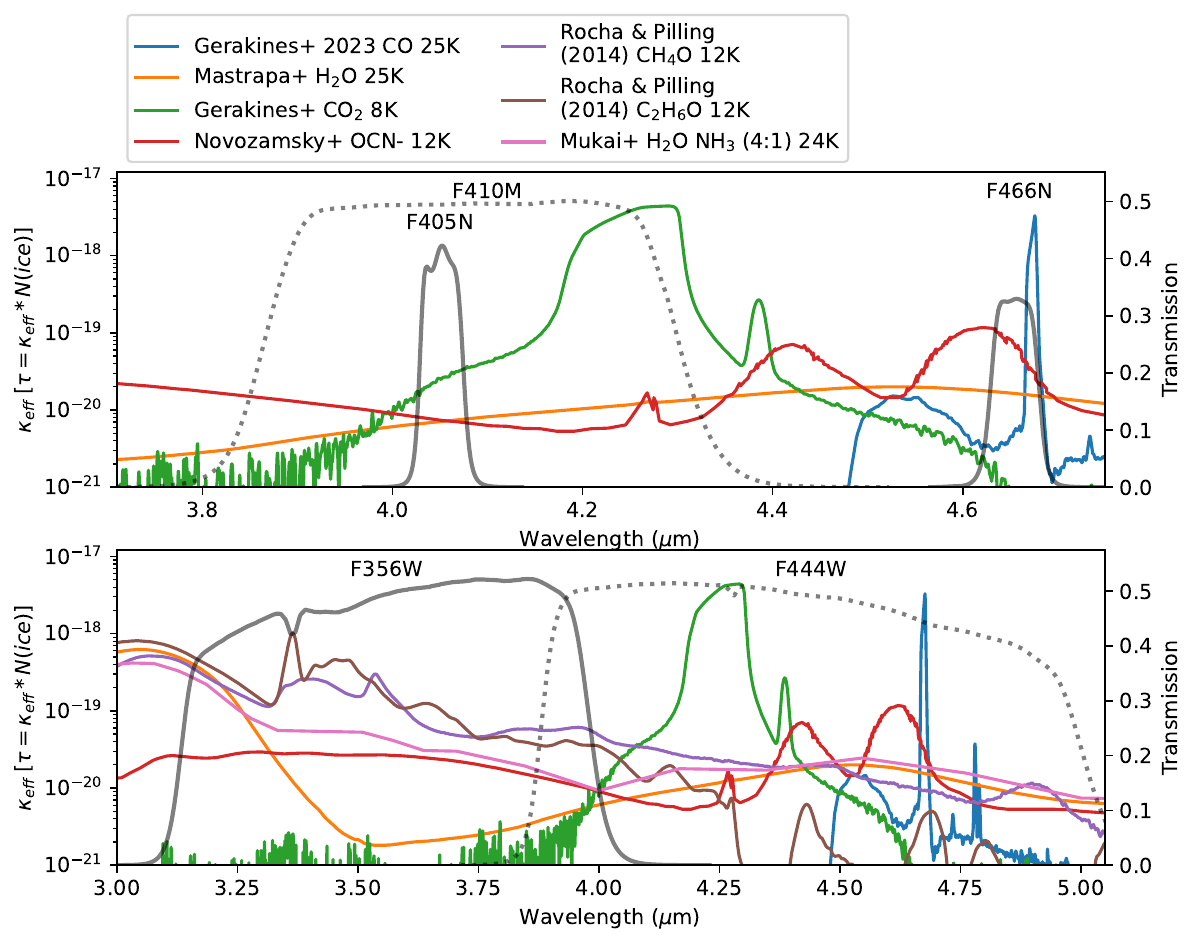}
    \caption{(top) Selected laboratory model ice opacities overlaid on the F405N, F410M, and F466N filter transmission profiles.
    The same opacity curves as in Figure \ref{fig:F466Nplusopacities} are shown, and CO$_2$ at 8 K from \citet{Gerakines2020} is added.
    (bottom) Selected model ice opacities overlaid on the wide-band filters F356W and F444W. 
    \referee{These opacity curves are shown on all remaining NIRCam filters in Appendix \ref{sec:moreabsorptionplots}}.
    }
    \label{fig:f410mplusopacities}
    \label{fig:opacities_on_widebands}
\end{figure*}

The F410M band overlaps with several ice species shown in Figure \ref{fig:f410mplusopacities}.
First we highlight CO$_2$, which covers roughly 20\% of the band.
CO$_2$ ice is expected to form wherever CO forms.
Its abundance appears to be lower than CO based on the fact that the F410M-F466N color trends blue (more on this in \S \ref{sec:colors466410}).
The OCN$^-$ opacities appear to significantly affect this band in Figure~\ref{fig:f410mplusopacities}, but \citet{Novozamsky2001} identifies the features in the 3.8-4.4 \um range with NH$_4^
+$, \referee{i.e., the only available laboratory-measured opacity table including OCN- includes a contaminant molecule that affects this band}.  
The water line produces significant opacity across the F410M band, suggesting that it competes with CO$_2$ as the dominant absorber in F410M.

\subsection{F405N is \referee{weakly} absorbed by \water}
\label{sec:f405n}
Unlike F410M, F405N does not overlap with the CO$_2$ feature.
It is only moderately affected by the broad H$_2$O feature seen in Figure \ref{fig:f410mplusopacities}.
Amorphous HDO affects the F405N filter, but still less than the F410M filter, and even in extreme cases, the HDO opacity is expected to be relatively small \citep{Slavicinska2025}.
Both \methanol and H$_2$S may also affect both F405N and F410M.


\subsection{F356W is absorbed by a mix of ices}
\label{sec:f356w}
The F356W band is \referee{absorbed by the `red wing' of the 3 micron water ice feature.}
\referee{
    This feature has been extensively observed in molecular clouds.
    \citet{Dartois2024} showed that the feature can be well-modeled by grains with a large maximum size $a_{max}\approx1$ \um with a mix of \methanol and \ammonia ice.
    As they note, ``methanol explains \textit{some} excess extinction in the 3 μm red wing'' (emphasis ours), indicating that the absorption in this band is more complex than can be explained with a single species.
}

Figure \ref{fig:opacities_on_widebands} shows selected opacity curves from CO, CO$_2$, H$_2$O, OCN$^-$, \referee{NH$_3$}, methanol, and ethanol overlaid on both F356W and F444W.
While the 3.05 \micron \water ice feature overlaps with F356W, it covers only a small fraction of the bandwidth and therefore has a limited effect on the transmitted flux.
\referee{NH$_3$ ice is potentially a substantial absorber, but it is generally expected and observed that NH$_3$ is among the last species to freeze out \citep{Caselli2022}.}
The \referee{C-H} vibrational feature, from either complex molecules with methyl (CH$_3$) groups or aliphatic hydrocarbons, are more likely absorbers.

Methyl-bearing species are known to absorb this range.
\citet{Slavicinska2025} show that \methanol ice affects this band (their Figure B5).
\citet{An2017} showed that at least one massive young stellar object (MYSO) in the Galactic Center has prominent \methanol absorption.
In Appendix \ref{sec:moreabsorptionplots}, we show the opacity curves of a few mixtures of \water and \methanol that prominently absorb this band.

More generally, vibrational modes in aliphatic \referee{C-H} bonds affect this band, \referee{but they have been ruled out as a significant absorber in local dense clouds}.
Many ISO spectra show a broad `wing' feature at 3.2-3.8 \um that would also explain the F356W absorption \citep{Gibb2004}.
The absorption in F356W may be caused by the aliphatic \referee{C-H} stretching feature at about 3.4 \um \citep{Gunay2018,Gunay2020,Gunay2025}.
\citet{Jang2022} showed that absorption in this band is likely a mix of the \referee{C-H} stretches and \methanol and that separating their contributions is difficult.
\referee{However, \citet{Dartois2024} showed that the aliphatic C-H stretching mode is limited to be a small fraction ($<1/3$ in their specific case) of the total absorption at 3.4 \um in the Chamaeleon I cloud, suggesting that this is not an important absorber toward the GC clouds.}

The spectra in Figure \ref{fig:opacities_on_widebands} show minimal absorption in F356W.
The HOPS YSO does not show any clear absorption, but that is in part because the continuum level is difficult to identify.
The background star shows a long line wing feature extending from 3.2-3.8 \um, \referee{but this feature is weaker than in NIR38 from \citet{McClure2023} \& \citet{Dartois2024}}. 
In both cases, the absorption is much weaker than the extremes observed in the Brick --- we will revisit this in \S \ref{sec:f356wf444w}.


\subsection{F444W is only weakly absorbed by ice}
As shown in Figure \ref{fig:opacities_on_widebands}, F444W is absorbed by several ices, but because the most abundant species (CO$_2$ and CO) have narrow bandwidths, the total effect is small.
Water ice and \ammonia ice both absorb broadly in the F444W band, but with lower amplitude than F356W.

\subsection{Short-wavelength bands are unaffected by ice}
\label{sec:shortwavelength}
At wavelengths short of about 2.5 microns, ices have very little effect.
Water exhibits some opacity peaks at 1.5 and 2.0 microns, but with absorbance per molecule three orders of magnitude below the peak.
To date these features have only been detected in the lab and in solar system objects \citep{2008AJ....135...55B,2024NatCo..15.8247P,2025PSJ.....6..154M}.

\subsection{Others: F277W, F300M, F323N, F335M, F360M, F430M, F460M, F470N, F480M}
We briefly discuss additional colors that have been used in other projects.
Most JWST bands longward of 2 microns have some overlap with ice bands.
We limit our exploration of these bands to those we and others have observed, but we note that our models can easily be applied to any JWST filter using the \texttt{icemodels} package.

Appendix \ref{sec:moreabsorptionplots} shows these filters with a few relevant opacities overlaid.
F277W, \referee{F300M}, F323N, and F335M all lie atop the strongest water ice absorption feature and are very sensitive to the presence of water ice.
The F360M filter lies on a minimum in the lab-measured \water ice feature, suggesting that it should be a good reference filter against which to measure other ice features, but the \referee{CH$_3$} ice feature (discussed in \S \referee{\ref{sec:f356w} and} \ref{sec:f356wf444w}) at this wavelength suggests that deeper knowledge of the ice composition is needed before trusting this filter as a reference.
F480M is relatively unaffected by ice absorption, with effective opacity from common ices down by 1-2 orders of magnitude from their peak; the F480M filter covers the CO absorption feature, but \referee{that feature} is too narrow to have substantial effect on F480M colors.

F430M, F460M are not used in any Galactic Center projects.
F430M \referee{is very sensitive to CO$_2$, which covers $>50\%$ of the band}.
F460M is moderately sensitive to CO:\referee{ like F466N, it completely covers the CO absorption feature, but the absorption is diluted by the broader bandwidth of this filter.}
Our modeling of these filters in this work is limited to the rough estimates in Table \ref{tab:nircam_ice_absorption} and \ref{tab:nircam_ice_absorption_19}.

\begin{table*}[ht]
\centering
\caption{Ice molecules that significantly absorb NIRCam filters [N(ice)=10$^{18}$ cm$^{-2}$]}
\label{tab:nircam_ice_absorption}
\begin{tabular}{|l|p{10cm}|p{6cm}|}
\hline
\textbf{Filter} & \textbf{Ice Molecules} & \textbf{$\Delta$mag Values} \\
\hline
F277W & H$_2$O & 0.16 \\
F300M & H$_2$O, CH$_3$OH & 0.5, 0.26 \\
F323N & H$_2$O, HCN, CH$_3$OH & 0.45, 0.31, 0.2 \\
F322W2 & CH$_3$OH, H$_2$O & 0.11, 0.11 \\
F335M & CH$_3$OH, HCN, H$_2$O & 0.23, 0.14, 0.13 \\
F356W & CH$_3$OH & 0.14 \\
F410M & CO$_2$ & 0.29 \\
F430M & CO$_2$ & 0.9 \\
F444W & CO$_2$ & 0.17 \\
F466N & CO & 0.25 \\
F480M & OCS & 0.58 \\
\hline
\end{tabular}
\par Molecules that absorb NIRCam filters by at least 0.1 mag when their column density is 10$^{18}$ cm$^{-2}$.  This table is not comprehensive, since some molecules are potentially much more abundant (e.g., \water), and the more complex molecules are likely to be rarer.  NIRCam filters excluded from this table do not have significant ($>0.1$ mag) ice absorption at N(ice)=10$^{18}$ cm$^{-2}$.  Several molecules in the ice database are excluded because they have not been reported in the ISM, including NH$_4$CN, N$_2$H$_4$, and HC$_3$N.  
\end{table*}

\section{The effects of ice absorption on stellar colors}
\label{sec:colors}

\referee{In this section, we show how specific colors, i.e., differences in magnitudes between filters or flux ratios between filters, can be used to identify and measure ices.}
\referee{While this approach is quite general, we focus on} the filters used to observe the Brick in projects 1182 and 2221. 

Our aim is to determine which ices are responsible for the observed colors that cannot be explained by dust.
\referee{We attempt to isolate ice-affected filters by selecting colors in which one filter is not absorbed, or only weakly absorbed, by ices, while the other is strongly absorbed.}
We progress from narrow to broad filters, since the narrow filters have more specific diagnostic power.

In each subsection below, we show color-color diagrams including \referee{photometry} from the observational programs overlaid with a series of models picked to illustrate key points.
We evaluated a much wider range of models than shown here: for example, we evaluated models of pure CO, of \water and CO with no CO$_2$, and of pure \water.
Each of this class of models was effectively ruled out by one of the \referee{photometric} data sets below; e.g., \S \ref{sec:f405nf410m} demonstrates that CO$_2$ must be present in the ice mixture.
We therefore further considered only mixtures of ices that \referee{are consistent with the data.
We do not, however, require that the models reproduce colors including F356W because we found no suitable fits to those colors, which is in part because there is inadequate laboratory data for the ice mixes that could account for this band's absorption.}

In each of the color-color diagrams (Figures \ref{fig:colorcolor}-\ref{fig:icevsice_466356}), we plot along the X-axis a color that represents mostly or entirely dust extinction.
For the 2221 \referee{photometry}, that preferred color is F182M-F212N, selected because \referee{all filters with wavelength longer than $\lambda \gtrsim2.5 \um$ are}, to some degree, absorbed by ices.
For the 1182 \referee{photometry}, the F115W-F200W color is unaffected by ice, but it also excludes many sources detected at longer wavelengths but too extincted to be seen at 1.15 \um, so we also show F200W-F356W and F200W-F444W, even though they may be `contaminated' by ice.

The Y-axes selected are chosen to be ice-absorbed, except where otherwise noted.
To a rough first approximation, vertical position in the plots indicates degree of ice absorption; the extinction vectors (yellow arrow) show the expected color change due to dust, and it is most strongly along the X-axis for most of these CCDs.
However, the model ice curves are not vertical lines because they encode an assumption that the column density of a modeled ice is proportional to the column density of hydrogen, which in turn implies that it is proportional to A$_V$.

The plotted models effectively assume that 100\% of the molecular species plotted is in ice form.
That is not a correct model, but is useful for these plots, as most of the spread of the data occurs at higher column density and at higher freezeout fraction.
Models incorporating a freezeout fraction that progresses from 0\% at the cloud edge to 100\% in the interior would have the same endpoints as the plotted models (assuming that the highest measured density is 100\% frozen-out), but would bend slightly at lower A$_V$ to follow the extinction vector.
We treat the cloud edge as universally starting at A$_V=17$ under the assumption that this is foreground dust that does not shield the cloud and does not facilitate ice formation (see Appendix \ref{sec:foregroundav}).
We adopt the \citet{Chiar2006} extinction curve as implemented by \citet{Gordon2021}, which is appropriate for the Galactic Center; we test the effect of this assumption in Appendix \ref{sec:extinctioncurve}.

We selected model curves to display to demonstrate the effects of different assumptions.
No single model fits all of the data, but in order to infer column densities in \S \ref{sec:measuringicecolumn}, we adopt \water:CO:CO$_2$ = 10:1:1.
We show similar mixes, varying only \water$=(1, 2, 3, 5, 10, 15, 20)$, because the lower ratios (\water $<3$) match local clouds, while higher ratios (\water $> 5$) are required to match the Galactic Center and inner galaxy data, as we will show below.
One mix with \water:CO:CO$_2$ = 10:1:0.5 is included to show that factor-of-two variations in CO:CO$_2$ have relatively little impact.

We use the NIRSpec spectra as a well-characterized control sample.
We plot synthetic photometry derived from the NIRSpec spectra as red \texttt{x}'s, \referee{blue $\tridownmarker$'s,} and purple +'s in Figure \ref{fig:colorcolor} (see \S \ref{sec:spectra}).
In all of the currently-available JWST archival spectra of Galactic sources in PRISM mode, the sources do not follow the path of the Brick data, i.e., the NIRSpec sources are either red, not blue, in the F410M-F466N colors, or they are much less blue at a given $A_V$.
Comparing  the synthetic colors (Fig. \ref{fig:colorcolor}) to the spectra (Fig. \ref{fig:sample_spectra_medbands}) illustrates two key results: red sources in the F410M-F466N color are disk-dominated YSOs, and background sources behind Galactic disk clouds show a qualitatively similar but quantitatively different behavior to the Galactic Center background stars.
The rise from 4-5 \um in the YSOs (purple in Fig \ref{fig:colorcolor}, spectrum in Fig. \ref{fig:sample_spectra_medbands}a) is caused by warm circumstellar dust around YSOs and is not seen in stellar spectra. 
While these YSOs have strong CO, H$_2$O, and XCN features visible in the spectra, their colors in the NIRCam bands are dominated by the rising continuum.
By contrast, the red \texttt{x} points, which are background sources behind the IRAS 16293 cloud, exhibit flat spectra that are consistent with reddened stellar photospheres with ice features on top.
These stars exhibit the same blue-ing effect that is seen toward the Brick sources, though the effect is substantially weaker even at comparable extinction. 

The observed colors require an ice ratio in the Galactic Center clouds that is different from that in local clouds.
Several model curves are plotted in Figure \ref{fig:colorcolor} showing different combinations of \water, CO$_2$, and CO ice. 
\citet[][project 1309]{Smith2025} measured the H$_2$O:CO:CO$_2$ ratios using background stars in the Chameleon I cloud, finding H$_2$O:CO $\sim2.5$ and CO:CO$_2\sim$ 1:0.8, as shown by the blue dashed line in Figure \ref{fig:colorcolor}.
Such an ice mixture is incompatible with the Galactic Center data, but matches reasonably well both the IRAS 16293 and Serpens data sets.


\subsection{F410M-F466N and F405N-F466N}
\label{sec:colors466410}

\begin{figure*}
\centering
\includegraphics[width=\linewidth]{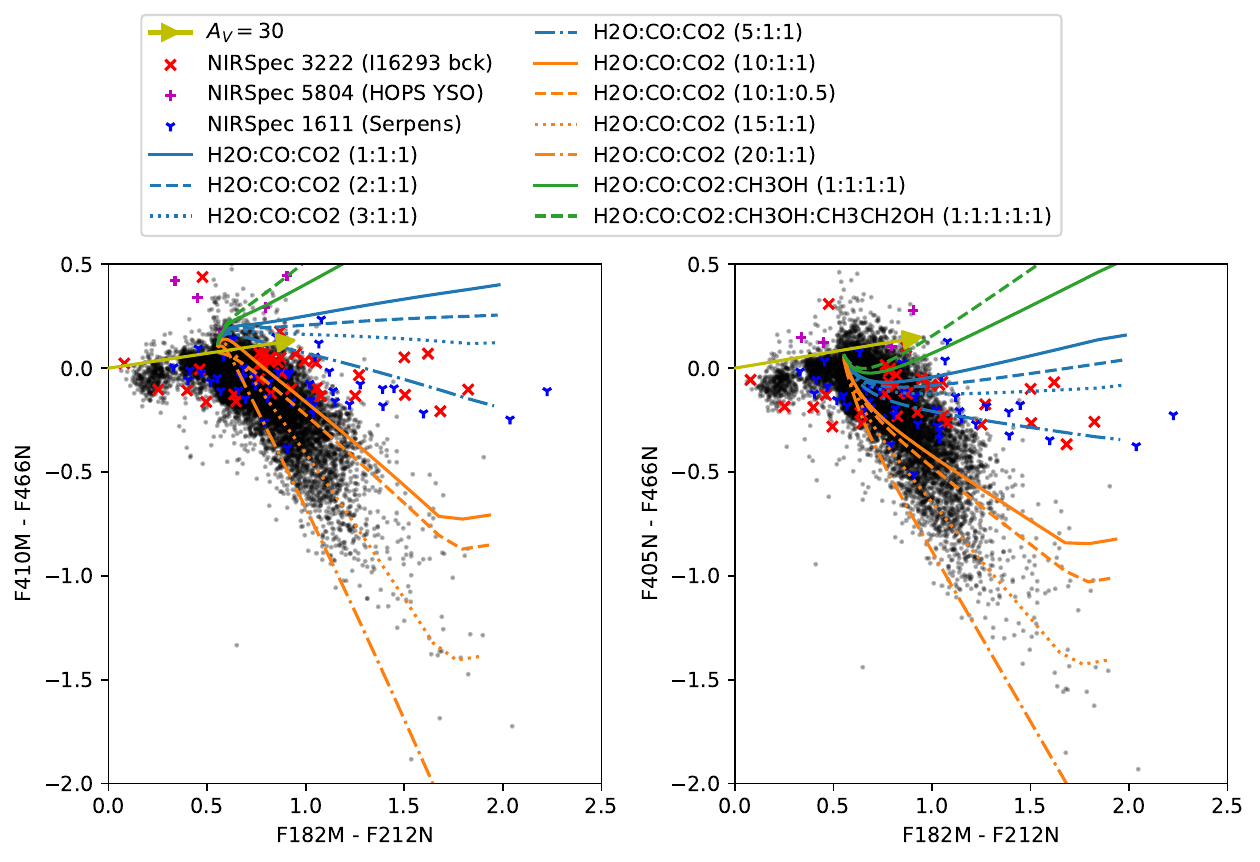}
\caption{
\referee{The distributions of Brick sources in the [F405N]$-$[F466N] and [F410M]$-$[F466N] vs [F182M]$-$[F212N] color-color spaces compared to the intrinsic estimated colors of ice mixtures as described in \S \ref{sec:colors} and indicated with curves of different colors as shown above. }
These diagrams show \referee{photometry from} narrow- and medium-band filters used in JWST program 2221.
The model curves show the effect of adding the labeled ice mixtures to the absorber starting at A$_V=17$ with a fixed CO/H$_2=2.5\times10^{-4}$ up to a maximum column density of N(H$_2)=5\times10^{22}$ \persc.
As indicated in the legend, synthetic photometry has been computed for background stars behind local clouds in the I16923 observations from JWST project 3222 and the Serpens cloud from project 1611, and for YSOs from the Orion region from project 5804.
These are plotted as red x's, blue y's, and purple +'s, respectively.
}
\label{fig:colorcolor}
\end{figure*}


Figure \ref{fig:colorcolor} was shown in \citet{Ginsburg2023} as evidence of the effect of ice, in that this color becomes bluer instead of redder with increasing extinction.
The effect is abundantly clear here, where data follow along the extinction vector until roughly 0.75 magnitudes on the X-axis, then they start trending downward.
This trend occurs in spite of ice opacity affecting both bands; evidently, the absorption in F466N from CO dominates over the absorption from CO$_2$ in F410M (see \S \ref{sec:f405nf410m} for further validation), and F405N is unaffected by CO$_2$.

A wide range of water plus CO ice mixtures match the data well, though these models put constraints on the adopted ice mixture.
The colored curves in Figure \ref{fig:colorcolor}, and in future color-color diagrams, are computed by assuming CO/C = 0.25 and C/H = $2.5\times10^{-4}$.
Each curve is shown from a column density of effectively zero up to a maximum total column density N(ice) $=2\times10^{20}$~\persc. 

\subsubsection{The 3 kpc arm cloud}

\begin{figure*}
\centering
\includegraphics[width=\linewidth]{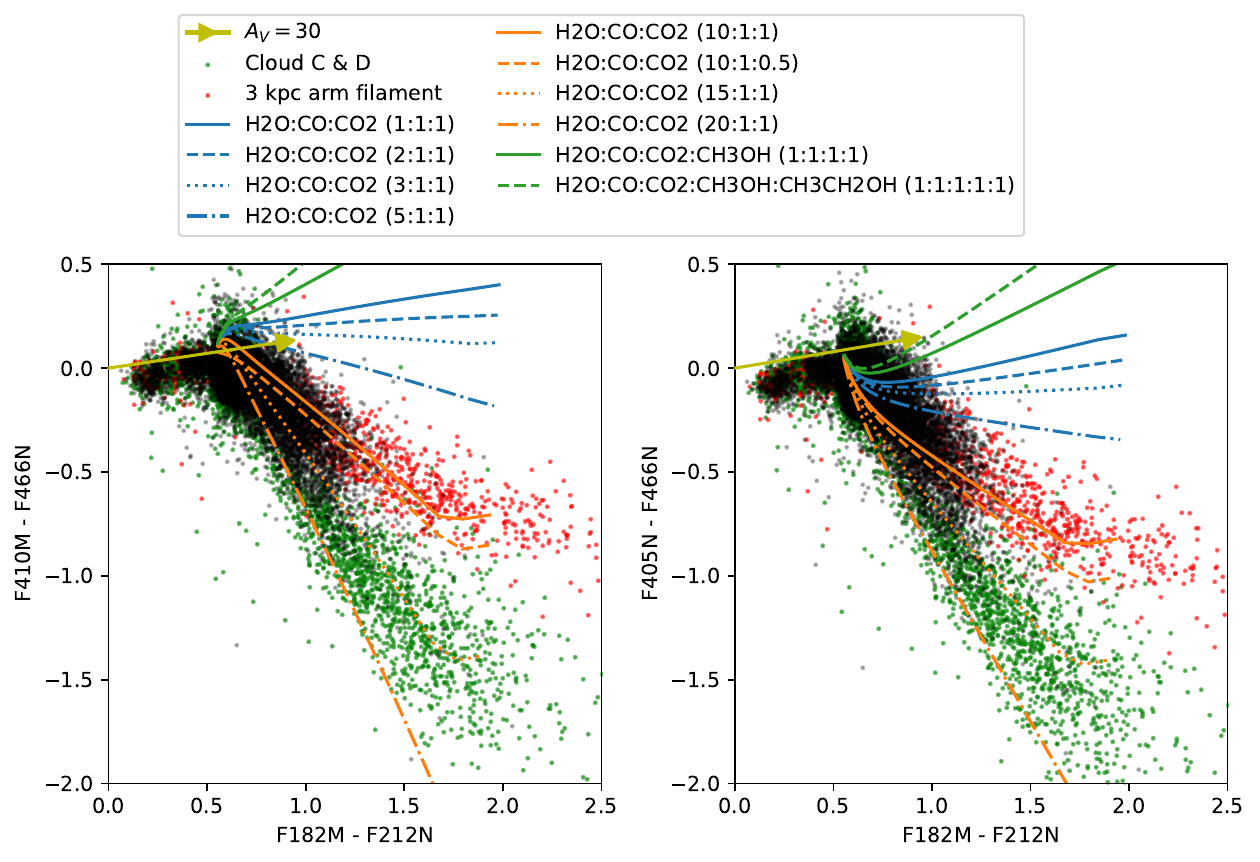}
\caption{
\referee{The distributions of Brick, Cloud C, and Cloud D sources in the [F405N]$-$[F466N] and [F410M]$-$[F466N] vs [F182M]$-$[F212N] color-color spaces compared to the intrinsic estimated colors of ice mixtures as described in \S \ref{sec:colors} and indicated with curves of different colors as shown above.}
The legend and models are the same as in figures \ref{fig:colorcolor} and \ref{fig:ccd_f405nmf410m}.
}
\label{fig:colorcolorwithcloudc}
\end{figure*}

The difference between the spectroscopic data from local clouds and the photometric measurements from the Galactic Center already indicates that there is a range of ice properties in our Galaxy.
As further validation of both this physical interpretation and of our methodology, we show an additional comparison between objects in the same data set.
A foreground cloud was serendipitously observed in project 2221 \citep{Gramze2025}, and it shows a clear difference from the Galactic Center clouds.
Figure \ref{fig:colorcolorwithcloudc} adds two additional data sets to Figure \ref{fig:colorcolor}: Cloud C, a dust ridge cloud, is shown with green points that fall along roughly the same locus drawn by extrapolating the Brick data.
By contrast, the 3 kpc arm cloud shows less of the blue-ing effect in these colors than the Galactic Center clouds, closer to what is seen in the Serpens and IRAS 16293 cloud stars, but nevertheless still different.
A roughly 5:1:1 \water:CO:CO$_2$ ratio matches the 3 kpc arm cloud, while 10:1:1 is a better match to the Galactic Center clouds.

\subsection{F405N-F410M}
\label{sec:f405nf410m}
\begin{figure}
    \centering
    \includegraphics[width=1\linewidth]{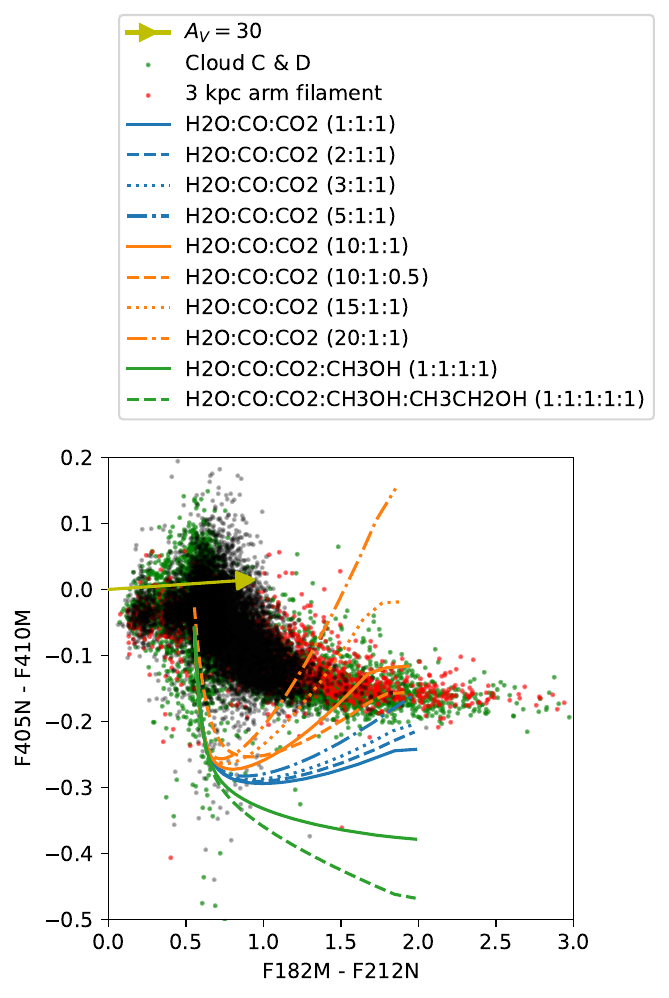}
    \caption{
\referee{The distributions of Brick, Cloud C, and Cloud D sources in the [F405N]$-$[F410M] vs [F182M]$-$[F212N] color-color space compared to the intrinsic estimated colors of ice mixtures as described in \S \ref{sec:colors} and indicated with curves of different colors as shown above. }
    }
    \label{fig:ccd_f405nmf410m}
\end{figure}

The F405N-F410M color is sensitive primarily to the 4.3~\um CO$_2$ feature, which absorbs F410M but not F405N.
Figure \ref{fig:ccd_f405nmf410m} shows this effect empirically, plotting F405N-F410M color against the extinction-sensitive F182M-F212N color.
The F405N excess (negative Y-axis values) at X-axis values above F182M-F212N $>0.5$ shows that CO$_2$ ice is forming at roughly the same column density as CO ice and confirms that it is present.

Figure \ref{fig:ccd_f405nmf410m} shows that the simple models adopted here, which have N(ice)$\propto$N(H$_2$) above some fixed $\mathrm{N}(\mathrm{H}_2)$ threshold, do not match the data.
They all predict the correct general behavior at low column densities, but they over-predict the relative color at higher column densities and predict a turnaround that is not observed.
This `turnaround' behavior is caused by the main CO$_2$ line saturating, after which the line wing becomes a significant source of absorption, and this wing overlaps the F405N filter (Fig. \ref{fig:f410mplusopacities}a).

The lack of the `turnaround' feature in the data suggests that CO$_2$ does not reach total saturation and does not affect the F405N filter substantially.
The spectra (like Figure \ref{fig:sample_spectra_medbands}b) corroborate this interpretation, showing nearly 100\% absorption within the CO$_2$ band but no absorption at 4.05 \um.
From Figure \ref{fig:f410mplusopacities}b, this observation implies that N(CO$_2$) $<10^{18}$ \persc.
That upper limit is reached relatively quickly, at [F182M]-[F212N] = 1.0, or N(H$_2$) = $1.5\times10^{22}$ \persc (adopting our standard assumptions), above which it remains constant.
This behavior appears common to both the Galactic Center data and the Galactic disk data; unlike in colors involving F466N, the photometric measurements largely overlap with the spectroscopic measurements (not shown).
The abundance of CO$_2$ relative to H$_2$ therefore appears to decrease at higher column density, which may be because it is chemically converted into more complex species.\footnote{\referee{An alternative possibility is that the dust grains grow large enough that they can be treated as solid bodies, i.e., into the geometric optics regime where $a_{grain} > \lambda/2 \pi \sim 0.8$ \um, which would hide the ice features.  In such a case, however, \emph{all} ice features would be reduced at higher column densities, which is not observed.}}

We note that this `saturation' problem is a general feature of the modeling approach we have adopted.
In the color-color diagrams, all model curves are drawn assuming that the ice increases linearly with hydrogen column density.
If this assumption does not hold --- e.g., if the species has entirely frozen out at some column density --- the curves change shape.
More sophisticated models of the freezeout process and molecular abundances are likely to present better matches to these data.




\subsection{F356W\text{-}F444W}
\label{sec:f356wf444w}
\begin{figure*}
\centering
\includegraphics[width=\linewidth]{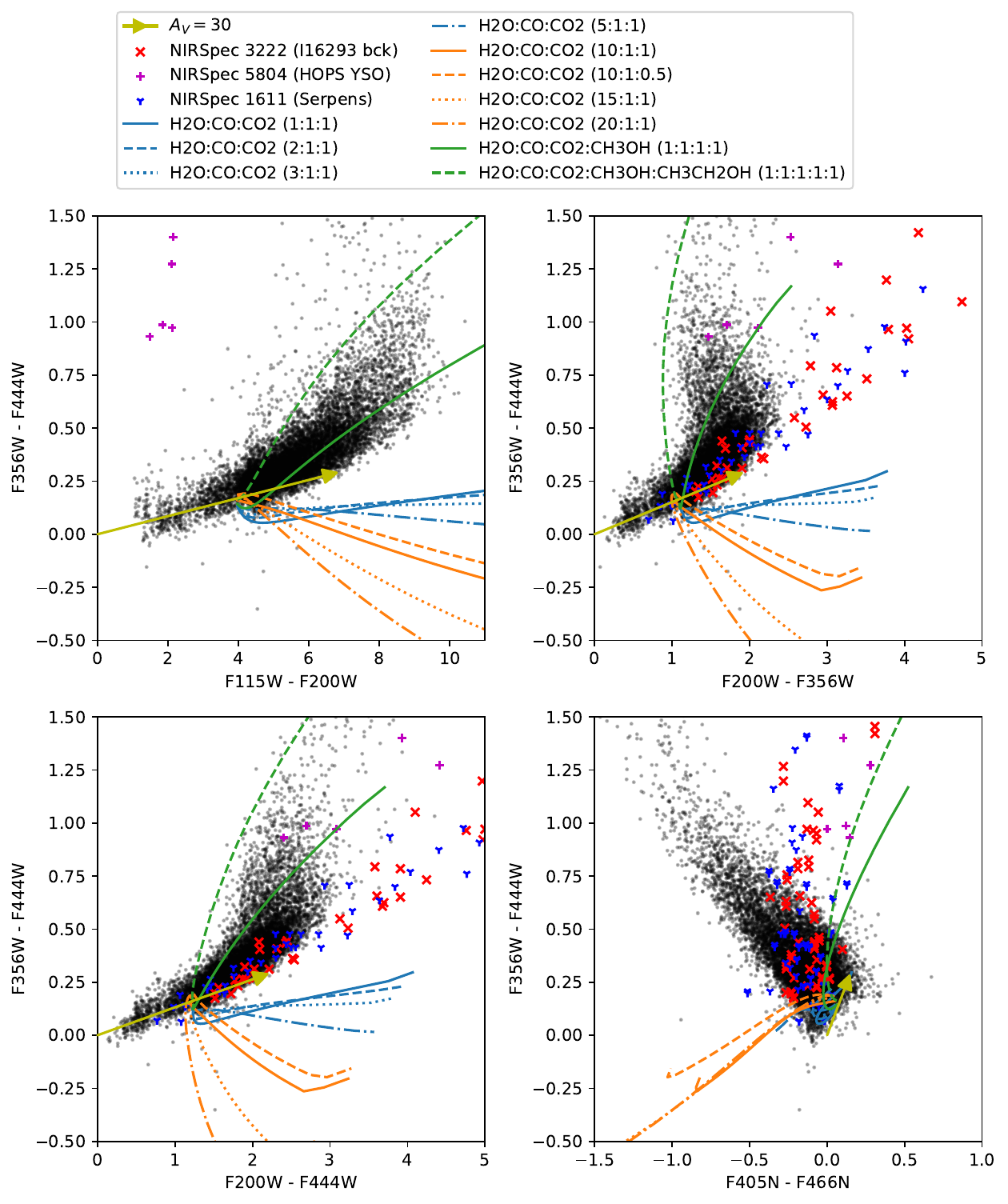}
\caption{Color-color diagram like in Figure \ref{fig:colorcolor} but for the wide-band filters.
The F356W-F444W color shows excess reddening compared to dust, and that excess is only reproduced if the ice is predominantly ethanol or some other complex methylated molecule (CH$_3$CHO, CH$_3$OCH$_3$, etc.).
The bottom-right panel shows a color-color diagram comparing the F405N-F466N color to the F356W-F444W color.
The F405N-F466N color is sensitive to CO ice, while the F356W-F444W color is apparently sensitive to \methanol ice.
Unlike the other color-color diagrams, both axes are sensitive to ices and weakly sensitive to extinction.
\referee{The excess absorption in F356W produces redder F356W-F444W colors, while the excess absorption in F466N produces bluer F405N-F466N colors.}
The good (anti)correlation between these colors indicates that similar conditions are producing both features --- i.e., most likely, colder and denser gas and dust.
}
\label{fig:colorcolorwide}
\label{fig:icevsice_466356}
\end{figure*}

In project 1182, the wide-band filters F115W, F200W, F356W, and F444W were observed.
The F356W-F444W vs. F115W-F200W color-color diagram (Figure \ref{fig:colorcolorwide}) shows that there is excess reddening in the F356W-F444W color compared to expectation from extinction.
This excess reddening is caused by excess absorption in the F356W band, as color-color diagrams with the other filters show linear correlation.
In the {F200W-F356W} vs. {F356W-F444W} color-color diagram (Figure \ref{fig:colorcolorwide}b), a blue population in F200W-F356W is apparent above F356W-F444W~$>0.75$.
Spatially, these sources occur at the edge of the cloud, roughly in the same locations as the maximum observed CO ice (i.e., bluest F405N-F466N colors), but more concentrated toward the cloud center (see \S \ref{sec:spatial}).
In Figure \ref{fig:icevsice_466356}, we show F356W-F444W vs. F405N-F466N colors to demonstrate that the F356W excess absorption is correlated with the F466N excess absorption, i.e., the F356W absorption is correlated with CO ice column density.


We examined the color effects of all of the ices in the online ice databases (LIDA, OCDB) and found that the only species that specifically absorb F356W are methyl-bearing species, e.g., methanol (\methanol) and ethanol (\ethanol).
Other species either do not affect these bands or preferentially extinguish F444W, producing the opposite of what we observe (e.g., \water).
Water mixed with \methanol ice may produce a similar signature to \methanol, preferentially absorbing F356W, though that behavior appears only in extremely high-abundance \methanol mixes (i.e., \water:\methanol $\lesssim 2$).
While there are ices that absorb F356W, we found none that combine to match the correlation seen in Fig. \ref{fig:icevsice_466356}.

The Galactic Center data exhibit colors distinct from those seen in the spectroscopic surveys.
While many of the YSOs from HEFE show F356W-F444W colors as red as seen toward the Brick, inspection of their spectra shows that this is caused by red excess emission, not by ice absorption: the HOPS sources all rise consistently from 2-5 \um.
There are only two HOPS sources with high signal-to-noise and colors similar to the GC sources, HOPS-383 and HOPS-394, both of which have consistently rising spectra across the entire spectral band; these sources are moderately red in [F405N]-[F466N] colors, even though they emit in Br$\alpha$, while all the GC sources are blue in these colors, thus the possibility that the GC sources are all similar YSOs is excluded.
No sources with high signal-to-noise ($m<20$) from projects 1611 or 3222 exhibit colors similar to the Galactic Center sources.
These results combine to suggest that there is a very large excess of \methanol in the ices in the Galactic Center at high column densities.
\citet{Jang2022} measured \methanol and \water ice toward a handful of sources, but found \water:\methanol $\sim4-8$:1, lower than required to explain the more extreme observed colors, though they also found \methanol:CO$_2$ $\approx2.9$, indicating a very high abundance of \methanol.
The conclusion that the F356W absorption excess is caused by \methanol will need to be tested with spectroscopy, as we cannot be as confident about ice contributions in the broad-band filters as in the narrow.
Nevertheless, with a clear identification in hand from spectroscopy, or corroboration from MIRI observations of \methanol bands at $>5$ \um, it may be possible to quantify and spatially resolve the \methanol ice abundance.



%

\section{Discussion: Measuring ice column density}
\label{sec:measuringicecolumn}

The color excess in bands affected by ice absorption is related to the column density of that ice.
\referee{In this Section, we analyze the color excess and column density for CO ice in particular because its absorption profile} is well-matched to the F466N band and less blended with other ice features than CO$_2$, \water, or \methanol.

\subsection{Measuring CO abundance}
We measure the column density of CO ice by computing the color excess ($\Delta m$) of CO in the F466N band.
To infer that \referee{the observed flux represents absorption}, we need to know \referee{a given} star's intrinsic brightness.
We use F405N as the reference filter, as it is the closest in color that is not strongly affected by other abundant ices (see \S \ref{sec:f405n} above); F410M is also a good reference, but it is subject to substantial CO$_2$ absorption (see \S \ref{sec:f405nf410m}).
F405N can have excess emission from Br$\alpha$, but such emission is rare in stars.
From the equations in \S \ref{sec:rteqns}, we compute the $\Delta m$ as a function of column density.
We then measure the F405N-F466N color and de-redden it using the F182M-F212N color, adopting the \citet{Chiar2006} extinction curve \referee{assuming all stars have zero intrinsic color} (see \S \ref{sec:extinctioncurve} for discussion of the uncertainty).

\begin{figure}
    \centering
    \includegraphics[width=\linewidth]{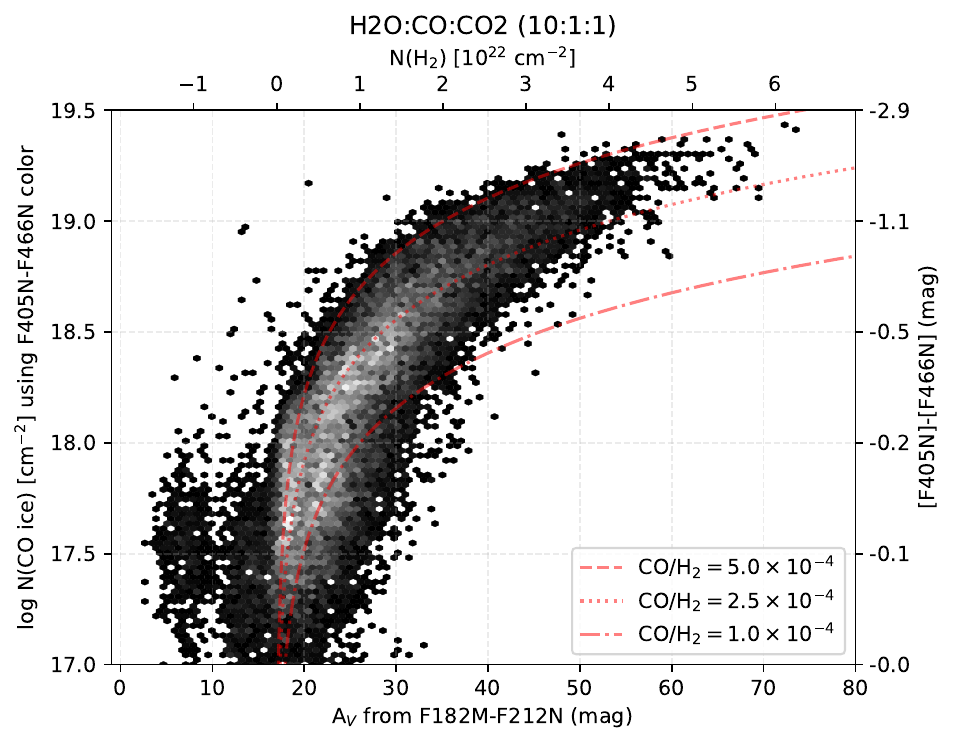}
    \caption{Reproduction of Figure 9 of \citet{Ginsburg2023}, but with an assumed mix of H$_2$O:CO:CO$_2$ = 10:1:1 and using F405N in place of F410M,
    an assumed foreground $A_V=17$, and corrected color zero points for each included filter.
    The red lines indicate CO/H$_2$ fractions of $5\times10^{-4}$, $2.5\times10^{-4}$, and $1\times10^{-4}$ in dashed, dotted, and dash-dotted, respectively.
    The right axis shows the dereddened measured color in the F405N-F466N filter; it is not linear.
    }
    \label{fig:fig9colvsav}
\end{figure}

While no single track through Figure \ref{fig:colorcolor} can accommodate all of the data, and no plausible tracks explain any of the wide-band colors shown in \ref{fig:colorcolorwide}, the H$_2$O:CO:CO$_2$ = 10:1:1 line roughly traverses most of the Brick data from project 2221.
We therefore adopt this ratio of ice species to infer the column densities of the observed molecules.

Figure \ref{fig:fig9colvsav} shows the inferred ice column density as a function of $A_V$ and $N(H_2)$ as in \citet{Ginsburg2023}.
It corrects for some systematic errors in that work's color calculations that result in all data being shifted to lower $A_V$.
It also includes a foreground, ice-free $A_V$ component (up to $A_V=17$), which better matches the data.
Sources in this plot with N(CO) $<10^{18}$ \persc have $\Delta m < 0.04$ and therefore are dominated by noise; the points to the left of the curve at low column are foreground sources with no CO detection but for which measurement error has created a spurious signal.
A detailed analysis of the uncertainties in this plot is provided in Appendix \ref{appendix:columnuncertainties}.
In brief, there is systematic uncertainty of up to $\sim0.3\times$ in N(CO) from the combination of choice of extinction curve and ice mixture, which amounts to a fairly narrow allowed range of parameter space.

What is the origin of the scatter in Figure \ref{fig:fig9colvsav}?
A likely explanation for the failure of any single line to cover all data is that the $A_V$ at which ice starts to form varies.
This is expected if there is not one single line-of-sight into the cloud, but several different depths into one cloud or several distict clouds \referee{- i.e., the Brick is `porous'} \citep{Henshaw2019}.
Within the Galactic center, there are likely some ice-rich and ice-free components of the line of sight.
At the highest column densities, it is more likely that the gas density is also higher, which can drive the transformation of CO into more complex species.
In this case, the ice mixture as a function of column density is likely not to be constant, as we have assumed.

\subsection{The fraction of carbon in ice}
\label{sec:carbonisinice}
One of the startling results of \citet{Ginsburg2023} \referee{and this work} is that the abundance of CO ice matches or even exceeds the maximum allowed based on \referee{locally} measured carbon abundances.
This excess is particularly significant because it ignores the gas-phase CO, which is observed and is \referee{generally assumed} to be the dominant phase of CO in most of the volume of molecular clouds \citep{Bergin2007}.

The evidence for a high fraction of carbon in CO ice is given in Figure \ref{fig:fig9colvsav}.
This figure shows curves of constant CO ice to H$_2$ ratio.
About half of the highly-extincted (A$_V>20$) stars are over the CO/H$_2 = 2.5 \times10^{-4}$ line, and nearly all are above CO/H$_2 = 1 \times10^{-4}$.
Clearly, the Galactic Center has higher CO abundance than the solar neighborhood.

To provide a constraint on the carbon, we assume that carbon is the limiting ingredient in CO, since the observed oxygen abundance in all environments exceeds that of carbon.
Gas-phase carbon abundance in the solar neighborhood is typically measured to be [C/H+12]=8.3-8.5, i.e., C/H $\approx2-3\times10^{-4}$ \citep{Nieva2012,Arellano-Cordova2020}.
If one were to adopt this Solar Neighborhood abundance C/H=$2\times10^{-4}$ ([C/12+H]=8.3), the fraction of carbon in CO ice would exceed 100\% for many lines-of-sight, which confirms that the Galactic Center carbon abundance is enhanced.

The Galactic Center has higher-metallicity gas than the solar neighborhood.
If we adopt $Z_{GC}=2.5Z_\odot$, on the high end of what has been observed or inferred \citep{Do2015,Nandakumar2018,Arellano-Cordova2020}, and that the carbon abundance scales with metallicity, we constrain the fraction of CO frozen into ice.
The data show that a large fraction of background stars have CO/H$_2>2.5\times10^{-4}$, however, implying either that a larger fraction of carbon is in CO in the Galactic Center or that carbon abundance is higher than implied by metallicity alone.
The CO/H$_2=5\times10^{-4}$ line, which traces the upper envelope of the data in Figure \ref{fig:fig9colvsav}, shows the consequence of assuming 100\% of carbon is in CO ice if carbon scales linearly with metallicity \referee{\citep[though][show that the extrapolated C/O ratio in the GC could be higher than the solar neighborhood by as little as 10\% or as much as $2.5\times$]{Mendez-Delgado2022}}.
Adopting that line as the upper limit of what is plausible, the implied fraction of carbon in CO ice is at least $\gtrsim50\%$ above A$_V\gtrsim30$.

\subsubsection{GC ice is different: Comparison to a local cloud (Chameleon I)}
\label{sec:chameleon}
In this section, we compare the Galactic Center cloud measurements to local cold clouds.
While the freezeout fraction of CO is similar, the ice composition is significantly different.

\citet{Smith2025} reported JWST NIRCam WFSS spectroscopic measurements toward 44 background stars behind the Chameleon I molecular cloud, providing gold-standard measurements of the CO, CO$_2$, and H$_2$O abundances, which we can compare to our photometric observations.
Figure \ref{fig:fig9withother} includes their measured CO column density from NIRCam WFSS spectroscopy as a function of H$_2$ column density.
They inferred N(H$_2$) from \textit{Herschel} PACS/SPIRE spectral energy distribution fits rather than from extinction, as we have.
In the plot, we have converted this N(H$_2$) to A$_V$ and shifted by +17 to match the Galactic Center data.
Only data with CO column uncertainty $<20\%$ are plotted.
The \citet{Smith2025} data imply 10--50\% of carbon (20--100\% of CO) is in CO ice along each line of sight probed, even at low $A_V$ where it might naively be expected that a larger fraction of CO is in the gas phase. 
Even with such large fractions of frozen-out CO, their data fall far below the Galactic Center and 3 kpc arm clouds.

\citet{Smith2025} found that the ratio between CO, CO$_2$, and H$_2$O ices is small, with fitted H$_2$O:CO:CO$_2$ $\approx$ 2.6:1:0.8.
Figure \ref{fig:colorcolor} shows that such ratios are incompatible with the Galactic Center ices; the dashed blue line approximates the Smith+ abundance with H$_2$O:CO:CO$_2$ = 3:1:1, and that line misses most of the Galactic Center data.
The GC data show higher \water fractions, which may indicate a higher O/C ratio in GC gas.
Appendix \ref{appendix:icemixes} shows that assuming lower \water:CO ratio for the GC gas would increase the CO abundance, widening the discrepancy with the \citet{Smith2025} data, so that is not a viable explanation for the excess GC abundance.
Appendix \ref{app:kp5} compares our GC data to the KP5 \citep{Pontoppidan2024} model, which was fitted to local YSOs, and shows that it is also a poor fit \referee{to the Galactic Center data we present}, though it reasonably matches the local cloud background stars.

%

\subsection{Ice abundance variations within the Galactic Center}
\label{sec:icevariations}

\begin{figure}
    \centering
    \includegraphics[width=\linewidth]{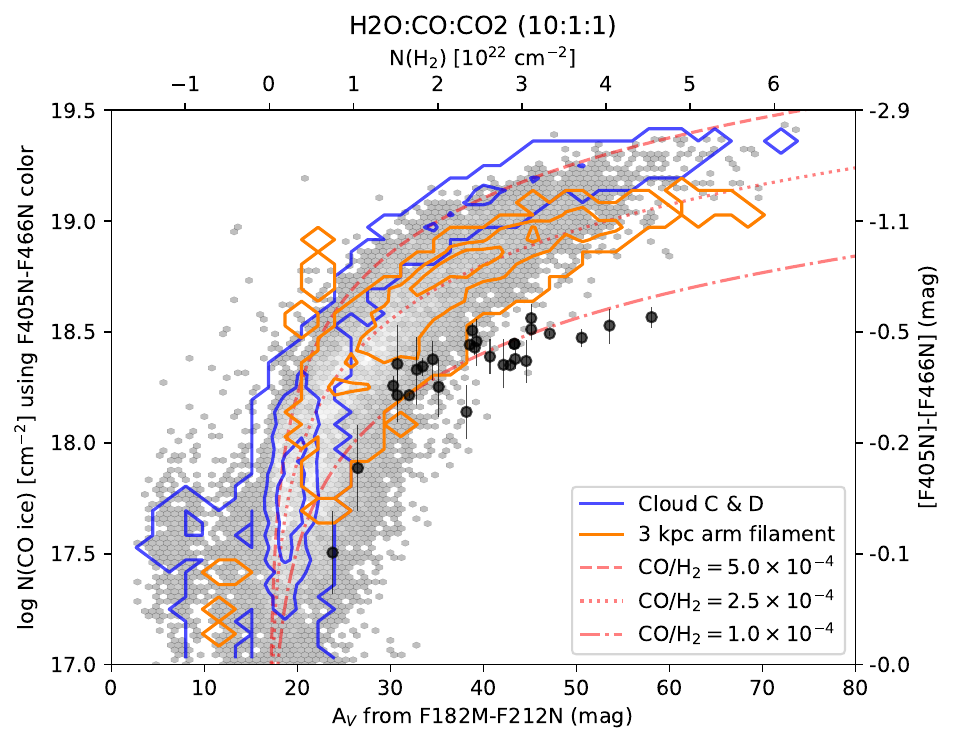}
    \caption{
    \referee{As Figure \ref{fig:fig9colvsav}, but  with the analogous measurements for Galactic Center Clouds C and D and the foreground 3\,kpc arm cloud overlaid for comparison.}    
    The black data points show measurements from \citet{Smith2025} of the Chameleon cloud.  They have been artificially shifted to the right by A$_V=17$ to match the assumed foreground in the Galactic Center data.
    The right axis shows the dereddened measured color in the F405N-F466N filter; it is not linear.
    }
    \label{fig:fig9withother}
\end{figure}

The measurements from project 2221 include an additional region, the Cloud C/D component of the Galactic Center's dust ridge, and a foreground infrared dark cloud.
These data are presented in \citet{Gramze2025}.
Figure \ref{fig:fig9withother} shows these data overlaid on the Brick data.
There are evidently environmental differences, both within the Galactic Center (i.e., between Cloud C and the Brick) and between the GC and clouds outside the center.
Cloud C has marginally higher ($\sim0.1$ dex) ice column density at all A$_V$, while the 3 kpc arm cloud has substantially lower ($\sim0.3$ dex) values, especially at high column density.
Notably, all probed regions have higher CO ice abundance than the local Chameleon cloud (\S \ref{sec:chameleon}).

\subsection{Metallicity is the best explanation}
\label{sec:measuringmetals}
Section \ref{sec:carbonisinice} demonstrated that, throughout much of the Brick lit by background stars, the fraction of carbon in CO ice exceeds 50\%.
Section \ref{sec:icevariations} confirmed that this is true throughout the Galactic Center dust ridge and even well into the Galactic disk.
Furthermore, it showed that there are clear environmental differences, with CO ice fractions in the GC larger than those in the 3 kpc arm and both larger than the solar neighborhood.
We evaluate the different assumptions below and conclude that the most parsimonious interpretation is that the carbon abundance with respect to hydrogen, and in turn the abundance of CO, is significantly and measurably higher in the Galactic Center.
We evaluate several alternative hypotheses.

First, we check whether the apparent excess CO could be caused by incorrect ice opacity assumptions.
Indeed, as discussed in Appendix \ref{sec:corrections}, \citet{Ginsburg2023} adopted incorrect assumptions.
However, we performed extensive checks, and found that there is no available parameter space that allows this possibility:
the laboratory measurements of CO ice are close enough across different laboratories and a wide range of temperature assumptions that changes to the opacity cannot explain the CO excess.
Allowing other molecules to supplant CO in the opacity curve results in color changes that do not fit the data, as hinted in Figures \ref{fig:colorcolor} and \ref{fig:colorcolorwide}.
We note, however, that there are several opacity curves \citep[e.g., those from ][]{Ehrenfreund1996a,Ehrenfreund1996b,Ehrenfreund1997} that were not available in the online archives any longer\footnote{Some opacity files are recoverable from the internet archive, \url{https://web.archive.org/web/*/https://www.strw.leidenuniv.nl/~ehrenfreund/isodb*}, but others, like the \texttt{E75\_SR.NK} opacity table used in KP5 (Appendix \ref{app:kp5}), are not.}; these modified opacities still should not substantively change the inferred CO abundance, but we were not able to evaluate them quantitatively.
\referee{The structure of ice, whether CO is embedded in water or CO$_2$ ice, can substantially affect the ice opacity profile \citep{Bergner2024,Bergner2026}; we have not accounted for the possible effects yet but will report them in a future work.}

Second, we consider whether the apparent excess in CO abundance is caused by variation in the gas-to-dust ratio (GDR) rather than an increase in the total CO.
It is clear that this explanation cannot completely account for the observations: we directly measure only the CO-to-dust ratio, which is evidently higher in the GC, and only infer the CO-to-hydrogen abundance from the dust.
Nevertheless, our abundance measurements could be systematically affected by our assumed GDR.
A lesser GDR and a greater C abundance, $X(C)$, have the same effect of increasing CO abundance at fixed $A_V$.
Assuming $X(C)$ is proportional to metallicity ($Z$) while the gas-to-dust ratio is inversely proportional to $Z$,
i.e., ${N(\mathrm{H})\propto GDR\propto 1/Z}$ and ${N(\mathrm{CO}) \propto X(\mathrm{C}) \propto Z}$,
\begin{equation}
    X(\mathrm{CO}) \equiv \frac{N(\mathrm{CO})}{N(\mathrm{H})} \sim Z^2.
\end{equation}
The actual relation is unlikely to be linear in both variables, since we show evidence that there is CO excess with respect to dust, but still, the CO abundance should be very strongly sensitive to metallicity.

\referee{
Third, we consider whether varying availability of carbon in the gas phase could be the driving effect.
We noted in \S \ref{sec:introduction} that the total gas-phase carbon abundance in diffuse gas is consistent with a wide range of values, from 50\%-100\%, based on \citet{Jenkins2009} Table 4.
This range technically allows for Galactic Center gas to preferentially form CO, at up to $\sim$twice the rate in the solar neighborhood, without requiring a higher total carbon abundance.
If this scenario held, it would require a significant modification of the assumed dust opacity, since it would require that grains in the GC are entirely non-carbonaceous.  
While not explicitly ruled out by the data, this scenario seems exceedingly unlikely.
}

Finally, we examine whether the choice of extinction curve might be a primary bias.
The main concern is that a different conversion from color (e.g., F182M-F212N) to A$_V$ holds in the GC and the Galactic disk, such that the GC data would appear as a high-A$_V$ extrapolation of the disk data rather than vertically offset in Figures \ref{fig:fig9withother}.
The extinction curve toward the Galactic Center has only limited measurements \citep[][]{Chiar2006,Fritz2011,BravoFerres2025}.
It is particularly challenging because the line of sight to the Galactic Center is long and likely populated with different grains between 0--7 kpc and between 7--8 kpc, i.e., the inner kpc is very likely to have qualitatively different dust properties than the Galactic disk.
We do not have any evidence for a systematic variation in the extinction curve, and indeed find that alternative extinction curves are incompatible with the data (see Appendix \ref{sec:extinctioncurve}).

\subsection{Correlations of CO ice with metallicity and galactocentric radius}
\label{sec:metallicityvscoice}
\begin{figure}
    \centering
    \includegraphics[width=1.0\linewidth]{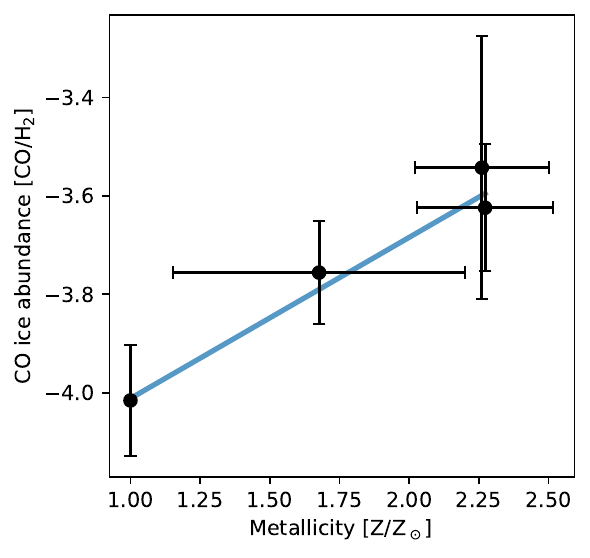}
    \includegraphics[width=1.0\linewidth]{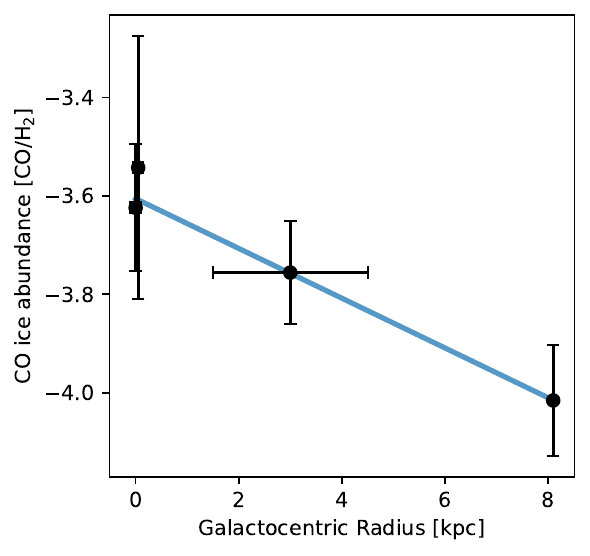}
    \caption{Plot of CO ice abundance vs metallicity (top) and Galactocentric radius (bottom).
    See \S \ref{sec:metallicityvscoice} for details.
    }
    \label{fig:coicevsmetallicity}
\end{figure}

Given the hints that CO ice abundance is correlated with total metallicity, we quantify that relation in this section.
We suggest a relation that may be used to infer metallicity from CO ice abundance measurements, though we caution that it needs further validation.
The results are summarized in Figure \ref{fig:coicevsmetallicity}.

We measure cloud-average CO abundances and compare them to the expected environmental metallicity.
The CO ice abundance we report is the average of the binned medians at A$_V>22$, i.e., 5 magnitudes into the cloud, for the Galactic Center clouds (the Brick and dust ridge clouds) and the 3 kpc arm cloud.
For the local Chameleon cloud, we use the weighted mean and RMS of the data with S/N$>5$ from \citet{Smith2025}, which closely match the standard local assumed CO abundance assuming a high freezout fraction, $\log\mathrm{(CO/H}_2) = -4.02 \pm 0.11$.

For metallicity, we adopt a gradient $-0.044\pm0.05$ dex kpc$^{-1}$, i.e., assuming a linear gradient in log(Z) from Z($R_{gal}=0$)=$2.25\pm0.25$ to Z($R_{gal}=8.1$)=1 in units of $Z_\odot$.
\citet{Mendez-Delgado2022} have found this gradient to fit well the oxygen abundance values for Galactic radii greater than about 5 kpc; measurements at smaller radii are sparse.
We assume the dust ridge and Brick clouds are at the Galactic center and the 3 kpc arm cloud is at 3 kpc from the Galactic center, but note that the uncertainty on that cloud's position is $\gtrsim1$ kpc: it's definitely in front of the Galactic center, and almost certainly within the bar-dominated radius, but the actual distance remains unknown; we adopt an uncertainty of 1.5 kpc for model fitting below.
We assume the Chameleon cloud is at 8.1 kpc from the Galactic center.

We fit a line to the data, finding the CO ice abundance varies as
\begin{equation}
    [\mathrm{CO}/\mathrm{H}_2] = 0.23 (Z/Z_\odot) - 4.25 \\
\end{equation}
The uncertainties on the metallicity are large and systematic, while the uncertainties on the CO ice abundance arise from the data and reflect real measurement uncertainty.
As a more direct fit to the data, we also fit the CO ice abundance as a function of Galactocentric radius, finding
\begin{equation}
    [\mathrm{CO}/\mathrm{H}_2] = -0.04 R_{gal} -3.74
\end{equation}

We also infer the metallicity by adopting a different set of assumptions, specifically that the freezeout fraction in the GC and disk clouds are the same at the same relative depth into a cloud.
Under this assumption, by examining Figure \ref{fig:fig9withother}, the GC metallicity appears to be 2.5-5 times  solar.
We do not measure a single value; as emphasized in Appendix \ref{app:coabundance}, the CO ice abundance and the ratio of ice column densities between clouds varies as a function of depth.
Nevertheless, the average star is at a CO ice abundance roughly 2.5$\times$ solar, which is consistent with other measurements of GC metallicity.
This measurement may be translatable into a more precise measurement with statistical uncertainties once we have a greater understanding of what produces the correlation between abundance and column density, and that understanding may come through improved chemical models, spectroscopy, or both.

While these approaches need further validation by direct measurement of metallicity and CO abundance in the same environment, they have the potential to provide a powerful new tool for metallicity measurement.
Stellar color measurements within and behind dense molecular clouds may be inverted to determine the ice-phase metallicity in the environments most closely linked to star and planet formation.

It is even plausible that this method could be used toward extragalactic cold molecular clouds if the individual stars can be resolved or if the inevitable foreground population can be adequately subtracted.
We note, however, that in extragalactic observations, there is a greater risk of blending surrounding ISM emission lines into the photometric bands (e.g., from recombination lines or PAH emission features) and therefore `filling in' the ice absorption, and therefore added caution should be exercised before using purely photometric data as we have here.

%

\subsection{The spatial distribution of ices}
\label{sec:spatial}
\begin{figure}
    \centering
    \includegraphics[width=0.9\linewidth]{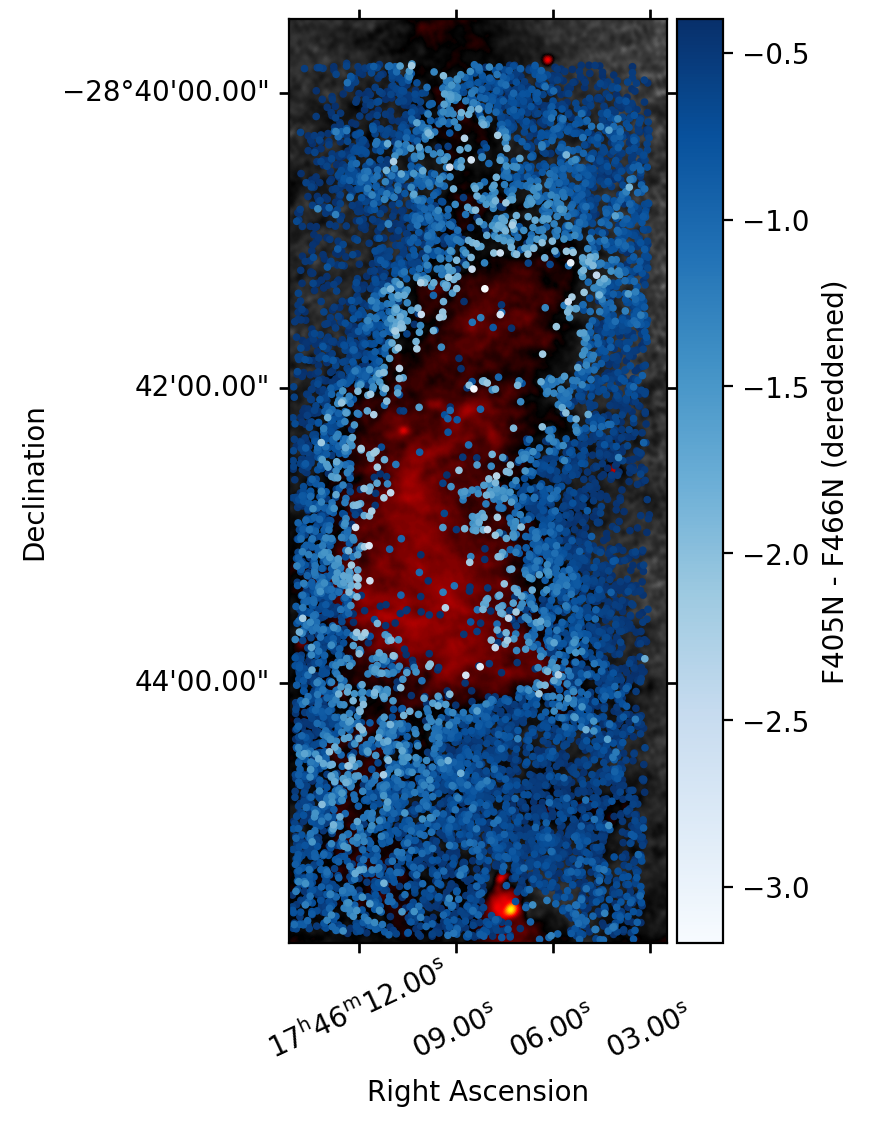}
    \includegraphics[width=0.9\linewidth]{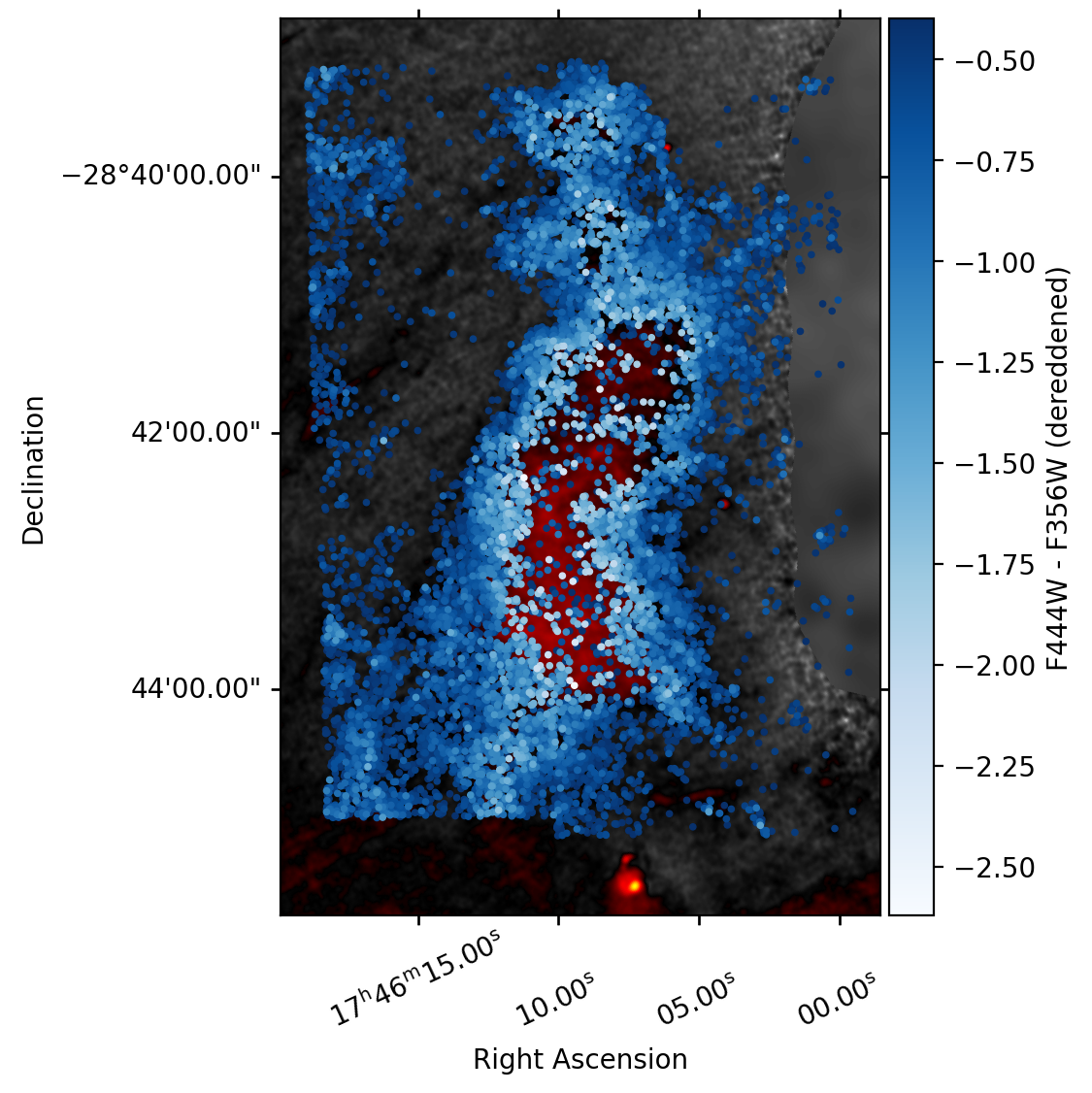}
    \caption{Spatial distribution of the stars selected to have dereddened F405N-F466N $< -0.4$ and F356W-F444W $> 0.4$, respectively.
    The background image is the ACES + MUSTANG-2 image \citep{Ginsburg2026} that combines ALMA with Green Bank Telescope data to show the emission primarily from dust, which roughly traces the column density of the gas.
    The fields-of-view in these images are not the same because projects 2221 and 1182, which covered the narrow- and wide-bands, respectively, had different footprints.
    }
    \label{fig:spatialdistribution}
\end{figure}

The observations we report include spatial information, which we have ignored to this point but now return to.
The Brick data discussed here are limited to A$_V \lesssim 80$ \referee{(N(H$_2$)$\approx1\times10^{23} \persc$)}, but the Brick \referee{contains gas at} column densities factors of several larger \citep{Rathborne2014a}: we simply do not detect stars in F466N in the innermost extincted part of the cloud.
While there is a clear correlation between the F466N and F356W excess extinction, as shown in Figure \ref{fig:icevsice_466356}, the wide-band exposures went substantially deeper and therefore penetrate much further into the cloud.
Figure \ref{fig:spatialdistribution} shows this with two figures showing the highly-ice-absorbed stars overlaid on an emission map from ALMA and the GBT \citep[][Appendix \ref{app:spatialplotstwo} shows the ice-absorbed stars on a JWST background]{Ginsburg2025}.
There are highly excess-reddened F356W-F444W stars covering most of the area of the Brick, while the inner area is nearly blank, indicating no stars are detected, in F405N-F466N colors.

The F356W-F444W excess sources are more tightly coupled to the dense gas.
Unlike the CO traced by F405N-F466N, which is detected across the whole field-of-view, the F356W excesses are limited to regions with dust emission detected \referee{at millimeter wavelengths} --- i.e., \referee{parts of the cloud with} higher column densities.
This feature hints at a chemical processing explanation: CO and \water ice are likely present in dust throughout the Galactic Center even at relatively low column densities, while the methylated features that produce excess absorption in F356W only appear at higher column densities.
These hints match prior conclusions from gas observations in which substantial depletion was inferred in the cloud interior based on lack of correlation between dust and gas emission \citep{Rathborne2014b}.
That implies that there is fairly rapid hydrogenation of CO on dust grain surfaces once adequate density and/or shielding from external UV radiation is reached, in at least rough qualitative agreement with chemical models.

The extensive CO ice seen throughout the cloud, and especially in the outskirts where the column densities are fairly low, implies that CO freezeout is happening early in the cloud's evolution.
The Brick is not necessarily collapsing, and instead may be undergoing a simultaneous crushing and stretching by tidal forces \citep{Kruijssen2019,Henshaw2019}, such that its density and temperature evolution has followed a more complex path than usually considered in chemical evolution models.
We have assumed in much of the above that CO is 100\% frozen out, since that is the most conservative assumption when using CO as a metallicity tracer; any additional CO in the gas phase would require an even higher metallicity.
However, this total freezeout is itself an interesting possibility, as it may imply that the cosmic ray ionization rate in the Brick is lower than previously considered \citep{Clarke2024}.

\section{Conclusions}
\label{sec:conclusion}
We report a photometric study of the Brick combined with modeling based on laboratory measurements and comparison spectroscopic and photometric data sets from other Galactic regions.
We demonstrate the use of the \texttt{icemodels} package to predict JWST photometry using laboratory transmission data, which are intended for general use.

We report several observational results:
\begin{itemize}
    \item There is excess absorption in the F466N filter as compared with F212N, F405N, and F410M, indicating the presence of CO and \water ice.
    \item There is excess absorption in the F410M filter as compared with the F405N filter, indicating the presence of CO$_2$ ice.
    \item There is excess absorption in the F356W filter as compared with the F444W filter.  This excess \referee{is likely due to ice mixtures that include \methanol, but} is not fully explained.
    \item These excesses are correlated with one another, suggesting they are all attributed to increased ice accumulation on dust grains.
    \item Color-color diagrams including the above filters were compared to laboratory ice models.  An ice mixture of \water:CO:CO$_2$ $\sim$10:1:1 best fits the observations of the narrow-band filters, though the uncertainty or physical variation in this ratio may be large.  This is a higher ratio of \water:CO than observed in local clouds.
\end{itemize}

Using these observed excesses, we measure the column density of CO across the Brick.
We find that the CO ice abundance exceeds local values, and the CO ice abundance exceeds the total CO abundance adopting solar neighborhood values.
We attribute the higher CO ice abundance to higher metallicity.
Comparing the Brick to local clouds, a cloud in the 3 kpc arm, and the Galactic Center dust ridge, we see a clear gradient in CO abundance from the solar neighborhood to the Galactic Center.
We use the abundance difference to infer metallicity with two different approaches.
By assuming the freezeout fraction is independent of environment, we obtain $Z_{GC}\sim2.5-5Z_\odot$.
By calibrating against other metallicity measurements, we derive a relation between the CO ice abundance and metallicity,  $ [\mathrm{CO_{ice}}/\mathrm{H}_2] = 0.23 (Z / Z_\odot) - 4.27 $.
This relation may be used to measure metallicity in cold molecular clouds using JWST NIRCam photometry.

The unexplained excess absorption in F356W, its strong correlation with F466N absorption, and its increase toward the center of the Brick cloud suggest that extraordinary chemical activity is occurring in this region.
We suggest that this activity is, at least, rapid and substantial formation of \methanol ice, and possibly formation of substantially more complex molecules.
Followup spectroscopy is needed to say more.


Dust in the Galactic Center is icier and more water-rich than the solar neighborhood.
\section*{Acknowledgments} 
We thank the referee for a detailed series of repports that helped improve the clarity of this work.
    This work is based in part on observations made with the NASA/ESA/CSA James Webb Space Telescope. The data were obtained from the Mikulski Archive for Space Telescopes at the Space Telescope Science Institute, which is operated by the Association of Universities for Research in Astronomy, Inc., under NASA contract NAS 5-03127 for JWST. These observations are associated with program \#2221, \#1182, \#1186, \#3222, and \#5804.
    The data are located at DOI 10.17909/w3sv-fj59
    Support for programs \#2221 and \#5365 was provided by NASA through a grant from the Space Telescope Science Institute, which is operated by the Association of Universities for Research in Astronomy, Inc., under NASA contract NAS 5-03127.
    AG acknowledges support from the NSF under grants AAG 2206511 and CAREER 2142300.
    CB gratefully  acknowledges  funding  from  National  Science  Foundation  under  Award  Nos. 2108938, 2206510, and CAREER 2145689, as well as from the National Aeronautics and Space Administration through the Astrophysics Data Analysis Program under Award ``3-D MC: Mapping Circumnuclear Molecular Clouds from X-ray to Radio,” Grant No. 80NSSC22K1125.
    This research has made use of the SVO Filter Profile Service ``Carlos Rodrigo", funded by MCIN/AEI/10.13039/501100011033/ through grant PID2023-146210NB-I00.
    Optical constants data were retrieved from the Optical Constants database.
    Optical constants were also retrieved from the Leiden Ice Database LIDA.

    The authors acknowledge University of Florida Research Computing for providing computational resources on the HiPerGator supercomputer and support that have contributed to the research results reported in this publication. URL: \url{http://www.rc.ufl.edu}.


\software{\texttt{astropy} \citep{2013A&A...558A..33A}, \texttt{scipy} \citep{2020SciPy-NMeth}, \texttt{numpy} \citep{2020NumPy-Array}, the SVO Filter Profile Service \citep{2012ivoa.rept.1015R,2020sea..confE.182R}, \texttt{icemodels}, and \texttt{stpsf/webbpsf} for JWST PSF modeling \citep{2023PASP..135d8001R}.}
\facility{JWST (NIRCAM, NIRSPEC)} 


\bibliographystyle{abbrvnat} 
\bibliography{bib.bib}

@ARTICLE{Nogueras-Lara2020,
       author = {{Nogueras-Lara}, F. and {Sch{\"o}del}, R. and {Neumayer}, N. and {Gallego-Cano}, E. and {Shahzamanian}, B. and {Gallego-Calvente}, A.~T. and {Najarro}, F.},
        title = "{GALACTICNUCLEUS: A high angular-resolution JHK$_{s}$ imaging survey of the Galactic centre. III. Evidence for wavelength-dependence of the extinction curve in the near-infrared}",
      journal = {\aap},
         year = 2020,
        month = sep,
       volume = {641},
          eid = {A141},
        pages = {A141},
          doi = {10.1051/0004-6361/202038606},
archivePrefix = {arXiv},
       eprint = {2007.04401},
 primaryClass = {astro-ph.GA},
       adsurl = {https://ui.adsabs.harvard.edu/abs/2020A&A...641A.141N}
}

@ARTICLE{Kruijssen2019,
       author = {{Kruijssen}, J.~M.~D. and {Dale}, J.~E. and {Longmore}, S.~N. and {Walker}, D.~L. and {Henshaw}, J.~D. and {Jeffreson}, S.~M.~R. and {Petkova}, M.~A. and {Ginsburg}, A. and {Barnes}, A.~T. and {Battersby}, C.~D. and {Immer}, K. and {Jackson}, J.~M. and {Keto}, E.~R. and {Krieger}, N. and {Mills}, E.~A.~C. and {S{\'a}nchez-Monge}, {\'A}. and {Schmiedeke}, A. and {Suri}, S.~T. and {Zhang}, Q.},
        title = "{The dynamical evolution of molecular clouds near the Galactic Centre - II. Spatial structure and kinematics of simulated clouds}",
      journal = {\mnras},
         year = 2019,
        month = apr,
       volume = {484},
       number = {4},
        pages = {5734-5754},
          doi = {10.1093/mnras/stz381},
archivePrefix = {arXiv},
       eprint = {1902.01860},
 primaryClass = {astro-ph.GA},
       adsurl = {https://ui.adsabs.harvard.edu/abs/2019MNRAS.484.5734K}
}

@ARTICLE{Robitaille2017,
       author = {{Robitaille}, T.~P.},
        title = "{A modular set of synthetic spectral energy distributions for young stellar objects}",
      journal = {\aap},
         year = 2017,
        month = apr,
       volume = {600},
          eid = {A11},
        pages = {A11},
          doi = {10.1051/0004-6361/201425486},
archivePrefix = {arXiv},
       eprint = {1703.05765},
 primaryClass = {astro-ph.SR},
       adsurl = {https://ui.adsabs.harvard.edu/abs/2017A&A...600A..11R}
}

@ARTICLE{Rathborne2014a,
       author = {{Rathborne}, J.~M. and {Longmore}, S.~N. and {Jackson}, J.~M. and {Kruijssen}, J.~M.~D. and {Alves}, J.~F. and {Bally}, J. and {Bastian}, N. and {Contreras}, Y. and {Foster}, J.~B. and {Garay}, G. and {Testi}, L. and {Walsh}, A.~J.},
        title = "{Turbulence Sets the Initial Conditions for Star Formation in High-pressure Environments}",
      journal = {\apjl},
         year = 2014,
        month = nov,
       volume = {795},
       number = {2},
          eid = {L25},
        pages = {L25},
          doi = {10.1088/2041-8205/795/2/L25},
archivePrefix = {arXiv},
       eprint = {1409.0935},
 primaryClass = {astro-ph.GA},
       adsurl = {https://ui.adsabs.harvard.edu/abs/2014ApJ...795L..25R}
}

@ARTICLE{Rathborne2014b,
       author = {{Rathborne}, J.~M. and {Longmore}, S.~N. and {Jackson}, J.~M. and {Foster}, J.~B. and {Contreras}, Y. and {Garay}, G. and {Testi}, L. and {Alves}, J.~F. and {Bally}, J. and {Bastian}, N. and {Kruijssen}, J.~M.~D. and {Bressert}, E.},
        title = "{G0.253+0.016: A Centrally Condensed, High-mass Protocluster}",
      journal = {\apj},
         year = 2014,
        month = may,
       volume = {786},
       number = {2},
          eid = {140},
        pages = {140},
          doi = {10.1088/0004-637X/786/2/140},
archivePrefix = {arXiv},
       eprint = {1403.0996},
 primaryClass = {astro-ph.GA},
       adsurl = {https://ui.adsabs.harvard.edu/abs/2014ApJ...786..140R}
}

@ARTICLE{An2011,
       author = {{An}, Deokkeun and {Ram{\'\i}rez}, Solange V. and {Sellgren}, Kris and {Arendt}, Richard G. and {Adwin Boogert}, A.~C. and {Robitaille}, Thomas P. and {Schultheis}, Mathias and {Cotera}, Angela S. and {Smith}, Howard A. and {Stolovy}, Susan R.},
        title = "{Massive Young Stellar Objects in the Galactic Center. I. Spectroscopic Identification from Spitzer Infrared Spectrograph Observations}",
      journal = {\apj},
         year = 2011,
        month = aug,
       volume = {736},
       number = {2},
          eid = {133},
        pages = {133},
          doi = {10.1088/0004-637X/736/2/133},
archivePrefix = {arXiv},
       eprint = {1104.4788},
 primaryClass = {astro-ph.GA},
       adsurl = {https://ui.adsabs.harvard.edu/abs/2011ApJ...736..133A}
}

@ARTICLE{Jang2022,
       author = {{Jang}, Dajeong and {An}, Deokkeun and {Sellgren}, Kris and {Ram{\'\i}rez}, Solange V. and {Boogert}, A.~C. Adwin and {Schultheis}, Mathias},
        title = "{Massive Young Stellar Objects in the Galactic Center. II. Seeing Through the Ice-rich Envelopes}",
      journal = {\apj},
         year = 2022,
        month = may,
       volume = {930},
       number = {1},
          eid = {16},
        pages = {16},
          doi = {10.3847/1538-4357/ac5d51},
archivePrefix = {arXiv},
       eprint = {2203.16833},
 primaryClass = {astro-ph.GA},
       adsurl = {https://ui.adsabs.harvard.edu/abs/2022ApJ...930...16J}
}

@ARTICLE{An2017,
       author = {{An}, Deokkeun and {Sellgren}, Kris and {Boogert}, A.~C. Adwin and {Ram{\'\i}rez}, Solange V. and {Pyo}, Tae-Soo},
        title = "{Abundant Methanol Ice toward a Massive Young Stellar Object in the Central Molecular Zone}",
      journal = {\apjl},
         year = 2017,
        month = jul,
       volume = {843},
       number = {2},
          eid = {L36},
        pages = {L36},
          doi = {10.3847/2041-8213/aa7cfe},
archivePrefix = {arXiv},
       eprint = {1707.03120},
 primaryClass = {astro-ph.GA},
       adsurl = {https://ui.adsabs.harvard.edu/abs/2017ApJ...843L..36A}
}

@ARTICLE{Onaka2022,
       author = {{Onaka}, Takashi and {Sakon}, Itsuki and {Shimonishi}, Takashi},
        title = "{Near-infrared spectroscopy of a massive young stellar object in the direction toward the Galactic Center: XCN and aromatic C-D features}",
      journal = {arXiv e-prints},
         year = 2022,
        month = dec,
          eid = {arXiv:2212.11424},
        pages = {arXiv:2212.11424},
archivePrefix = {arXiv},
       eprint = {2212.11424},
 primaryClass = {astro-ph.GA},
       adsurl = {https://ui.adsabs.harvard.edu/abs/2022arXiv221211424O}
}

@ARTICLE{Chiar2006,
       author = {{Chiar}, J.~E. and {Tielens}, A.~G.~G.~M.},
        title = "{Pixie Dust: The Silicate Features in the Diffuse Interstellar Medium}",
      journal = {\apj},
         year = 2006,
        month = feb,
       volume = {637},
       number = {2},
        pages = {774-785},
          doi = {10.1086/498406},
archivePrefix = {arXiv},
       eprint = {astro-ph/0510156},
 primaryClass = {astro-ph},
       adsurl = {https://ui.adsabs.harvard.edu/abs/2006ApJ...637..774C}
}

@ARTICLE{Henshaw2019,
       author = {{Henshaw}, J.~D. and {Ginsburg}, A. and {Haworth}, T.~J. and {Longmore}, S.~N. and {Kruijssen}, J.~M.~D. and {Mills}, E.~A.~C. and {Sokolov}, V. and {Walker}, D.~L. and {Barnes}, A.~T. and {Contreras}, Y. and {Bally}, J. and {Battersby}, C. and {Beuther}, H. and {Butterfield}, N. and {Dale}, J.~E. and {Henning}, T. and {Jackson}, J.~M. and {Kauffmann}, J. and {Pillai}, T. and {Ragan}, S. and {Riener}, M. and {Zhang}, Q.},
        title = "{`The Brick' is not a brick: a comprehensive study of the structure and dynamics of the central molecular zone cloud G0.253+0.016}",
      journal = {\mnras},
         year = 2019,
        month = may,
       volume = {485},
       number = {2},
        pages = {2457-2485},
          doi = {10.1093/mnras/stz471},
archivePrefix = {arXiv},
       eprint = {1902.02793},
 primaryClass = {astro-ph.GA},
       adsurl = {https://ui.adsabs.harvard.edu/abs/2019MNRAS.485.2457H}
}

@ARTICLE{Nogueras-Lara2021,
       author = {{Nogueras-Lara}, F. and {Sch{\"o}del}, R. and {Neumayer}, N.},
        title = "{GALACTICNUCLEUS: A high-angular-resolution JHK$_{s}$ imaging survey of the Galactic centre. IV. Extinction maps and de-reddened photometry}",
      journal = {\aap},
         year = 2021,
        month = sep,
       volume = {653},
          eid = {A133},
        pages = {A133},
          doi = {10.1051/0004-6361/202140996},
archivePrefix = {arXiv},
       eprint = {2107.00021},
 primaryClass = {astro-ph.GA},
       adsurl = {https://ui.adsabs.harvard.edu/abs/2021A&A...653A.133N}
}

@ARTICLE{Fritz2011,
       author = {{Fritz}, T.~K. and {Gillessen}, S. and {Dodds-Eden}, K. and {Lutz}, D. and {Genzel}, R. and {Raab}, W. and {Ott}, T. and {Pfuhl}, O. and {Eisenhauer}, F. and {Yusef-Zadeh}, F.},
        title = "{Line Derived Infrared Extinction toward the Galactic Center}",
      journal = {\apj},
         year = 2011,
        month = aug,
       volume = {737},
       number = {2},
          eid = {73},
        pages = {73},
          doi = {10.1088/0004-637X/737/2/73},
archivePrefix = {arXiv},
       eprint = {1105.2822},
 primaryClass = {astro-ph.GA},
       adsurl = {https://ui.adsabs.harvard.edu/abs/2011ApJ...737...73F}
}

@ARTICLE{Gordon2021,
       author = {{Gordon}, Karl D. and {Misselt}, Karl A. and {Bouwman}, Jeroen and {Clayton}, Geoffrey C. and {Decleir}, Marjorie and {Hines}, Dean C. and {Pendleton}, Yvonne and {Rieke}, George and {Smith}, J.~D.~T. and {Whittet}, D.~C.~B.},
        title = "{Milky Way Mid-Infrared Spitzer Spectroscopic Extinction Curves: Continuum and Silicate Features}",
      journal = {\apj},
         year = 2021,
        month = jul,
       volume = {916},
       number = {1},
          eid = {33},
        pages = {33},
          doi = {10.3847/1538-4357/ac00b7},
archivePrefix = {arXiv},
       eprint = {2105.05087},
 primaryClass = {astro-ph.GA},
       adsurl = {https://ui.adsabs.harvard.edu/abs/2021ApJ...916...33G}
}

@ARTICLE{Rocha2021,
       author = {{Rocha}, W.~R.~M. and {Perotti}, G. and {Kristensen}, L.~E. and {J{\o}rgensen}, J.~K.},
        title = "{Fitting infrared ice spectra with genetic modelling algorithms. Presenting the ENIIGMA fitting tool}",
      journal = {\aap},
         year = 2021,
        month = oct,
       volume = {654},
          eid = {A158},
        pages = {A158},
          doi = {10.1051/0004-6361/202039360},
archivePrefix = {arXiv},
       eprint = {2107.08555},
 primaryClass = {astro-ph.IM},
       adsurl = {https://ui.adsabs.harvard.edu/abs/2021A&A...654A.158R}
}

@ARTICLE{McClure2023,
       author = {{McClure}, M.~K. and {Rocha}, W.~R.~M. and {Pontoppidan}, K.~M. and {Crouzet}, N. and {Chu}, L.~E.~U. and {Dartois}, E. and {Lamberts}, T. and {Noble}, J.~A. and {Pendleton}, Y.~J. and {Perotti}, G. and {Qasim}, D. and {Rachid}, M.~G. and {Smith}, Z.~L. and {Sun}, Fengwu and {Beck}, Tracy L and {Boogert}, A.~C.~A. and {Brown}, W.~A. and {Caselli}, P. and {Charnley}, S.~B. and {Cuppen}, Herma M. and {Dickinson}, H. and {Drozdovskaya}, M.~N. and {Egami}, E. and {Erkal}, J. and {Fraser}, H. and {Garrod}, R.~T. and {Harsono}, D. and {Ioppolo}, S. and {Jimenez-Serra}, I. and {Jin}, M. and {J{\o}rgensen}, J.~K. and {Kristensen}, L.~E. and {Lis}, D.~C. and {McCoustra}, M.~R.~S. and {McGuire}, Brett A. and {Melnick}, G.~J. and {Oberg}, Karin I. and {Palumbo}, M.~E. and {Shimonishi}, T. and {Sturm}, J.~A. and {van Dishoeck}, E.~F. and {Linnartz}, H.},
        title = "{An Ice Age JWST inventory of dense molecular cloud ices}",
      journal = {arXiv e-prints},
         year = 2023,
        month = jan,
          eid = {arXiv:2301.09140},
        pages = {arXiv:2301.09140},
archivePrefix = {arXiv},
       eprint = {2301.09140},
 primaryClass = {astro-ph.GA},
       adsurl = {https://ui.adsabs.harvard.edu/abs/2023arXiv230109140M}
}

@ARTICLE{Ginsburg2023,
       author = {{Ginsburg}, Adam and {Barnes}, Ashley T. and {Battersby}, Cara D. and {Bulatek}, Alyssa and {Gramze}, Savannah and {Henshaw}, Jonathan D. and {Jeff}, Desmond and {Lu}, Xing and {Mills}, E.~A.~C. and {Walker}, Daniel L.},
        title = "{JWST reveals widespread CO ice and gas absorption in the Galactic Center cloud G0.253+0.016}",
      journal = {arXiv e-prints},
         year = 2023,
        month = aug,
          eid = {arXiv:2308.16050},
        pages = {arXiv:2308.16050},
          doi = {10.48550/arXiv.2308.16050},
archivePrefix = {arXiv},
       eprint = {2308.16050},
 primaryClass = {astro-ph.GA},
       adsurl = {https://ui.adsabs.harvard.edu/abs/2023arXiv230816050G}
}

@ARTICLE{Boogert2008,
       author = {{Boogert}, A.~C.~A. and {Pontoppidan}, K.~M. and {Knez}, C. and {Lahuis}, F. and {Kessler-Silacci}, J. and {van Dishoeck}, E.~F. and {Blake}, G.~A. and {Augereau}, J. -C. and {Bisschop}, S.~E. and {Bottinelli}, S. and {Brooke}, T.~Y. and {Brown}, J. and {Crapsi}, A. and {Evans}, II, N.~J. and {Fraser}, H.~J. and {Geers}, V. and {Huard}, T.~L. and {J{\o}rgensen}, J.~K. and {{\"O}berg}, K.~I. and {Allen}, L.~E. and {Harvey}, P.~M. and {Koerner}, D.~W. and {Mundy}, L.~G. and {Padgett}, D.~L. and {Sargent}, A.~I. and {Stapelfeldt}, K.~R.},
        title = "{The c2d Spitzer Spectroscopic Survey of Ices around Low-Mass Young Stellar Objects. I. H$_{2}$O and the 5-8 {\ensuremath{\mu}}m Bands}",
      journal = {\apj},
         year = 2008,
        month = may,
       volume = {678},
       number = {2},
        pages = {985-1004},
          doi = {10.1086/533425},
archivePrefix = {arXiv},
       eprint = {0801.1167},
 primaryClass = {astro-ph},
       adsurl = {https://ui.adsabs.harvard.edu/abs/2008ApJ...678..985B}
}

@ARTICLE{Chiar2002,
       author = {{Chiar}, J.~E. and {Adamson}, A.~J. and {Pendleton}, Y.~J. and {Whittet}, D.~C.~B. and {Caldwell}, D.~A. and {Gibb}, E.~L.},
        title = "{Hydrocarbons, Ices, and ``XCN'' in the Line of Sight toward the Galactic Center}",
      journal = {\apj},
         year = 2002,
        month = may,
       volume = {570},
       number = {1},
        pages = {198-209},
          doi = {10.1086/339570},
       adsurl = {https://ui.adsabs.harvard.edu/abs/2002ApJ...570..198C}
}

@ARTICLE{Hudgins1993,
       author = {{Hudgins}, D.~M. and {Sandford}, S.~A. and {Allamandola}, L.~J. and {Tielens}, A.~G.~G.~M.},
        title = "{Mid- and Far-Infrared Spectroscopy of Ices: Optical Constants and Integrated Absorbances}",
      journal = {\apjs},
         year = 1993,
        month = jun,
       volume = {86},
        pages = {713},
          doi = {10.1086/191796},
       adsurl = {https://ui.adsabs.harvard.edu/abs/1993ApJS...86..713H}
}

@ARTICLE{Novozamsky2001,
       author = {{Novozamsky}, J.~H. and {Schutte}, W.~A. and {Keane}, J.~V.},
        title = "{Further evidence for the assignment of the XCN band in astrophysical ice analogs to OCN$^{-}$. Spectroscopy and deuterium shift}",
      journal = {\aap},
         year = 2001,
        month = nov,
       volume = {379},
        pages = {588-591},
          doi = {10.1051/0004-6361:20011332},
       adsurl = {https://ui.adsabs.harvard.edu/abs/2001A&A...379..588N}
}

@article{Ginsburg2025,
author = {{Ginsburg}, A. and {Walker}, D. and {Longmore}, S. and {Lu}, X. and {Hsieh}, P.Y. and {ACES data reduction team}},
title = "{ACES II: Continuum}",
year = {2025}
}

@ARTICLE{Budaiev2025,
       author = {{Budaiev}, Nazar and {Ginsburg}, Adam and {Barnes}, Ashley T. and {Jeff}, Desmond and {Yoo}, Taehwa and {Battersby}, Cara and {Bulatek}, Alyssa and {Lu}, Xing and {Mills}, Elisabeth A.~C. and {Walker}, Daniel L.},
        title = "{JWST's first view of the most vigorously star-forming cloud in the Galactic center -- Sagittarius B2}",
      journal = {arXiv e-prints},
         year = 2025,
        month = sep,
          eid = {arXiv:2509.11771},
        pages = {arXiv:2509.11771},
          doi = {10.48550/arXiv.2509.11771},
archivePrefix = {arXiv},
       eprint = {2509.11771},
 primaryClass = {astro-ph.GA},
       adsurl = {https://ui.adsabs.harvard.edu/abs/2025arXiv250911771B}
}

@ARTICLE{Clarke2024,
       author = {{Clarke}, S.~D. and {Makeev}, V.~A. and {S{\'a}nchez-Monge}, {\'A}. and {Williams}, G.~M. and {Tang}, Y. -W. and {Walch}, S. and {Higgins}, R. and {N{\"u}rnberger}, P.~C. and {Suri}, S.},
        title = "{GMF G214.5-1.8 as traced by CO: I - cloud-scale CO freeze-out as a result of a low cosmic-ray ionization rate}",
      journal = {\mnras},
         year = 2024,
        month = feb,
       volume = {528},
       number = {2},
        pages = {1555-1572},
          doi = {10.1093/mnras/stae117},
archivePrefix = {arXiv},
       eprint = {2401.04992},
 primaryClass = {astro-ph.GA},
       adsurl = {https://ui.adsabs.harvard.edu/abs/2024MNRAS.528.1555C}
}

@ARTICLE{Gerakines2023,
       author = {{Gerakines}, Perry A. and {Materese}, Christopher K. and {Hudson}, Reggie L.},
        title = "{Carbon monoxide ices - a semicentennial review and update for crystalline CO along with the first IR spectrum and band strength for amorphous CO}",
      journal = {\mnras},
         year = 2023,
        month = jun,
       volume = {522},
       number = {2},
        pages = {3145-3162},
          doi = {10.1093/mnras/stad1164},
       adsurl = {https://ui.adsabs.harvard.edu/abs/2023MNRAS.522.3145G}
}

@ARTICLE{Megeath2012,
       author = {{Megeath}, S.~T. and {Gutermuth}, R. and {Muzerolle}, J. and {Kryukova}, E. and {Flaherty}, K. and {Hora}, J.~L. and {Allen}, L.~E. and {Hartmann}, L. and {Myers}, P.~C. and {Pipher}, J.~L. and {Stauffer}, J. and {Young}, E.~T. and {Fazio}, G.~G.},
        title = "{The Spitzer Space Telescope Survey of the Orion A and B Molecular Clouds. I. A Census of Dusty Young Stellar Objects and a Study of Their Mid-infrared Variability}",
      journal = {\aj},
         year = 2012,
        month = dec,
       volume = {144},
       number = {6},
          eid = {192},
        pages = {192},
          doi = {10.1088/0004-6256/144/6/192},
archivePrefix = {arXiv},
       eprint = {1209.3826},
 primaryClass = {astro-ph.GA},
       adsurl = {https://ui.adsabs.harvard.edu/abs/2012AJ....144..192M}
}

@ARTICLE{Guver2009,
       author = {{G{\"u}ver}, Tolga and {{\"O}zel}, Feryal},
        title = "{The relation between optical extinction and hydrogen column density in the Galaxy}",
      journal = {\mnras},
         year = 2009,
        month = dec,
       volume = {400},
       number = {4},
        pages = {2050-2053},
          doi = {10.1111/j.1365-2966.2009.15598.x},
archivePrefix = {arXiv},
       eprint = {0903.2057},
 primaryClass = {astro-ph.GA},
       adsurl = {https://ui.adsabs.harvard.edu/abs/2009MNRAS.400.2050G}
}

@ARTICLE{Smith2025,
       author = {{Smith}, Z.~L. and {Dickinson}, H.~J. and {Fraser}, H.~J. and {McClure}, M.~K. and {Noble}, J.~A. and {Boogert}, A.~C.~A. and {Sun}, F. and {Egami}, E. and {Dartois}, E. and {Erkal}, J. and {Shimonishi}, T. and {Beck}, T.~L. and {Bergner}, J.~B. and {Caselli}, P. and {Charnley}, S.~B. and {Chu}, L. and {Drozdovskaya}, M.~N. and {Garrod}, R. and {Harsono}, D. and {Ioppolo}, S. and {Jimenez-Serra}, I. and {J{\o}rgensen}, J.~K. and {Melnick}, G.~J. and {{\~A}-berg}, K.~I. and {Palumbo}, M.~E. and {Pendleton}, Y.~J. and {Perotti}, G. and {Pontoppidan}, K.~M. and {Qasim}, D. and {Rocha}, W.~R.~M. and {Sturm}, J.~A. and {Taillard}, A. and {Urso}, R.~G. and {van Dishoeck}, E.~F.},
        title = "{Cospatial ice mapping of H$_{2}$O with CO$_{2}$ and CO across a molecular cloud with JWST/NIRCam}",
      journal = {Nature Astronomy},
         year = 2025,
        month = mar,
          doi = {10.1038/s41550-025-02511-z},
       adsurl = {https://ui.adsabs.harvard.edu/abs/2025NatAs.tmp...76S}
}

@ARTICLE{Arellano-Cordova2020,
       author = {{Arellano-C{\'o}rdova}, K.~Z. and {Esteban}, C. and {Garc{\'\i}a-Rojas}, J. and {M{\'e}ndez-Delgado}, J.~E.},
        title = "{The Galactic radial abundance gradients of C, N, O, Ne, S, Cl, and Ar from deep spectra of H II regions}",
      journal = {\mnras},
         year = 2020,
        month = aug,
       volume = {496},
       number = {2},
        pages = {1051-1076},
          doi = {10.1093/mnras/staa1523},
archivePrefix = {arXiv},
       eprint = {2005.11372},
 primaryClass = {astro-ph.GA},
       adsurl = {https://ui.adsabs.harvard.edu/abs/2020MNRAS.496.1051A}
}

@ARTICLE{Nieva2012,
       author = {{Nieva}, M. -F. and {Przybilla}, N.},
        title = "{Present-day cosmic abundances. A comprehensive study of nearby early B-type stars and implications for stellar and Galactic evolution and interstellar dust models}",
      journal = {\aap},
         year = 2012,
        month = mar,
       volume = {539},
          eid = {A143},
        pages = {A143},
          doi = {10.1051/0004-6361/201118158},
archivePrefix = {arXiv},
       eprint = {1203.5787},
 primaryClass = {astro-ph.SR},
       adsurl = {https://ui.adsabs.harvard.edu/abs/2012A&A...539A.143N}
}

@ARTICLE{Asplund2009,
       author = {{Asplund}, Martin and {Grevesse}, Nicolas and {Sauval}, A. Jacques and {Scott}, Pat},
        title = "{The Chemical Composition of the Sun}",
      journal = {\araa},
         year = 2009,
        month = sep,
       volume = {47},
       number = {1},
        pages = {481-522},
          doi = {10.1146/annurev.astro.46.060407.145222},
archivePrefix = {arXiv},
       eprint = {0909.0948},
 primaryClass = {astro-ph.SR},
       adsurl = {https://ui.adsabs.harvard.edu/abs/2009ARA&A..47..481A}
}

@ARTICLE{Slavicinska2025,
       author = {{Slavicinska}, Katerina and {Tychoniec}, {\L}ukasz and {Navarro}, Mar{\'\i}a Gabriela and {van Dishoeck}, Ewine F. and {Tobin}, John J. and {van Gelder}, Martijn L. and {Chen}, Yuan and {Boogert}, A.~C. Adwin and {Blake Drechsler}, W. and {Beuther}, Henrik and {Garatti}, Alessio Caratti o and {Megeath}, S. Thomas and {Klaassen}, Pamela and {Looney}, Leslie W. and {Kavanagh}, Patrick J. and {Brunken}, Nashanty G.~C. and {Sheehan}, Patrick and {Fischer}, William J.},
        title = "{HDO ice detected toward an isolated low-mass protostar with JWST}",
      journal = {arXiv e-prints},
         year = 2025,
        month = may,
          eid = {arXiv:2505.14686},
        pages = {arXiv:2505.14686},
          doi = {10.48550/arXiv.2505.14686},
archivePrefix = {arXiv},
       eprint = {2505.14686},
 primaryClass = {astro-ph.SR},
       adsurl = {https://ui.adsabs.harvard.edu/abs/2025arXiv250514686S}
}

@ARTICLE{Rocha2022,
       author = {{Rocha}, W.~R.~M. and {Rachid}, M.~G. and {Olsthoorn}, B. and {van Dishoeck}, E.~F. and {McClure}, M.~K. and {Linnartz}, H.},
        title = "{LIDA: The Leiden Ice Database for Astrochemistry}",
      journal = {\aap},
         year = 2022,
        month = dec,
       volume = {668},
          eid = {A63},
        pages = {A63},
          doi = {10.1051/0004-6361/202244032},
archivePrefix = {arXiv},
       eprint = {2208.12211},
 primaryClass = {astro-ph.IM},
       adsurl = {https://ui.adsabs.harvard.edu/abs/2022A&A...668A..63R}
}

@ARTICLE{Rodrigo2024,
       author = {{Rodrigo}, Carlos and {Cruz}, Patricia and {Aguilar}, John F. and {Aller}, Alba and {Solano}, Enrique and {G{\'a}lvez-Ortiz}, Maria Cruz and {Jim{\'e}nez-Esteban}, Francisco and {Mas-Buitrago}, Pedro and {Bayo}, Amelia and {Cort{\'e}s-Contreras}, Miriam and {Murillo-Ojeda}, Raquel and {Bonoli}, Silvia and {Cenarro}, Javier and {Dupke}, Renato and {L{\'o}pez-Sanjuan}, Carlos and {Mar{\'\i}n-Franch}, Antonio and {de Oliveira}, Claudia Mendes and {Moles}, Mariano and {Taylor}, Keith and {Varela}, Jes{\'u}s and {Rami{\'o}}, H{\'e}ctor V{\'a}zquez},
        title = "{Photometric segregation of dwarf and giant FGK stars using the SVO Filter Profile Service and photometric tools}",
      journal = {\aap},
         year = 2024,
        month = sep,
       volume = {689},
          eid = {A93},
        pages = {A93},
          doi = {10.1051/0004-6361/202449998},
archivePrefix = {arXiv},
       eprint = {2406.03310},
 primaryClass = {astro-ph.SR},
       adsurl = {https://ui.adsabs.harvard.edu/abs/2024A&A...689A..93R}
}

@ARTICLE{Bernstein2005,
       author = {{Bernstein}, Max P. and {Cruikshank}, Dale P. and {Sandford}, Scott A.},
        title = "{Near-infrared laboratory spectra of solid H $_{2}$O/CO $_{2}$ and CH $_{3}$OH/CO $_{2}$ ice mixtures}",
      journal = {\icarus},
         year = 2005,
        month = dec,
       volume = {179},
       number = {2},
        pages = {527-534},
          doi = {10.1016/j.icarus.2005.07.009},
       adsurl = {https://ui.adsabs.harvard.edu/abs/2005Icar..179..527B}
}

@ARTICLE{Oberg2007,
       author = {{{\"O}berg}, K.~I. and {Fraser}, H.~J. and {Boogert}, A.~C.~A. and {Bisschop}, S.~E. and {Fuchs}, G.~W. and {van Dishoeck}, E.~F. and {Linnartz}, H.},
        title = "{Effects of CO$_{2}$ on H\_2O band profiles and band strengths in mixed H\_2O:CO$_{2}$ ices}",
      journal = {\aap},
         year = 2007,
        month = feb,
       volume = {462},
       number = {3},
        pages = {1187-1198},
          doi = {10.1051/0004-6361:20065881},
archivePrefix = {arXiv},
       eprint = {astro-ph/0610751},
 primaryClass = {astro-ph},
       adsurl = {https://ui.adsabs.harvard.edu/abs/2007A&A...462.1187O}
}

@ARTICLE{Hudson2020,
       author = {{Hudson}, Reggie L. and {Loeffler}, Mark J. and {Ferrante}, Robert F. and {Gerakines}, Perry A. and {Coleman}, Falvia M.},
        title = "{Testing Densities and Refractive Indices of Extraterrestrial Ice Components Using Molecular Structures{\textemdash}Organic Compounds and Molar Refractions}",
      journal = {\apj},
         year = 2020,
        month = mar,
       volume = {891},
       number = {1},
          eid = {22},
        pages = {22},
          doi = {10.3847/1538-4357/ab6efa},
       adsurl = {https://ui.adsabs.harvard.edu/abs/2020ApJ...891...22H}
}

@ARTICLE{Satorre2008,
       author = {{Satorre}, M. {\'A}. and {Domingo}, M. and {Mill{\'a}n}, C. and {Luna}, R. and {Vilaplana}, R. and {Santonja}, C.},
        title = "{Density of CH$_{4}$, N$_{2}$ and CO$_{2}$ ices at different temperatures of deposition}",
      journal = {\planss},
         year = 2008,
        month = nov,
       volume = {56},
       number = {13},
        pages = {1748-1752},
          doi = {10.1016/j.pss.2008.07.015},
       adsurl = {https://ui.adsabs.harvard.edu/abs/2008P&SS...56.1748S}
}

@ARTICLE{Richardson2024,
       author = {{Richardson}, Theo and {Ginsburg}, Adam and {Indebetouw}, R{\'e}my and {Robitaille}, Thomas P.},
        title = "{An Updated Modular Set of Synthetic Spectral Energy Distributions for Young Stellar Objects}",
      journal = {\apj},
         year = 2024,
        month = feb,
       volume = {961},
       number = {2},
          eid = {188},
        pages = {188},
          doi = {10.3847/1538-4357/ad072d},
archivePrefix = {arXiv},
       eprint = {2401.12810},
 primaryClass = {astro-ph.SR},
       adsurl = {https://ui.adsabs.harvard.edu/abs/2024ApJ...961..188R}
}

@ARTICLE{Ginsburg2019,
       author = {{Ginsburg}, Adam and {Sip{\H{o}}cz}, Brigitta M. and {Brasseur}, C.~E. and {Cowperthwaite}, Philip S. and {Craig}, Matthew W. and {Deil}, Christoph and {Guillochon}, James and {Guzman}, Giannina and {Liedtke}, Simon and {Lian Lim}, Pey and {Lockhart}, Kelly E. and {Mommert}, Michael and {Morris}, Brett M. and {Norman}, Henrik and {Parikh}, Madhura and {Persson}, Magnus V. and {Robitaille}, Thomas P. and {Segovia}, Juan-Carlos and {Singer}, Leo P. and {Tollerud}, Erik J. and {de Val-Borro}, Miguel and {Valtchanov}, Ivan and {Woillez}, Julien and {Astroquery Collaboration} and {a subset of astropy Collaboration}},
        title = "{astroquery: An Astronomical Web-querying Package in Python}",
      journal = {\aj},
         year = 2019,
        month = mar,
       volume = {157},
       number = {3},
          eid = {98},
        pages = {98},
          doi = {10.3847/1538-3881/aafc33},
archivePrefix = {arXiv},
       eprint = {1901.04520},
 primaryClass = {astro-ph.IM},
       adsurl = {https://ui.adsabs.harvard.edu/abs/2019AJ....157...98G}
}

@ARTICLE{Crowe2025,
       author = {{Crowe}, Samuel and {Fedriani}, Rub{\'e}n and {Tan}, Jonathan C. and {Kinman}, Alva and {Zhang}, Yichen and {Andersen}, Morten and {Bravo Ferres}, Luc{\'\i}a and {Nogueras-Lara}, Francisco and {Sch{\"o}del}, Rainer and {Bally}, John and {Ginsburg}, Adam and {Cheng}, Yu and {Yang}, Yao-Lun and {Kendrew}, Sarah and {Law}, Chi-Yan and {Armstrong}, Joseph and {Li}, Zhi-Yun},
        title = "{The JWST-NIRCam View of Sagittarius C. I. Massive Star Formation and Protostellar Outflows}",
      journal = {\apj},
         year = 2025,
        month = apr,
       volume = {983},
       number = {1},
          eid = {19},
        pages = {19},
          doi = {10.3847/1538-4357/ad8889},
archivePrefix = {arXiv},
       eprint = {2410.09253},
 primaryClass = {astro-ph.GA},
       adsurl = {https://ui.adsabs.harvard.edu/abs/2025ApJ...983...19C}
}

@ARTICLE{Gordon2023,
       author = {{Gordon}, Karl D. and {Clayton}, Geoffrey C. and {Decleir}, Marjorie and {Fitzpatrick}, E.~L. and {Massa}, Derck and {Misselt}, Karl A. and {Tollerud}, Erik J.},
        title = "{One Relation for All Wavelengths: The Far-ultraviolet to Mid-infrared Milky Way Spectroscopic R(V)-dependent Dust Extinction Relationship}",
      journal = {\apj},
         year = 2023,
        month = jun,
       volume = {950},
       number = {2},
          eid = {86},
        pages = {86},
          doi = {10.3847/1538-4357/accb59},
archivePrefix = {arXiv},
       eprint = {2304.01991},
 primaryClass = {astro-ph.GA},
       adsurl = {https://ui.adsabs.harvard.edu/abs/2023ApJ...950...86G}
}

@ARTICLE{Mastrapa2009,
       author = {{Mastrapa}, R.~M. and {Sandford}, S.~A. and {Roush}, T.~L. and {Cruikshank}, D.~P. and {Dalle Ore}, C.~M.},
        title = "{Optical Constants of Amorphous and Crystalline H$_{2}$O-ice: 2.5-22 {\ensuremath{\mu}}m (4000-455 cm$^{-1}$) Optical Constants of H$_{2}$O-ice}",
      journal = {\apj},
         year = 2009,
        month = aug,
       volume = {701},
       number = {2},
        pages = {1347-1356},
          doi = {10.1088/0004-637X/701/2/1347},
       adsurl = {https://ui.adsabs.harvard.edu/abs/2009ApJ...701.1347M}
}

@ARTICLE{Gerakines2020,
       author = {{Gerakines}, Perry A. and {Hudson}, Reggie L.},
        title = "{A Modified Algorithm and Open-source Computational Package for the Determination of Infrared Optical Constants Relevant to Astrophysics}",
      journal = {\apj},
         year = 2020,
        month = sep,
       volume = {901},
       number = {1},
          eid = {52},
        pages = {52},
          doi = {10.3847/1538-4357/abad39},
       adsurl = {https://ui.adsabs.harvard.edu/abs/2020ApJ...901...52G}
}

@ARTICLE{Gunay2025,
       author = {{G{\"u}nay}, Burcu and {Gordon}, Karl D. and {Peek}, Joshua E.~G. and {Decleir}, Marjorie and {Van De Putte}, Dries and {Tchernyshyov}, Kirill and {Burton}, Michael G.},
        title = "{Photometric Mapping of Carbonaceous/Siliceous Dust and Water Ice in the ISM with JWST: Applications to the Dense Sightlines}",
      journal = {arXiv e-prints},
         year = 2025,
        month = jul,
          eid = {arXiv:2507.17550},
        pages = {arXiv:2507.17550},
archivePrefix = {arXiv},
       eprint = {2507.17550},
 primaryClass = {astro-ph.IM},
       adsurl = {https://ui.adsabs.harvard.edu/abs/2025arXiv250717550G}
}

@ARTICLE{Gibb2004,
       author = {{Gibb}, E.~L. and {Whittet}, D.~C.~B. and {Boogert}, A.~C.~A. and {Tielens}, A.~G.~G.~M.},
        title = "{Interstellar Ice: The Infrared Space Observatory Legacy}",
      journal = {\apjs},
         year = 2004,
        month = mar,
       volume = {151},
       number = {1},
        pages = {35-73},
          doi = {10.1086/381182},
       adsurl = {https://ui.adsabs.harvard.edu/abs/2004ApJS..151...35G}
}

@ARTICLE{Gunay2020,
       author = {{G{\"u}nay}, B. and {Burton}, M.~G. and {Af{\c{s}}ar}, M. and {Schmidt}, T.~W.},
        title = "{A method for mapping the aliphatic hydrocarbon content of interstellar dust towards the Galactic Centre}",
      journal = {\mnras},
         year = 2020,
        month = mar,
       volume = {493},
       number = {1},
        pages = {1109-1119},
          doi = {10.1093/mnras/staa288},
archivePrefix = {arXiv},
       eprint = {2002.04610},
 primaryClass = {astro-ph.GA},
       adsurl = {https://ui.adsabs.harvard.edu/abs/2020MNRAS.493.1109G}
}

@ARTICLE{Gunay2018,
       author = {{G{\"u}nay}, B. and {Schmidt}, T.~W. and {Burton}, M.~G. and {Af{\c{s}}ar}, M. and {Krechkivska}, O. and {Nauta}, K. and {Kable}, S.~H. and {Rawal}, A.},
        title = "{Aliphatic hydrocarbon content of interstellar dust}",
      journal = {\mnras},
         year = 2018,
        month = oct,
       volume = {479},
       number = {4},
        pages = {4336-4344},
          doi = {10.1093/mnras/sty1582},
archivePrefix = {arXiv},
       eprint = {2002.05116},
 primaryClass = {astro-ph.GA},
       adsurl = {https://ui.adsabs.harvard.edu/abs/2018MNRAS.479.4336G}
}

@ARTICLE{Mendez-Delgado2022,
       author = {{M{\'e}ndez-Delgado}, J.~E. and {Amayo}, A. and {Arellano-C{\'o}rdova}, K.~Z. and {Esteban}, C. and {Garc{\'\i}a-Rojas}, J. and {Carigi}, L. and {Delgado-Inglada}, G.},
        title = "{Gradients of chemical abundances in the Milky Way from H II regions: distances derived from Gaia EDR3 parallaxes and temperature inhomogeneities}",
      journal = {\mnras},
         year = 2022,
        month = mar,
       volume = {510},
       number = {3},
        pages = {4436-4455},
          doi = {10.1093/mnras/stab3782},
archivePrefix = {arXiv},
       eprint = {2112.12600},
 primaryClass = {astro-ph.GA},
       adsurl = {https://ui.adsabs.harvard.edu/abs/2022MNRAS.510.4436M}
}

@ARTICLE{Do2015,
       author = {{Do}, Tuan and {Kerzendorf}, Wolfgang and {Winsor}, Nathan and {St{\o}stad}, Morten and {Morris}, Mark R. and {Lu}, Jessica R. and {Ghez}, Andrea M.},
        title = "{Discovery of Low-metallicity Stars in the Central Parsec of the Milky Way}",
      journal = {\apj},
         year = 2015,
        month = aug,
       volume = {809},
       number = {2},
          eid = {143},
        pages = {143},
          doi = {10.1088/0004-637X/809/2/143},
archivePrefix = {arXiv},
       eprint = {1506.07891},
 primaryClass = {astro-ph.GA},
       adsurl = {https://ui.adsabs.harvard.edu/abs/2015ApJ...809..143D}
}

@ARTICLE{Nandakumar2018,
       author = {{Nandakumar}, G. and {Ryde}, N. and {Schultheis}, M. and {Thorsbro}, B. and {J{\"o}nsson}, H. and {Barklem}, P.~S. and {Rich}, R.~M. and {Fragkoudi}, F.},
        title = "{Chemical characterization of the inner Galactic bulge:North-South symmetry}",
      journal = {\mnras},
         year = 2018,
        month = aug,
       volume = {478},
       number = {4},
        pages = {4374-4389},
          doi = {10.1093/mnras/sty1255},
archivePrefix = {arXiv},
       eprint = {1805.05037},
 primaryClass = {astro-ph.GA},
       adsurl = {https://ui.adsabs.harvard.edu/abs/2018MNRAS.478.4374N}
}

@ARTICLE{2013A&A...558A..33A,
       author = {{Astropy Collaboration} and {Robitaille}, Thomas P. and {Tollerud}, Erik J. and {Greenfield}, Perry and {Droettboom}, Michael and {Bray}, Erik and {Aldcroft}, Tom and {Davis}, Matt and {Ginsburg}, Adam and {Price-Whelan}, Adrian M. and {Kerzendorf}, Wolfgang E. and {Conley}, Alexander and {Crighton}, Neil and {Barbary}, Kyle and {Muna}, Demitri and {Ferguson}, Henry and {Grollier}, Frédéric and {Parikh}, Madhura M. and {Nair}, Prasanth H. and {Günther}, Hans M. and {Deil}, Christoph and {Woillez}, Julien and {Conseil}, Simon and {Kramer}, Roban and {Turner}, James E. H. and {Singer}, Leo and {Fox}, Ryan and {Weaver}, Benjamin A. and {Zabalza}, Victor and {Edwards}, Zachary I. and {Azalee Bostroem}, K. and {Burke}, D. J. and {Casey}, Andrew R. and {Crawford}, Steven M. and {Dencheva}, Nadia and {Ely}, Justin and {Jenness}, Tim and {Labrie}, Kathleen and {Lim}, Pey Lian and {Pierfederici}, Francesco and {Pontzen}, Andrew and {Ptak}, Andy and {Refsdal}, Brian and {Servillat}, Mathieu and {Streicher}, Ole},
        title = "{Astropy: A community Python package for astronomy}",
      journal = {\aap},
         year = 2013,
        month = oct,
       volume = {558},
          eid = {A33},
        pages = {A33},
          doi = {10.1051/0004-6361/201322068},
archivePrefix = {arXiv},
       eprint = {1307.6212},
 primaryClass = {astro-ph.IM},
       adsurl = {https://ui.adsabs.harvard.edu/abs/2013A&A...558A..33A}
}

@ARTICLE{2020SciPy-NMeth,
       author = {{Virtanen}, Pauli and {Gommers}, Ralf and {Oliphant}, Travis E. and {Haberland}, Matt and {Reddy}, Tyler and {Cournapeau}, David and {Burovski}, Evgeni and {Peterson}, Pearu and {Weckesser}, Warren and {Bright}, Jonathan and {van der Walt}, Stéfan J. and {Brett}, Matthew and {Wilson}, Joshua and {Millman}, K. Jarrod and {Mayorov}, Nikolay and {Nelson}, Andrew R. J. and {Jones}, Eric and {Kern}, Robert and {Larson}, Eric and {Carey}, C J and {Polat}, Ilhan and {Feng}, Yu and {Moore}, Eric W. and {VanderPlas}, Jake and {Laxalde}, Denis and {Perktold}, Josef and {Cimrman}, Robert and {Henriksen}, Ian and {Quintero}, E. A. and {Harris}, Charles R. and {Archibald}, Anne M. and {Ribeiro}, Antônio H. and {Pedregosa}, Fabian and {van Mulbregt}, Paul and {SciPy 1.0 Contributors}},
        title = "{SciPy 1.0: Fundamental Algorithms for Scientific Computing in Python}",
      journal = {Nature Methods},
         year = 2020,
        month = feb,
       volume = {17},
        pages = {261-272},
          doi = {10.1038/s41592-019-0686-2},
       adsurl = {https://ui.adsabs.harvard.edu/abs/2020NatMt..17..261V}
}

@ARTICLE{2020NumPy-Array,
       author = {{Harris}, Charles R. and {Millman}, K. Jarrod and {van der Walt}, Stéfan J and {Gommers}, Ralf and {Virtanen}, Pauli and {Cournapeau}, David and {Wieser}, Eric and {Taylor}, Julian and {Berg}, Sebastian and {Smith}, Nathaniel J. and {Kern}, Robert and {Picus}, Matti and {Hoyer}, Stephan and {van Kerkwijk}, Marten H. and {Brett}, Matthew and {Haldane}, Allan and {Fernández del Río}, Jaime and {Wiebe}, Mark and {Peterson}, Pearu and {Gérard-Marchant}, Pierre and {Sheppard}, Kevin and {Reddy}, Tyler and {Weckesser}, Warren and {Abbasi}, Hameer and {Gohlke}, Christoph and {Oliphant}, Travis E.},
        title = "{Array programming with NumPy}",
      journal = {Nature},
         year = 2020,
        month = sep,
       volume = {585},
        pages = {357-362},
          doi = {10.1038/s41586-020-2649-2},
       adsurl = {https://ui.adsabs.harvard.edu/abs/2020Natur.585..357H}
}

@ARTICLE{2023PASP..135d8001R,
       author = {{Rigby}, Jane and {Perrin}, Marshall and {McElwain}, Michael and {Kimble}, Randy and {Friedman}, Scott and {Lallo}, Matt and {Doyon}, René and {Feinberg}, Lee and {Ferruit}, Pierre and {Glasse}, Alistair and {Rieke}, Marcia and {Rieke}, George and {Wright}, Gillian and {Willott}, Chris and others},
        title = "{The Science Performance of JWST as Characterized in Commissioning}",
      journal = {\pasp},
         year = 2023,
        month = apr,
       volume = {135},
       number = {1046},
          eid = {048001},
        pages = {048001},
          doi = {10.1088/1538-3873/acb293},
archivePrefix = {arXiv},
       eprint = {2207.05632},
 primaryClass = {astro-ph.IM},
       adsurl = {https://ui.adsabs.harvard.edu/abs/2023PASP..135d8001R}
}

@ARTICLE{2012ivoa.rept.1015R,
       author = {{Rodrigo}, C. and {Solano}, E. and {Bayo}, A.},
        title = "{SVO Filter Profile Service Version 1.0}",
      journal = {IVOA Working Draft 15 October 2012},
         year = 2012,
       adsurl = {https://ui.adsabs.harvard.edu/abs/2012ivoa.rept.1015R}
}

@ARTICLE{2020sea..confE.182R,
       author = {{Rodrigo}, C. and {Solano}, E.},
        title = "{The SVO Filter Profile Service}",
      journal = {The Spanish Virtual Observatory},
         year = 2020,
        pages = {182},
       adsurl = {https://ui.adsabs.harvard.edu/abs/2020sea..confE.182R}
}

@ARTICLE{Taillard2025,
       author = {{Taillard}, A. and {Mart{\'\i}n-Dom{\'e}nech}, R. and {Carrascosa}, H. and {Noble}, J.~A. and {Mu{\~n}oz Caro}, G.~M. and {Dartois}, E. and {Navarro-Almaida}, D. and {Escribano}, B. and {S{\'a}nchez-Monge}, {\'A}. and {Fuente}, A.},
        title = "{Predicting the detectability of sulphur-bearing molecules in the solid phase with simulated spectra of JWST instruments}",
      journal = {\aap},
         year = 2025,
        month = feb,
       volume = {694},
          eid = {A263},
        pages = {A263},
          doi = {10.1051/0004-6361/202452900},
archivePrefix = {arXiv},
       eprint = {2502.09384},
 primaryClass = {astro-ph.GA},
       adsurl = {https://ui.adsabs.harvard.edu/abs/2025A&A...694A.263T}
}

@ARTICLE{Moneti2001,
       author = {{Moneti}, Andrea and {Cernicharo}, Jos{\'e} and {Pardo}, Juan Ram{\'o}n},
        title = "{Cold H$_{2}$O and CO Ice and Gas toward the Galactic Center}",
      journal = {\apjl},
         year = 2001,
        month = mar,
       volume = {549},
       number = {2},
        pages = {L203-L207},
          doi = {10.1086/319168},
archivePrefix = {arXiv},
       eprint = {astro-ph/0012292},
 primaryClass = {astro-ph},
       adsurl = {https://ui.adsabs.harvard.edu/abs/2001ApJ...549L.203M}
}

@ARTICLE{Roser2021,
       author = {{Roser}, Joseph E. and {Ricca}, Alessandra and {Cartwright}, Richard J. and {Dalle Ore}, Cristina and {Cruikshank}, Dale P.},
        title = "{The Infrared Complex Refractive Index of Amorphous Ammonia Ice at 40 K (1.43-22.73 {\ensuremath{\mu}}m) and Its Relevance to Outer Solar System Bodies}",
      journal = {\psj},
         year = 2021,
        month = dec,
       volume = {2},
       number = {6},
          eid = {240},
        pages = {240},
          doi = {10.3847/PSJ/ac3336},
       adsurl = {https://ui.adsabs.harvard.edu/abs/2021PSJ.....2..240R}
}

@article{Rocha2014,
title = {Determination of optical constants n and k of thin films from absorbance data using Kramers-Kronig relationship},
journal = {Spectrochimica Acta Part A: Molecular and Biomolecular Spectroscopy},
volume = {123},
pages = {436-446},
year = {2014},
issn = {1386-1425},
doi = {https://doi.org/10.1016/j.saa.2013.12.075},
url = {https://www.sciencedirect.com/science/article/pii/S1386142513015060},
author = {W.R.M. Rocha and S. Pilling}
}

@ARTICLE{Mukai1986,
       author = {{Mukai}, T. and {Kraetschmer}, W.},
        title = "{Optical Constants of the Mixture of Ices}",
      journal = {Earth Moon and Planets},
         year = 1986,
        month = oct,
       volume = {36},
       number = {2},
        pages = {145-155},
          doi = {10.1007/BF00057607},
       adsurl = {https://ui.adsabs.harvard.edu/abs/1986EM&P...36..145M}
}

@ARTICLE{Meingast2025,
       author = {{Meingast}, Stefan},
        title = "{Mapping water ice with infrared broadband photometry}",
      journal = {arXiv e-prints},
         year = 2025,
        month = jul,
          eid = {arXiv:2507.18688},
        pages = {arXiv:2507.18688},
          doi = {10.48550/arXiv.2507.18688},
archivePrefix = {arXiv},
       eprint = {2507.18688},
 primaryClass = {astro-ph.GA},
       adsurl = {https://ui.adsabs.harvard.edu/abs/2025arXiv250718688M}
}

@misc{Bradley2025,
       author = {{Bradley}, Larry and {Sip{\H{o}}cz}, Brigitta and {Robitaille}, Thomas and {Tollerud}, Erik and {Vin{\'\i}cius}, Z{\'e} and {Deil}, Christoph and {Barbary}, Kyle and {Wilson}, Tom J and {Busko}, Ivo and {Donath}, Axel and {G{\"u}nther}, Hans Moritz and {Cara}, Mihai and {Lim}, P.~L. and {Me{\ss}linger}, Sebastian and {Burnett}, Zach and {Conseil}, Simon and {Droettboom}, Michael and {Bostroem}, Azalee and {Bray}, E.~M. and {Andersen Bratholm}, Lars and {Jamieson}, William and {Ginsburg}, Adam and {Barentsen}, Geert and {Craig}, Matt and {Pascual}, Sergio and {Rathi}, Shivangee and {Perrin}, Marshall and {Morris}, Brett M.},
        title = "{astropy/photutils: 2.2.0}",
         year = 2025,
        month = feb,
          eid = {10.5281/zenodo.596036},
          doi = {10.5281/zenodo.596036},
      version = {2.2.0},
    publisher = {Zenodo},
       adsurl = {https://ui.adsabs.harvard.edu/abs/2022zndo....596036B}
}

@INPROCEEDINGS{Perrin2012,
       author = {{Perrin}, Marshall D. and {Soummer}, R{\'e}mi and {Elliott}, Erin M. and {Lallo}, Matthew D. and {Sivaramakrishnan}, Anand},
        title = "{Simulating point spread functions for the James Webb Space Telescope with WebbPSF}",
    booktitle = {Space Telescopes and Instrumentation 2012: Optical, Infrared, and Millimeter Wave},
         year = 2012,
       editor = {{Clampin}, Mark C. and {Fazio}, Giovanni G. and {MacEwen}, Howard A. and {Oschmann}, Jr., Jacobus M.},
       series = {Society of Photo-Optical Instrumentation Engineers (SPIE) Conference Series},
       volume = {8442},
        month = sep,
          eid = {84423D},
        pages = {84423D},
          doi = {10.1117/12.925230},
       adsurl = {https://ui.adsabs.harvard.edu/abs/2012SPIE.8442E..3DP}
}

@INPROCEEDINGS{Perrin2014,
       author = {{Perrin}, Marshall D. and {Sivaramakrishnan}, Anand and {Lajoie}, Charles-Philippe and {Elliott}, Erin and {Pueyo}, Laurent and {Ravindranath}, Swara and {Albert}, Lo{\"\i}c.},
        title = "{Updated point spread function simulations for JWST with WebbPSF}",
    booktitle = {Space Telescopes and Instrumentation 2014: Optical, Infrared, and Millimeter Wave},
         year = 2014,
       editor = {{Oschmann}, Jr., Jacobus M. and {Clampin}, Mark and {Fazio}, Giovanni G. and {MacEwen}, Howard A.},
       series = {Society of Photo-Optical Instrumentation Engineers (SPIE) Conference Series},
       volume = {9143},
        month = aug,
          eid = {91433X},
        pages = {91433X},
          doi = {10.1117/12.2056689},
       adsurl = {https://ui.adsabs.harvard.edu/abs/2014SPIE.9143E..3XP}
}

@ARTICLE{Gerakines1999,
       author = {{Gerakines}, P.~A. and {Whittet}, D.~C.~B. and {Ehrenfreund}, P. and {Boogert}, A.~C.~A. and {Tielens}, A.~G.~G.~M. and {Schutte}, W.~A. and {Chiar}, J.~E. and {van Dishoeck}, E.~F. and {Prusti}, T. and {Helmich}, F.~P. and {de Graauw}, Th.},
        title = "{Observations of Solid Carbon Dioxide in Molecular Clouds with the Infrared Space Observatory}",
      journal = {\apj},
         year = 1999,
        month = sep,
       volume = {522},
       number = {1},
        pages = {357-377},
          doi = {10.1086/307611},
       adsurl = {https://ui.adsabs.harvard.edu/abs/1999ApJ...522..357G}
}

@ARTICLE{Pontoppidan2006,
       author = {{Pontoppidan}, K.~M.},
        title = "{Spatial mapping of ices in the Ophiuchus-F core. A direct measurement of CO depletion and the formation of CO\_2}",
      journal = {\aap},
         year = 2006,
        month = jul,
       volume = {453},
       number = {3},
        pages = {L47-L50},
          doi = {10.1051/0004-6361:20065569},
archivePrefix = {arXiv},
       eprint = {astro-ph/0605576},
 primaryClass = {astro-ph},
       adsurl = {https://ui.adsabs.harvard.edu/abs/2006A&A...453L..47P}
}

@ARTICLE{Harker1997,
       author = {{Harker}, D. and {Bregman}, J. and {Tielens}, A.~G.~G.~M. and {Temi}, P. and {Rank}, D.},
        title = "{The infrared reflection nebula around the embedded sources in S 140.}",
      journal = {\aap},
         year = 1997,
        month = aug,
       volume = {324},
        pages = {629-640},
       adsurl = {https://ui.adsabs.harvard.edu/abs/1997A&A...324..629H}
}

@ARTICLE{2025PSJ.....6..154M,
       author = {{Markwardt}, Larissa and {Wen Lin}, Hsing and {Holler}, Bryan J. and {Gerdes}, David W. and {Adams}, Fred C. and {Malhotra}, Renu and {Napier}, Kevin J.},
        title = "{From Colors to Spectra and Back Again: First Near-IR Spectroscopic Survey of Neptunian Trojans}",
      journal = {\psj},
         year = 2025,
        month = jul,
       volume = {6},
       number = {7},
          eid = {154},
        pages = {154},
          doi = {10.3847/PSJ/addecd},
archivePrefix = {arXiv},
       eprint = {2310.03998},
 primaryClass = {astro-ph.EP},
       adsurl = {https://ui.adsabs.harvard.edu/abs/2025PSJ.....6..154M}
}

@ARTICLE{2024NatCo..15.8247P,
       author = {{Protopapa}, Silvia and {Raut}, Ujjwal and {Wong}, Ian and {Stansberry}, John and {Villanueva}, Geronimo L. and {Cook}, Jason and {Holler}, Bryan and {Grundy}, William M. and {Brunetto}, Rosario and {Cartwright}, Richard J. and {Mamo}, Bereket and {Emery}, Joshua P. and {Parker}, Alex H. and {Guilbert-Lepoutre}, Aurelie and {Pinilla-Alonso}, Noemi and {Milam}, Stefanie N. and {Hammel}, Heidi B.},
        title = "{Detection of carbon dioxide and hydrogen peroxide on the stratified surface of Charon with JWST}",
      journal = {Nature Communications},
         year = 2024,
        month = dec,
       volume = {15},
       number = {1},
          eid = {8247},
        pages = {8247},
          doi = {10.1038/s41467-024-51826-4},
       adsurl = {https://ui.adsabs.harvard.edu/abs/2024NatCo..15.8247P}
}

@ARTICLE{2008AJ....135...55B,
       author = {{Barkume}, K.~M. and {Brown}, M.~E. and {Schaller}, E.~L.},
        title = "{Near-Infrared Spectra of Centaurs and Kuiper Belt Objects}",
      journal = {\aj},
         year = 2008,
        month = jan,
       volume = {135},
       number = {1},
        pages = {55-67},
          doi = {10.1088/0004-6256/135/1/55},
       adsurl = {https://ui.adsabs.harvard.edu/abs/2008AJ....135...55B}
}

@ARTICLE{Noble2017,
       author = {{Noble}, J.~A. and {Fraser}, H.~J. and {Pontoppidan}, K.~M. and {Craigon}, A.~M.},
        title = "{Two-dimensional ice mapping of molecular cores}",
      journal = {\mnras},
         year = 2017,
        month = jun,
       volume = {467},
       number = {4},
        pages = {4753-4762},
          doi = {10.1093/mnras/stx329},
archivePrefix = {arXiv},
       eprint = {1703.01182},
 primaryClass = {astro-ph.GA},
       adsurl = {https://ui.adsabs.harvard.edu/abs/2017MNRAS.467.4753N}
}

@ARTICLE{Kurucz1979,
       author = {{Kurucz}, R.~L.},
        title = "{Model atmospheres for G, F, A, B, and O stars.}",
      journal = {\apjs},
         year = 1979,
        month = may,
       volume = {40},
        pages = {1-340},
          doi = {10.1086/190589},
       adsurl = {https://ui.adsabs.harvard.edu/abs/1979ApJS...40....1K}
}

@ARTICLE{Husser2013,
       author = {{Husser}, T. -O. and {Wende-von Berg}, S. and {Dreizler}, S. and {Homeier}, D. and {Reiners}, A. and {Barman}, T. and {Hauschildt}, P.~H.},
        title = "{A new extensive library of PHOENIX stellar atmospheres and synthetic spectra}",
      journal = {\aap},
         year = 2013,
        month = may,
       volume = {553},
          eid = {A6},
        pages = {A6},
          doi = {10.1051/0004-6361/201219058},
archivePrefix = {arXiv},
       eprint = {1303.5632},
 primaryClass = {astro-ph.SR},
       adsurl = {https://ui.adsabs.harvard.edu/abs/2013A&A...553A...6H}
}

@ARTICLE{Eltareb2024,
       author = {{Eltareb}, Ali and {Lopez}, Gustavo E. and {Giovambattista}, Nicolas},
        title = "{A continuum of amorphous ices between low-density and high-density amorphous ice}",
      journal = {Communications Chemistry},
         year = 2024,
        month = feb,
       volume = {7},
       number = {1},
          eid = {36},
        pages = {36},
          doi = {10.1038/s42004-024-01117-2},
       adsurl = {https://ui.adsabs.harvard.edu/abs/2024CmChe...7...36E}
}

@ARTICLE{Luna2022,
       author = {{Luna}, Ram{\'o}n and {Mill{\'a}n}, Carlos and {Domingo}, Manuel and {Santonja}, Carmina and {Satorre}, Miguel {\'A}.},
        title = "{Density and Refractive Index of Carbon Monoxide Ice at Different Temperatures}",
      journal = {\apj},
         year = 2022,
        month = aug,
       volume = {935},
       number = {2},
          eid = {134},
        pages = {134},
          doi = {10.3847/1538-4357/ac8001},
       adsurl = {https://ui.adsabs.harvard.edu/abs/2022ApJ...935..134L}
}

@ARTICLE{Pontoppidan2024,
       author = {{Pontoppidan}, Klaus M. and {Evans}, Neal and {Bergner}, Jennifer and {Yang}, Yao-Lun},
        title = "{A Constrained Dust Opacity for Models of Dense Clouds and Protostellar Envelopes}",
      journal = {Research Notes of the American Astronomical Society},
         year = 2024,
        month = mar,
       volume = {8},
       number = {3},
          eid = {68},
        pages = {68},
          doi = {10.3847/2515-5172/ad303f},
       adsurl = {https://ui.adsabs.harvard.edu/abs/2024RNAAS...8...68P}
}

@ARTICLE{Chapman2009,
       author = {{Chapman}, Nicholas L. and {Mundy}, Lee G. and {Lai}, Shih-Ping and {Evans}, II, Neal J.},
        title = "{The Mid-Infrared Extinction Law in the Ophiuchus, Perseus, and Serpens Molecular Clouds}",
      journal = {\apj},
         year = 2009,
        month = jan,
       volume = {690},
       number = {1},
        pages = {496-511},
          doi = {10.1088/0004-637X/690/1/496},
archivePrefix = {arXiv},
       eprint = {0809.1106},
 primaryClass = {astro-ph},
       adsurl = {https://ui.adsabs.harvard.edu/abs/2009ApJ...690..496C}
}

@ARTICLE{Lacy2017,
       author = {{Lacy}, John H. and {Sneden}, Christopher and {Kim}, Hwihyun and {Jaffe}, Daniel T.},
        title = "{H$_{2}$, CO, and Dust Absorption through Cold Molecular Clouds}",
      journal = {\apj},
         year = 2017,
        month = mar,
       volume = {838},
       number = {1},
          eid = {66},
        pages = {66},
          doi = {10.3847/1538-4357/aa6247},
archivePrefix = {arXiv},
       eprint = {1703.09826},
 primaryClass = {astro-ph.GA},
       adsurl = {https://ui.adsabs.harvard.edu/abs/2017ApJ...838...66L}
}

@ARTICLE{Simpson2018,
       author = {{Simpson}, Janet P.},
        title = "{Spitzer Infrared Spectrograph Observations of the Galactic Center: Quantifying the Extreme Ultraviolet/Soft X-ray Fluxes}",
      journal = {\apj},
         year = 2018,
        month = apr,
       volume = {857},
       number = {1},
          eid = {59},
        pages = {59},
          doi = {10.3847/1538-4357/aab55b},
archivePrefix = {arXiv},
       eprint = {1803.02806},
 primaryClass = {astro-ph.GA},
       adsurl = {https://ui.adsabs.harvard.edu/abs/2018ApJ...857...59S}
}

@ARTICLE{Caselli2022,
       author = {{Caselli}, Paola and {Pineda}, Jaime E. and {Sipil{\"a}}, Olli and {Zhao}, Bo and {Redaelli}, Elena and {Spezzano}, Silvia and {Maureira}, Maria Jos{\'e} and {Alves}, Felipe and {Bizzocchi}, Luca and {Bourke}, Tyler L. and {Chac{\'o}n-Tanarro}, Ana and {Friesen}, Rachel and {Galli}, Daniele and {Harju}, Jorma and {Jim{\'e}nez-Serra}, Izaskun and {Keto}, Eric and {Li}, Zhi-Yun and {Padovani}, Marco and {Schmiedeke}, Anika and {Tafalla}, Mario and {Vastel}, Charlotte},
        title = "{The Central 1000 au of a Prestellar Core Revealed with ALMA. II. Almost Complete Freeze-out}",
      journal = {\apj},
         year = 2022,
        month = apr,
       volume = {929},
       number = {1},
          eid = {13},
        pages = {13},
          doi = {10.3847/1538-4357/ac5913},
archivePrefix = {arXiv},
       eprint = {2202.13374},
 primaryClass = {astro-ph.SR},
       adsurl = {https://ui.adsabs.harvard.edu/abs/2022ApJ...929...13C}
}

@ARTICLE{Rubinstein2024,
       author = {{Rubinstein}, Adam E. and {Evans}, Neal J. and {Tyagi}, Himanshu and {Narang}, Mayank and {Nazari}, Pooneh and {Gutermuth}, Robert and {Federman}, Samuel and {Manoj}, P. and {Green}, Joel D. and {Watson}, Dan M. and {Megeath}, S. Thomas and {Rocha}, Will R.~M. and {Brunken}, Nashanty G.~C. and {Slavicinska}, Katerina and {van Dishoeck}, Ewine F. and {Beuther}, Henrik and {Bourke}, Tyler L. and {Caratti o Garatti}, Alessio and {Hartmann}, Lee and {Klaassen}, Pamela and {Linz}, Hendrik and {Looney}, Leslie W. and {Muzerolle}, James and {Stanke}, Thomas and {Tobin}, John J. and {Wolk}, Scott J. and {Yang}, Yao-Lun},
        title = "{IPA: Class 0 Protostars Viewed in CO Emission Using JWST}",
      journal = {\apj},
         year = 2024,
        month = oct,
       volume = {974},
       number = {1},
          eid = {112},
        pages = {112},
          doi = {10.3847/1538-4357/ad6b92},
archivePrefix = {arXiv},
       eprint = {2312.07807},
 primaryClass = {astro-ph.SR},
       adsurl = {https://ui.adsabs.harvard.edu/abs/2024ApJ...974..112R}
}

@ARTICLE{Okoda2025,
       author = {{Okoda}, Yuki and {Yang}, Yao-Lun and {Evans}, II, Neal J. and {Kim}, Jaeyeong and {Jin}, Mihwa and {Garrod}, Robin T. and {Francis}, Logan and {Johnstone}, Doug and {Ceccarelli}, Cecilia and {Codella}, Claudio and {Chandler}, Claire J. and {Yamamoto}, Satoshi and {Sakai}, Nami},
        title = "{CORINOS. III. Outflow Shocked Regions of the Low-mass Protostellar Source IRAS 15398{\textendash}3359 with JWST and ALMA}",
      journal = {\apj},
         year = 2025,
        month = apr,
       volume = {982},
       number = {2},
          eid = {149},
        pages = {149},
          doi = {10.3847/1538-4357/adb83f},
archivePrefix = {arXiv},
       eprint = {2503.03050},
 primaryClass = {astro-ph.SR},
       adsurl = {https://ui.adsabs.harvard.edu/abs/2025ApJ...982..149O}
}

@ARTICLE{Yang2022,
       author = {{Yang}, Yao-Lun and {Green}, Joel D. and {Pontoppidan}, Klaus M. and {Bergner}, Jennifer B. and {Cleeves}, L. Ilsedore and {Evans}, II, Neal J. and {Garrod}, Robin T. and {Jin}, Miwha and {Kim}, Chul Hwan and {Kim}, Jaeyeong and {Lee}, Jeong-Eun and {Sakai}, Nami and {Shingledecker}, Christopher N. and {Shope}, Brielle and {Tobin}, John J. and {van Dishoeck}, Ewine F.},
        title = "{CORINOS. I. JWST/MIRI Spectroscopy and Imaging of a Class 0 Protostar IRAS 15398{\textendash}3359}",
      journal = {\apjl},
         year = 2022,
        month = dec,
       volume = {941},
       number = {1},
          eid = {L13},
        pages = {L13},
          doi = {10.3847/2041-8213/aca289},
archivePrefix = {arXiv},
       eprint = {2208.10673},
 primaryClass = {astro-ph.SR},
       adsurl = {https://ui.adsabs.harvard.edu/abs/2022ApJ...941L..13Y}
}

@ARTICLE{Federman2024,
       author = {{Federman}, Samuel A. and {Megeath}, S. Thomas and {Rubinstein}, Adam E. and {Gutermuth}, Robert and {Narang}, Mayank and {Tyagi}, Himanshu and {Manoj}, P. and {Anglada}, Guillem and {Atnagulov}, Prabhani and {Beuther}, Henrik and {Bourke}, Tyler L. and {Brunken}, Nashanty and {Caratti o Garatti}, Alessio and {Evans}, Neal J. and {Fischer}, William J. and {Furlan}, Elise and {Green}, Joel D. and {Habel}, Nolan and {Hartmann}, Lee and {Karnath}, Nicole and {Klaassen}, Pamela and {Linz}, Hendrik and {Looney}, Leslie W. and {Osorio}, Mayra and {Muzerolle Page}, James and {Nazari}, Pooneh and {Pokhrel}, Riwaj and {Rahatgaonkar}, Rohan and {Rocha}, Will R.~M. and {Sheehan}, Patrick and {Slavicinska}, Katerina and {Stanke}, Thomas and {Stutz}, Amelia M. and {Tobin}, John J. and {Tychoniec}, Lukasz and {Van Dishoeck}, Ewine F. and {Watson}, Dan M. and {Wolk}, Scott and {Yang}, Yao-Lun},
        title = "{Investigating Protostellar Accretion-driven Outflows across the Mass Spectrum: JWST NIRSpec Integral Field Unit 3{\textendash}5 {\ensuremath{\mu}}m Spectral Mapping of Five Young Protostars}",
      journal = {\apj},
         year = 2024,
        month = may,
       volume = {966},
       number = {1},
          eid = {41},
        pages = {41},
          doi = {10.3847/1538-4357/ad2fa0},
archivePrefix = {arXiv},
       eprint = {2310.03803},
 primaryClass = {astro-ph.SR},
       adsurl = {https://ui.adsabs.harvard.edu/abs/2024ApJ...966...41F}
}

@ARTICLE{Ehrenfreund1997,
       author = {{Ehrenfreund}, P. and {Boogert}, A.~C.~A. and {Gerakines}, P.~A. and {Tielens}, A.~G.~G.~M. and {van Dishoeck}, E.~F.},
        title = "{Infrared spectroscopy of interstellar apolar ice analogs}",
      journal = {\aap},
         year = 1997,
        month = dec,
       volume = {328},
        pages = {649-669},
       adsurl = {https://ui.adsabs.harvard.edu/abs/1997A&A...328..649E}
}

@ARTICLE{Ehrenfreund1996a,
       author = {{Ehrenfreund}, P. and {Boogert}, A.~C.~A. and {Gerakines}, P.~A. and {Jansen}, D.~J. and {Schutte}, W.~A. and {Tielens}, A.~G.~G.~M. and {van Dishoeck}, E.~F.},
        title = "{A laboratory database of solid CO and CO\_2\_ for ISO.}",
      journal = {\aap},
         year = 1996,
        month = nov,
       volume = {315},
        pages = {L341-L344},
       adsurl = {https://ui.adsabs.harvard.edu/abs/1996A&A...315L.341E}
}

@ARTICLE{Ehrenfreund1996b,
       author = {{Ehrenfreund}, P. and {Gerakines}, P.~A. and {Schutte}, W.~A. and {van Hemert}, M.~C. and {van Dishoeck}, E.~F.},
        title = "{Infrared properties of isolated water ice.}",
      journal = {\aap},
         year = 1996,
        month = aug,
       volume = {312},
        pages = {263-274},
       adsurl = {https://ui.adsabs.harvard.edu/abs/1996A&A...312..263E}
}

@ARTICLE{Gramze2025,
       author = {{Gramze}, Savannah and {Ginsburg}, Adam and {Budaiev}, Nazar and {Bulatek}, Alyssa and {Richardson}, Theo and {Barnes}, A.~T. and {Santa-Maria}, Miriam G. and {Sormani}, Mattia C. and {Lu}, Xing and {Nogueras-Lara}, Francisco and {Gaches}, Brandt A.~L. and {Battersby}, Cara D. and {Wallace}, Jennifer and {Walker}, Daniel L. and {Mills}, Elisabeth A.~C. and {Mattern}, Michael},
        title = "{Mapping CO Ice in a Star-Forming Filament in the 3 kpc Arm with JWST}",
      journal = {arXiv e-prints},
         year = 2025,
        month = sep,
          eid = {arXiv:2509.21763},
        pages = {arXiv:2509.21763},
archivePrefix = {arXiv},
       eprint = {2509.21763},
 primaryClass = {astro-ph.GA},
       adsurl = {https://ui.adsabs.harvard.edu/abs/2025arXiv250921763G}
}

@ARTICLE{Pecaut2013,
       author = {{Pecaut}, Mark J. and {Mamajek}, Eric E.},
        title = "{Intrinsic Colors, Temperatures, and Bolometric Corrections of Pre-main-sequence Stars}",
      journal = {\apjs},
         year = 2013,
        month = sep,
       volume = {208},
       number = {1},
          eid = {9},
        pages = {9},
          doi = {10.1088/0067-0049/208/1/9},
archivePrefix = {arXiv},
       eprint = {1307.2657},
 primaryClass = {astro-ph.SR},
       adsurl = {https://ui.adsabs.harvard.edu/abs/2013ApJS..208....9P}
}

@ARTICLE{Girardi2005,
       author = {{Girardi}, L. and {Groenewegen}, M.~A.~T. and {Hatziminaoglou}, E. and {da Costa}, L.},
        title = "{Star counts in the Galaxy. Simulating from very deep to very shallow photometric surveys with the TRILEGAL code}",
      journal = {\aap},
         year = 2005,
        month = jun,
       volume = {436},
       number = {3},
        pages = {895-915},
          doi = {10.1051/0004-6361:20042352},
archivePrefix = {arXiv},
       eprint = {astro-ph/0504047},
 primaryClass = {astro-ph},
       adsurl = {https://ui.adsabs.harvard.edu/abs/2005A&A...436..895G}
}

@ARTICLE{Dartois2024,
       author = {{Dartois}, E. and {Noble}, J.~A. and {Caselli}, P. and {Fraser}, H.~J. and {Jim{\'e}nez-Serra}, I. and {Mat{\'e}}, B. and {McClure}, M.~K. and {Melnick}, G.~J. and {Pendleton}, Y.~J. and {Shimonishi}, T. and {Smith}, Z.~L. and {Sturm}, J.~A. and {Taillard}, A. and {Wakelam}, V. and {Boogert}, A.~C.~A. and {Drozdovskaya}, M.~N. and {Erkal}, J. and {Harsono}, D. and {Herrero}, V.~J. and {Ioppolo}, S. and {Linnartz}, H. and {McGuire}, B.~A. and {Perotti}, G. and {Qasim}, D. and {Rocha}, W.~R.~M.},
        title = "{Spectroscopic sizing of interstellar icy grains with JWST}",
      journal = {Nature Astronomy},
         year = 2024,
        month = mar,
       volume = {8},
        pages = {359-367},
          doi = {10.1038/s41550-023-02155-x},
       adsurl = {https://ui.adsabs.harvard.edu/abs/2024NatAs...8..359D}
}

@ARTICLE{BravoFerres2025,
       author = {{Bravo Ferres}, Luc{\'\i}a and {Nogueras-Lara}, Francisco and {Sch{\"o}del}, Rainer and {Fedriani}, Rub{\'e}n and {Ginsburg}, Adam and {Crowe}, Samuel and {Tan}, Jonathan C. and {Andersen}, Morten and {Armstrong}, Joseph and {Cheng}, Yu and {Li}, Zhi-Yun},
        title = "{The JWST-NIRCam View of Sagittarius C. III. The Extinction Curve}",
      journal = {arXiv e-prints},
         year = 2025,
        month = oct,
          eid = {arXiv:2510.10749},
        pages = {arXiv:2510.10749},
          doi = {10.48550/arXiv.2510.10749},
archivePrefix = {arXiv},
       eprint = {2510.10749},
 primaryClass = {astro-ph.GA},
       adsurl = {https://ui.adsabs.harvard.edu/abs/2025arXiv251010749B}
}

@ARTICLE{Hensley2021,
       author = {{Hensley}, Brandon S. and {Draine}, B.~T.},
        title = "{Observational Constraints on the Physical Properties of Interstellar Dust in the Post-Planck Era}",
      journal = {\apj},
         year = 2021,
        month = jan,
       volume = {906},
       number = {2},
          eid = {73},
        pages = {73},
          doi = {10.3847/1538-4357/abc8f1},
archivePrefix = {arXiv},
       eprint = {2009.00018},
 primaryClass = {astro-ph.GA},
       adsurl = {https://ui.adsabs.harvard.edu/abs/2021ApJ...906...73H}
}

@ARTICLE{Jenkins2009,
       author = {{Jenkins}, Edward B.},
        title = "{A Unified Representation of Gas-Phase Element Depletions in the Interstellar Medium}",
      journal = {\apj},
         year = 2009,
        month = aug,
       volume = {700},
       number = {2},
        pages = {1299-1348},
          doi = {10.1088/0004-637X/700/2/1299},
archivePrefix = {arXiv},
       eprint = {0905.3173},
 primaryClass = {astro-ph.GA},
       adsurl = {https://ui.adsabs.harvard.edu/abs/2009ApJ...700.1299J}
}

@INPROCEEDINGS{Girardi2012,
       author = {{Girardi}, L{\'e}o and {Barbieri}, Mauro and {Groenewegen}, Martin A.~T. and {Marigo}, Paola and {Bressan}, Alessandro and {Rocha-Pinto}, Helio J. and {Santiago}, Bas{\'\i}lio X. and {Camargo}, Julio I.~B. and {da Costa}, Luiz N.},
        title = "{TRILEGAL, a TRIdimensional modeL of thE GALaxy: Status and Future}",
    booktitle = {Red Giants as Probes of the Structure and Evolution of the Milky Way},
         year = 2012,
       editor = {{Miglio}, Andrea and {Montalb{\'a}n}, Josefina and {Noels}, Arlette},
       series = {Astrophysics and Space Science Proceedings},
       volume = {26},
        month = jan,
        pages = {165},
          doi = {10.1007/978-3-642-18418-5_17},
       adsurl = {https://ui.adsabs.harvard.edu/abs/2012ASSP...26..165G}
}

@ARTICLE{Bock2026,
       author = {{Bock}, James J. and {Aboobaker}, Asad M. and {Adamo}, Joseph and {Akeson}, Rachel and {Alred}, John M. and {Alibay}, Farah and {Ashby}, Matthew L.~N. and {Bach}, Yoonsoo P. and {Bleem}, Lindsey E. and {Bolton}, Douglas and {Braun}, David F. and {Bruton}, Sean and {Bryan}, Sean A. and {Chang}, Tzu-Ching and {Chen}, Shuang-Shuang and {Cheng}, Yun-Ting and {Cheshire}, IV, James R. and {Chiang}, Yi-Kuan and {Choppin de Janvry}, Jean and {Condon}, Samuel and {Cook}, Walter R. and {Crill}, Brendan P. and {Cukierman}, Ari J. and {Dore}, Olivier and {Dowell}, C. Darren and {Dubois-Felsmann}, Gregory P. and {Everett}, Spencer and {Fabinsky}, Beth E. and {Faisst}, Andreas L. and {Fanson}, James L. and {Farrington}, Allen H. and {Fatahi}, Tamim and {Fazar}, Candice M. and {Feder}, Richard M. and {Frater}, Eric H. and {Grasshorn Gebhardt}, Henry S. and {Giri}, Utkarsh and {Goldina}, Tatiana and {Gorjian}, Varoujan and {Hart}, William G. and {Hora}, Joseph L. and {Huai}, Zhaoyu and {Hui}, Howard and {Jo}, Young-Soo and {Jeong}, Woong-Seob and {Kang}, Jae Hwan and {Kang}, Miju and {Kecman}, Branislav and {Kim}, Chul-Hwan and {Kim}, Jaeyeong and {Kim}, Minjin and {Kim}, Young-Jun and {Kim}, Yongjung and {Kirkpatrick}, J. Davy and {Korngut}, Phil M. and {Krause}, Elisabeth and {Lee}, Bomee and {Lee}, Ho-Gyu and {Lee}, Jae-Joon and {Lee}, Jeong-Eun and {Lisse}, Carey M. and {Mariani}, Giacomo and {Masters}, Daniel C. and {Mauskopf}, Philip D. and {Melnick}, Gary J. and {Minasyan}, Mary H. and {Mirocha}, Jordan and {Miyasaka}, Hiromasa and {Moore}, Anne and {Moore}, Bradley D. and {Murgia}, Giulia and {Naylor}, Bret J. and {Nelson}, Christina and {Nguyen}, Chi H. and {Noh}, Jinyoung K. and {Padin}, Stephen and {Paladini}, Roberta and {Penanen}, Konstantin I. and {Putnam}, Dustin S. and {Pyo}, Jeonghyun and {Ramachandra}, Nesar and {Ramanathan}, Keshav and {Reiley}, Daniel J. and {Rice}, Eric B. and {Rocca}, Jennifer M. and {Seok}, Ji Yeon and {Stober}, Jeremy and {Susca}, Sara and {Teplitz}, Harry I. and {Thelen}, Michael P. and {Tolls}, Volker and {Torrini}, Gabriela and {Trangsrud}, Amy R. and {Unwin}, Stephen and {Velicheti}, Phani and {Wang}, Pao-Yu and {Wen}, Robin Y. and {-Werner}, Michael-W. and {Williamson}, Ross and {Wincentsen}, James and {Yang}, Soung-Chul and {Yang}, Yujin and {Zemcov}, Michael},
        title = "{The SPHEREx Satellite Mission}",
      journal = {arXiv e-prints},
         year = 2025,
        month = nov,
          eid = {arXiv:2511.02985},
        pages = {arXiv:2511.02985},
          doi = {10.48550/arXiv.2511.02985},
archivePrefix = {arXiv},
       eprint = {2511.02985},
 primaryClass = {astro-ph.IM},
       adsurl = {https://ui.adsabs.harvard.edu/abs/2025arXiv251102985B}
}

@ARTICLE{Ashby2023,
       author = {{Ashby}, Matthew L.~N. and {Hora}, Joseph L. and {Lakshmipathaiah}, Kiran and {Vig}, Sarita and {Sai Subrahmanyam Gorthi}, Rama Krishna and {Kang}, Miju and {Tolls}, Volker and {Melnick}, Gary J. and {Werner}, Michael W. and {Crill}, Brendan P. and {Masters}, Daniel C. and {Pe{\~n}a}, Carlos Contreras and {Lee}, Jeong-Eun and {Kim}, Jaeyeong and {Lee}, Ho-Gyu and {Yoon}, Sung-Yong and {Yang}, Soung-Chul and {Flagey}, Nicolas and {Mennesson}, Bertrand},
        title = "{The SPHEREx Target List of Ice Sources (SPLICES)}",
      journal = {\apj},
         year = 2023,
        month = jun,
       volume = {949},
       number = {2},
          eid = {105},
        pages = {105},
          doi = {10.3847/1538-4357/acc86b},
archivePrefix = {arXiv},
       eprint = {2501.17797},
 primaryClass = {astro-ph.GA},
       adsurl = {https://ui.adsabs.harvard.edu/abs/2023ApJ...949..105A}
}

@ARTICLE{Oberg2008,
       author = {{{\"O}berg}, Karin I. and {Boogert}, A.~C. Adwin and {Pontoppidan}, Klaus M. and {Blake}, Geoffrey A. and {Evans}, Neal J. and {Lahuis}, Fred and {van Dishoeck}, Ewine F.},
        title = "{The c2d Spitzer Spectroscopic Survey of Ices around Low-Mass Young Stellar Objects. III. CH$_{4}$}",
      journal = {\apj},
         year = 2008,
        month = may,
       volume = {678},
       number = {2},
        pages = {1032-1041},
          doi = {10.1086/533432},
archivePrefix = {arXiv},
       eprint = {0801.1223},
 primaryClass = {astro-ph},
       adsurl = {https://ui.adsabs.harvard.edu/abs/2008ApJ...678.1032O}
}

@ARTICLE{Pontoppidan2008,
       author = {{Pontoppidan}, Klaus M. and {Boogert}, A.~C.~A. and {Fraser}, Helen J. and {van Dishoeck}, Ewine F. and {Blake}, Geoffrey A. and {Lahuis}, Fred and {{\"O}berg}, Karin I. and {Evans}, II, Neal J. and {Salyk}, Colette},
        title = "{The c2d Spitzer Spectroscopic Survey of Ices around Low-Mass Young Stellar Objects. II. CO$_{2}$}",
      journal = {\apj},
         year = 2008,
        month = may,
       volume = {678},
       number = {2},
        pages = {1005-1031},
          doi = {10.1086/533431},
archivePrefix = {arXiv},
       eprint = {0711.4616},
 primaryClass = {astro-ph},
       adsurl = {https://ui.adsabs.harvard.edu/abs/2008ApJ...678.1005P}
}

@ARTICLE{2026Melnick,
       author = {{Melnick}, G.~J. and {Ashby}, M.~L.~N. and {Tolls}, V. and {Hora}, J.~L. and
                  {Paladini}, R.},
        title = "{The SPHEREx Ices Investigation}",
      journal = {},
         year = 2026,
        month = jun,
       volume = {in preparation},
       number = {},
        pages = {},
}

@ARTICLE{Bergin2007,
       author = {{Bergin}, Edwin A. and {Tafalla}, Mario},
        title = "{Cold Dark Clouds: The Initial Conditions for Star Formation}",
      journal = {\araa},
         year = 2007,
        month = sep,
       volume = {45},
       number = {1},
        pages = {339-396},
          doi = {10.1146/annurev.astro.45.071206.100404},
archivePrefix = {arXiv},
       eprint = {0705.3765},
 primaryClass = {astro-ph},
       adsurl = {https://ui.adsabs.harvard.edu/abs/2007ARA&A..45..339B}
}

@ARTICLE{Aikawa2012,
       author = {{Aikawa}, Y. and {Kamuro}, D. and {Sakon}, I. and {Itoh}, Y. and {Terada}, H. and {Noble}, J.~A. and {Pontoppidan}, K.~M. and {Fraser}, H.~J. and {Tamura}, M. and {Kandori}, R. and {Kawamura}, A. and {Ueno}, M.},
        title = "{AKARI observations of ice absorption bands towards edge-on young stellar objects}",
      journal = {\aap},
         year = 2012,
        month = feb,
       volume = {538},
          eid = {A57},
        pages = {A57},
          doi = {10.1051/0004-6361/201015999},
       adsurl = {https://ui.adsabs.harvard.edu/abs/2012A&A...538A..57A}
}

@ARTICLE{Ehrenfreund1999,
       author = {{Ehrenfreund}, P. and {Kerkhof}, O. and {Schutte}, W.~A. and {Boogert}, A.~C.~A. and {Gerakines}, P.~A. and {Dartois}, E. and {D'Hendecourt}, L. and {Tielens}, A.~G.~G.~M. and {van Dishoeck}, E.~F. and {Whittet}, D.~C.~B.},
        title = "{Laboratory studies of thermally processed H\_2O-CH\_3OH-CO\_2 ice mixtures and their astrophysical implications}",
      journal = {\aap},
         year = 1999,
        month = oct,
       volume = {350},
        pages = {240-253},
       adsurl = {https://ui.adsabs.harvard.edu/abs/1999A&A...350..240E}
}

@ARTICLE{Ehrenfreund1996,
       author = {{Ehrenfreund}, P. and {Boogert}, A.~C.~A. and {Gerakines}, P.~A. and {Jansen}, D.~J. and {Schutte}, W.~A. and {Tielens}, A.~G.~G.~M. and {van Dishoeck}, E.~F.},
        title = "{A laboratory database of solid CO and CO\_2\_ for ISO.}",
      journal = {\aap},
         year = 1996,
        month = nov,
       volume = {315},
        pages = {L341-L344},
       adsurl = {https://ui.adsabs.harvard.edu/abs/1996A&A...315L.341E}
}

@ARTICLE{Bergner2026,
       author = {{Bergner}, Jennifer B. and {Arulanantham}, Nicole and {Dartois}, Emmanuel and {Drozdovskaya}, Maria N. and {Harsono}, Daniel and {McClure}, Melissa and {Noble}, Jennifer A. and {{\"O}berg}, Karin I. and {Pontoppidan}, Klaus M. and {Yang}, Yao-Lun and {Assani}, Korash and {Li}, Zhi-Yun and {Santos}, Julia C. and {Thompson}, Will E. and {Welzel}, Lukas and {Yunerman}, Elizabeth S. and {Arabhavi}, Aditya M. and {Booth}, Alice S. and {Mentzer}, Charles and {Narang}, Mayank and {Henning}, Thomas and {Kamp}, Inga and {Perotti}, Giulia and {Somigliana}, Alice},
        title = "{JWST Edge-on Disk Ice (JEDIce): Program overview and ice survey results}",
      journal = {arXiv e-prints},
         year = 2026,
        month = mar,
          eid = {arXiv:2603.18163},
        pages = {arXiv:2603.18163},
          doi = {10.48550/arXiv.2603.18163},
archivePrefix = {arXiv},
       eprint = {2603.18163},
 primaryClass = {astro-ph.EP},
       adsurl = {https://ui.adsabs.harvard.edu/abs/2026arXiv260318163B}
}

@ARTICLE{Bergner2024,
       author = {{Bergner}, Jennifer B. and {Sturm}, J.~A. and {Piacentino}, Elettra L. and {McClure}, M.~K. and {{\"O}berg}, Karin I. and {Boogert}, A.~C.~A. and {Dartois}, E. and {Drozdovskaya}, M.~N. and {Fraser}, H.~J. and {Harsono}, Daniel and {Ioppolo}, Sergio and {Law}, Charles J. and {Lis}, Dariusz C. and {McGuire}, Brett A. and {Melnick}, Gary J. and {Noble}, Jennifer A. and {Palumbo}, M.~E. and {Pendleton}, Yvonne J. and {Perotti}, Giulia and {Qasim}, Danna and {Rocha}, W.~R.~M. and {van Dishoeck}, E.~F.},
        title = "{JWST Ice Band Profiles Reveal Mixed Ice Compositions in the HH 48 NE Disk}",
      journal = {\apj},
         year = 2024,
        month = nov,
       volume = {975},
       number = {2},
          eid = {166},
        pages = {166},
          doi = {10.3847/1538-4357/ad79fc},
archivePrefix = {arXiv},
       eprint = {2409.08117},
 primaryClass = {astro-ph.EP},
       adsurl = {https://ui.adsabs.harvard.edu/abs/2024ApJ...975..166B}
}

@ARTICLE{Whittet2010,
       author = {{Whittet}, D.~C.~B.},
        title = "{Oxygen Depletion in the Interstellar Medium: Implications for Grain Models and the Distribution of Elemental Oxygen}",
      journal = {\apj},
         year = 2010,
        month = feb,
       volume = {710},
       number = {2},
        pages = {1009-1016},
          doi = {10.1088/0004-637X/710/2/1009},
archivePrefix = {arXiv},
       eprint = {0912.3298},
 primaryClass = {astro-ph.GA},
       adsurl = {https://ui.adsabs.harvard.edu/abs/2010ApJ...710.1009W}
}

@ARTICLE{Ginsburg2026,
       author = {{Ginsburg}, Adam and {Walker}, Daniel L. and {Barnes}, Ashley T. and {Lu}, Xing and {S{\'a}nchez-Monge}, {\'A}lvaro and {Pineda}, Jaime E. and {Pound}, Marc W. and {Hsieh}, Pei-Ying and {Immer}, Katharina and {Zhang}, Qizhou and {Budaiev}, Nazar and {Gramze}, Savannah R. and {Jeff}, Desmond and {Cook}, Claire and {Bulatek}, Alyssa and {Mills}, Elisabeth A.~C. and {Bally}, John and {Colzi}, Laura and {Garc{\'\i}a}, Pablo and {Henshaw}, Jonathan D. and {Jim{\'e}nez-Serra}, Izaskun and {Klessen}, Ralf S. and {Dicker}, Simon R. and {Longmore}, Steven N. and {Nogueras-Lara}, Francisco and {Rivilla}, V{\'\i}ctor M. and {Santa-Maria}, Miriam G. and {Wang}, Q. Daniel and {Xu}, Fengwei and {Battersby}, Cara and {Ho}, Paul T.~P. and {Kruijssen}, J.~M. Diederik and {Petkova}, Maya and {Sormani}, Mattia C. and {Tress}, Robin G. and {Wallace}, Jennifer and {Armijos-Abenda{\~n}o}, J. and {Armillotta}, Lucia and {Bijas}, N. and {Buddhacharya}, Rojita and {Busch}, Laura A. and {Butterfield}, Natalie O. and {Chevance}, M{\'e}lanie and {Crowe}, Samuel and {D{\'\i}az-Rodr{\'\i}guez}, Ana Karla and {Dutkowska}, Katarzyna M. and {Fedriani}, Rub{\'e}n and {Federrath}, Christoph and {Glover}, Simon C.~O. and {Gu}, Qi-Lao and {Houghton}, Rebecca J. and {Hu}, Yue and {Issac}, Namitha and {Karoly}, Janik and {Krumholz}, Mark R. and {Liang}, Fu-Heng and {Mart{\'\i}n}, Sergio and {Mazoochi}, Farideh and {Pan}, Xing and {Par{\'e}}, Dylan and {Pillai}, Thushara G.~S. and {Riquelme-V{\'a}squez}, Denise and {Schmiedeke}, Anika and {Sofue}, Yoshiaki and {Tolls}, Volker and {Williams}, Gwenllian M. and {Zhang}, Suinan and {Moravec}, Emily and {Romero}, Charles E. and {Mason}, Brian S. and {Orlowski-Scherer}, John and {Hatchfield}, H Perry},
        title = "{ALMA Central Molecular Zone Exploration Survey (ACES) II: 3mm continuum images}",
      journal = {arXiv e-prints},
         year = 2026,
        month = feb,
          eid = {arXiv:2602.20240},
        pages = {arXiv:2602.20240},
          doi = {10.48550/arXiv.2602.20240},
archivePrefix = {arXiv},
       eprint = {2602.20240},
 primaryClass = {astro-ph.GA},
       adsurl = {https://ui.adsabs.harvard.edu/abs/2026arXiv260220240G}
}

\clearpage
\appendix
\onecolumngrid

\section{Appendices: Overview}
\referee{
We present extensive appendices below.
For readability, and to keep the figures close to the text describing them, the appendices are split such that each starts on a fresh page.  
The appendices are as follows:
\begin{itemize}
    \item Appendix \ref{sec:icemodelscaveats}: Cautions and caveats with the \texttt{icemodels} approach
    \item Appendix \ref{sec:corrections}: Corrections to \citet{Ginsburg2023}
    \item Appendix \ref{appendix:colors405410}: Discussion of how CO$_2$ affects F405N-F410M colors
    \item Appendix \ref{appendix:columnuncertainties}: A deep dive on the uncertainties in the measurements presented in the main text
    \item Appendix \ref{app:coabundance}: Examination of the CO abundance growth with depth into the cloud
    \item Appendix \ref{app:kp5}: An examination of the KP5 extinction curve
    \item Appendix \ref{app:spatialplotstwo}: Overlays of the catalog on JWST images
    \item Appendix \ref{sec:1e19table}: Table of ice model opacities at higher column density than shown in the main text
    \item Appendix \ref{sec:isospectra}: Example ISO spectra of nearby targets
    \item Appendix \ref{sec:icemixturesredux}: Further examination of the difference between linear mixtures and lab mixtures of ices
    \item Appendix \ref{sec:moreabsorptionplots}: Ice opacity curves overlaid on NIRCam filter transmission curves not shown in the main text
    \item Appendix \ref{sec:stellarsedplots} shows a stellar SED with ice absorption features and extinction applied
    \item Appendix \ref{sec:colorvscolumn} shows the color as a function of A$_V$ and column density for the most analyzed color (F405N-F466N) for two sets of ice models.
\end{itemize}.
}

\clearpage
\section{Cautions and Caveats with the \texttt{icemodels} approach}
\label{sec:icemodelscaveats}
We outline some of the limitations and caveats of the \texttt{icemodels} modeling software.

By using laboratory-measured opacities, the absorption models are tied to those measurements and inherit both their strengths and weaknesses.
The lab measurements report opacity or absorbance values at all wavelengths at which the experiment was performed, whether or not there is real signal at that wavelength.
This means that there is an effective opacity floor set by the noise level of the lab measurements, and this noise floor is not consistent across reported opacity values.
The opacity floor plays an important role in determining the total opacity in filters, especially the broad-band filters, for which narrow absorption features (like CO) have little effect, and at high column densities where the opacity in ice peaks exceeds $\tau>>1$ such that further increase in the optical depth no longer produce measurable changes in the observed flux.
The noise floor is rarely the appropriate value to use; instead, for most of the NIRCam band at least, the broad \water ice opacity features set the relevant opacity floor.

Further, we are using only ices that have been measured in laboratory.
It is possible that there are ices, and certain that there are ice combinations, that exist in the ISM but have not been measured in lab.
For example, OCN$^-$ ice has only been measured in a specific mixture because of the photolysis required to produce it \citep{Novozamsky2001}.
\referee{As a further example, the profile of the CO ice feature depends strongly on the matrix in which the CO is embedded \citep{Ehrenfreund1996,Ehrenfreund1999,Bergner2024,Bergner2026}; \texttt{icemodels} does not currently include this dependency, but as the data become available or are recovered from archives, the capability will be added.}
Modeling performed with \texttt{icemodels} and the laboratory databases should therefore never be considered complete - any inadequacies in the laboratory measurements or databases will be reflected in models with this code.

The normalizations adopted in this code do not necessarily conform to conventions used by the ice chemistry community.
The core module \texttt{absorbed\_spectrum} computes the optical depth, and in turn the absorbed spectrum, based on an input \emph{column density} of ice, i.e., number of molecules per square centimeter.
For a pure ice, this number is unambiguous.
When modeling the absorbed spectrum as a function of column density of a specific ice species, as we do in this work, ice mixtures are more complicated.
The opacity in the laboratory measurement must be re-scaled before passing to the \texttt{absorbed\_spectrum} function, or the column density must be recomputed to be column of an average ice particle.

For example, if modeling CO ice in a 1:2:3 mixture of CO, CO$_2$, and H$_2$O, the mass of CO is
\begin{eqnarray}
    M(CO) & = & M(ice) \frac{m_{CO}}{m_{CO} + 2 m_{CO_2} + 3 m_{H_2O}}\\
          & = & M(ice) \frac{12}{12 + 88 + 54} \\
          & = & 0.078 M(ice)
\end{eqnarray}
The absorption constants $k$ are defined for the total ice, and therefore if the ice column is specified as N(CO), the constants should be scaled down by $M(CO)/M(ice)=0.078$ before passing to \texttt{absorbed\_spectrum}.

\clearpage
\section{Corrections to Ginsburg+ 2023}
\label{sec:corrections}
\begin{figure}[!h]
    \centering
    \includegraphics[width=\linewidth]{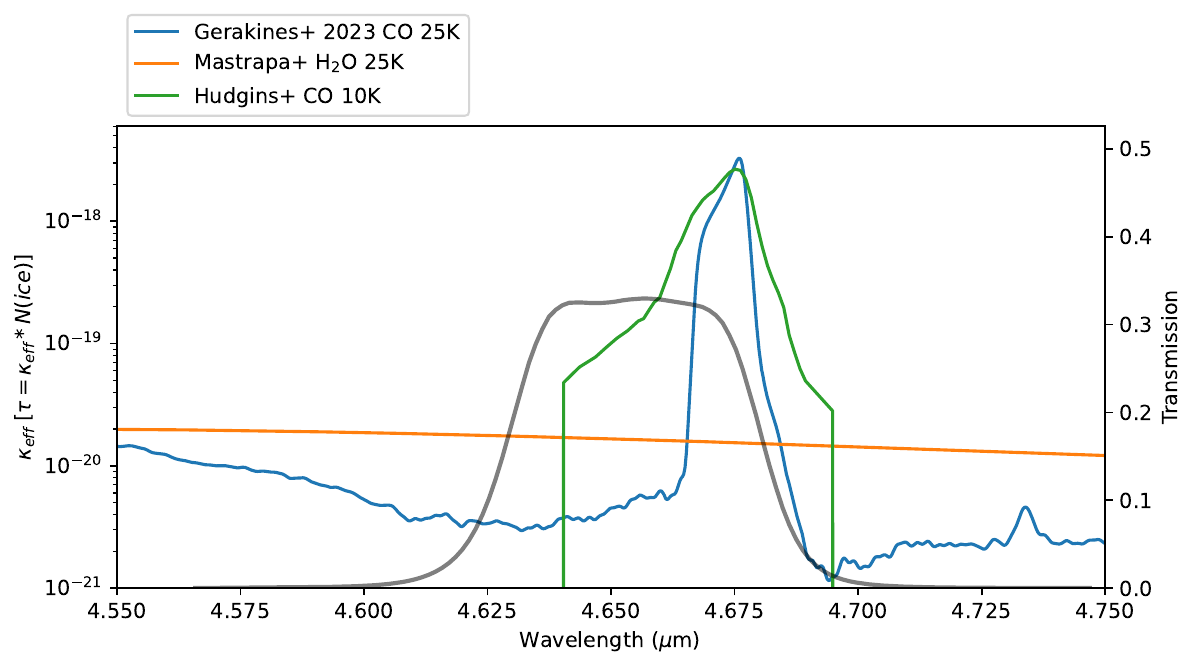}
    \caption{Opacities from several important ices (CO and \water) overlaid on the JWST transmission curve (thick grey line) in the F466N band.
    The \citet{Hudgins1993} curve was used in \citet{Ginsburg2023}, but was erroneously extrapolated from the cutoff edges seen in this figure.
    The other curves are from \citet{Mastrapa2009} and \citet{Gerakines2023}.
    }
    \label{fig:F466Nplusopacities_hudgins}
\end{figure}

\citet{Ginsburg2023} performed CO ice column measurements assuming the only ice present was pure CO.
That work was able to produce high absorptions, up to the measured $\sim75\%$, because of both a flawed assumption and a bad extrapolation.
Figure \ref{fig:F466Nplusopacities_hudgins} shows in green the opacity curve for CO ice from \citet{Hudgins1993}.
In \citet{Ginsburg2023}, the edges that drop to zero were instead assumed to stay flat at the last measured opacity - i.e., it was (incorrectly) assumed that pure CO ice has measured opacity at all wavelength with a constant $\kappa_{eff}\approx5\times10^{-20}$ cm$^2$.
Comparing to the \citet{Gerakines2023} curve, it is clear that that assumption overestimated the ice opacity across the F466N band by about an order of magnitude.
The CO ice abundances thus measured were therefore systematically incorrect.

\clearpage
\section{F405N-F410M behavior}
\label{appendix:colors405410}
\begin{figure}[!h]
    \includegraphics[width=0.7\linewidth]{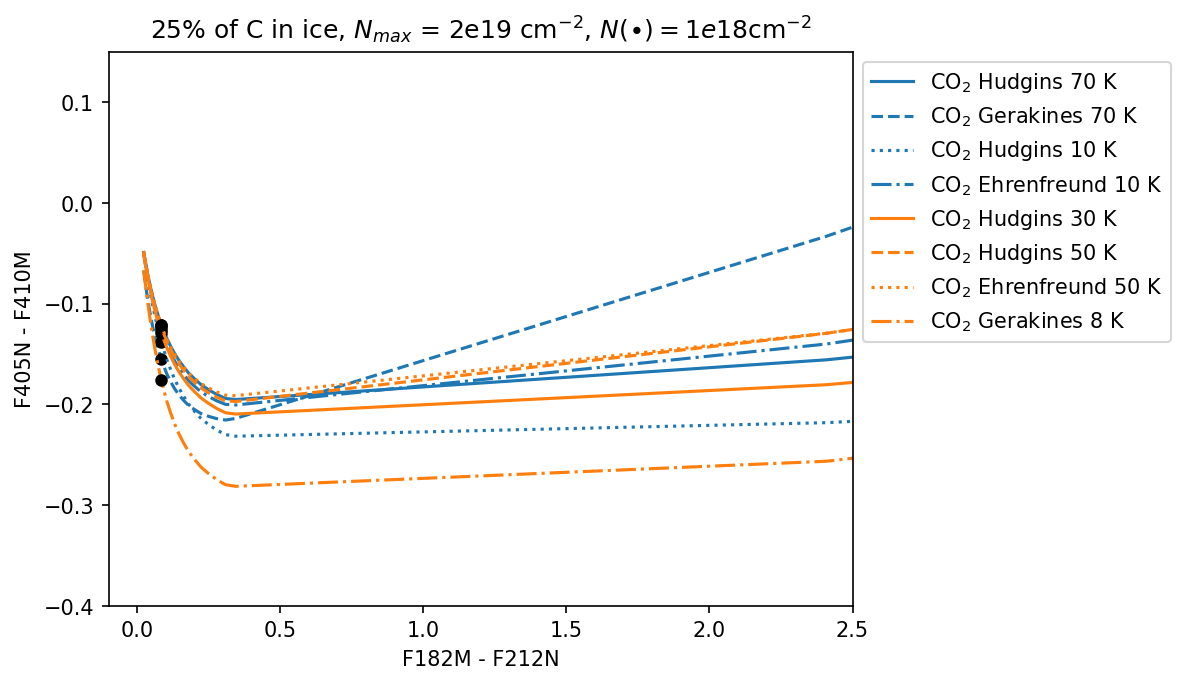}
     \caption{
     A demonstration of the effect of CO$_2$ ice on the F405N-F410M color.
     \referee{
        Both F405N and F410M are affected by CO$_2$ ice absorption, but F410M is more strongly affected because it is closer to the CO$_2$ ice absorption peak.
        See \S \ref{sec:f405nf410m}.}
     }
     \label{fig:f405nf410m_co2models}
\end{figure}

In \S \ref{sec:f405nf410m}, we discussed a `turnaround' feature seen in the F405N-F410M vs F182M-F212N color-color diagram.
\referee{This turnaround is caused by the absorption peak of the CO$_2$ feature saturating at moderately high column density (N(CO$_2$)$\sim10^{18}$ \persc).
At higher column densities, the wing of the CO$_2$ feature, which overlaps both the F405N and F410M filters, continues to increase in optical depth.}
In Figure \ref{fig:f405nf410m_co2models}, we show the curves for CO$_2$ ice from 0 to $2\times10^{19}$ \persc for a few different laboratory CO$_2$ ice measurements.
The black dots at $10^{18}$ \persc indicate roughly where the CO$_2$ line saturates and the turnaround point occurs.

\clearpage

\section{Uncertainties on CO column density and N(H$_2$) measurements}
\label{appendix:columnuncertainties}
In this appendix, we explore the uncertainties in Figure \ref{fig:fig9colvsav}.

\subsection{Foreground Extinction}
\label{sec:foregroundav}
\begin{figure*}[!h]
    \centering
    \includegraphics[width=0.32\linewidth]{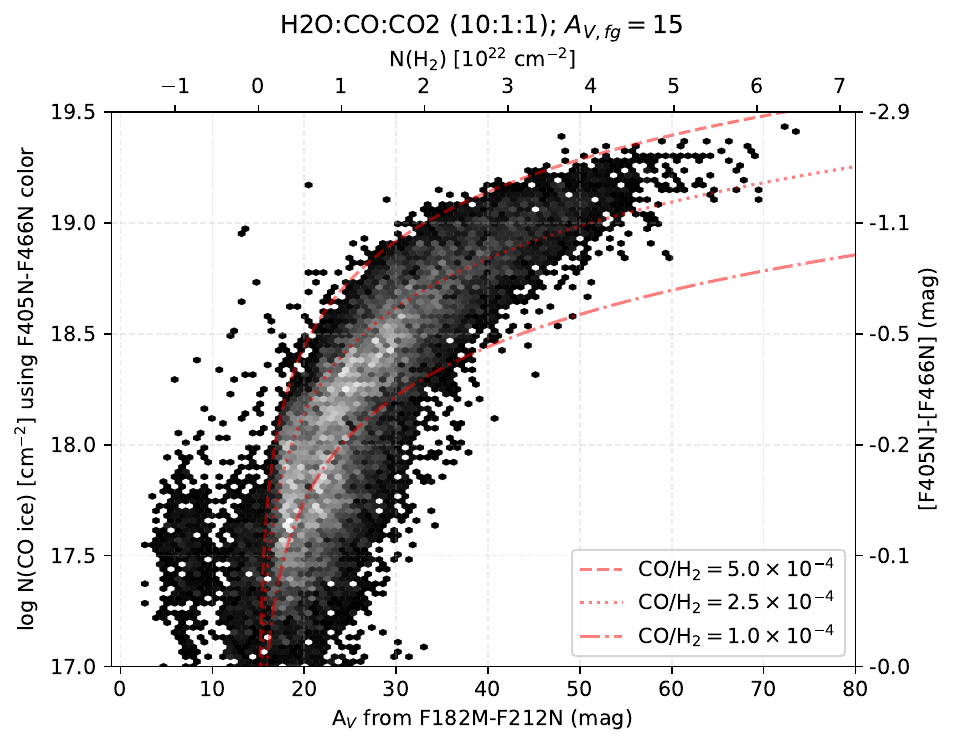}
    \includegraphics[width=0.32\linewidth]{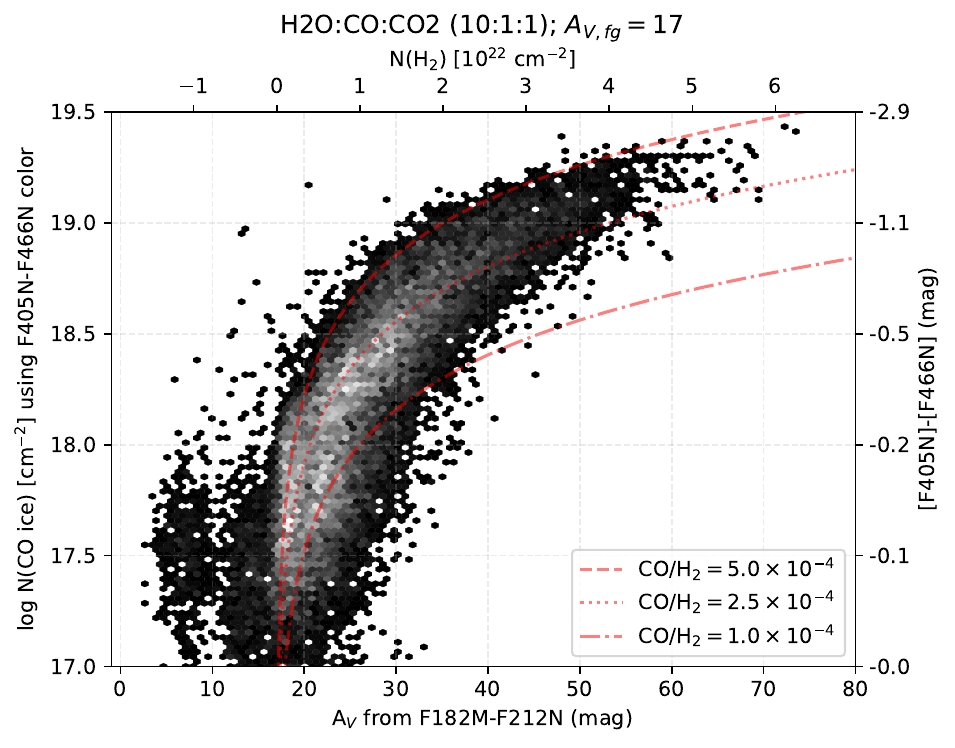}
    \includegraphics[width=0.32\linewidth]{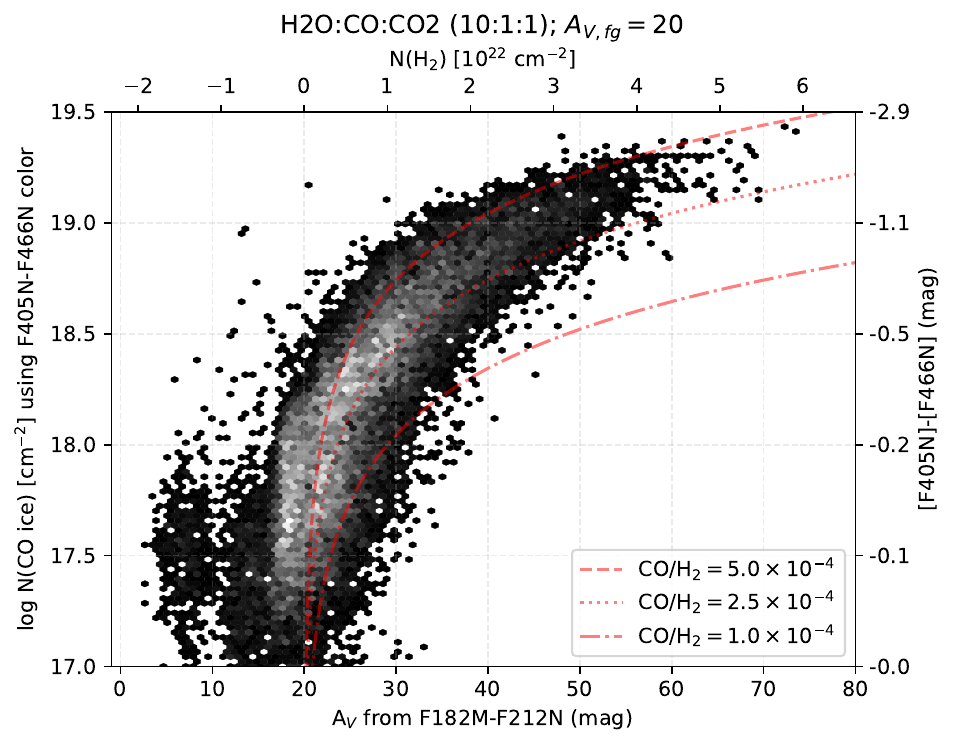}
    \caption{Comparison of different possible adopted foreground extinction values, all using the CT06 extinction curve.
    \referee{See \S \ref{sec:foregroundav}.}
    }
    \label{fig:extinction-uncertainty}
\end{figure*}

Figure \ref{fig:extinction-uncertainty} compares different possible choices of foreground extinction values adopting the CT06 extinction curve.
\citet{An2011} and \citet{Jang2022}, who examined Spitzer IRS spectra of icy sources in the Galactic Center, adopt A$_{V,fg}$=20.
\citet{Nogueras-Lara2021} adopt A$_{K,fg}=1.87$, or A$_{V,fg}\approx17$.
The leftmost figure shows that, for this adopted extinction curve, A$_V=15$ is too low - the red curves fall to the left of the sharp increase in star counts associated with the Galactic center.
The middle figure, A$_V=17$, closely matches the left edge of the star population, indicating that this is a good approximation for the purely foreground, ice-free extinction.
The right, with A$_V$=20, roughly bounds the right edge of the GC-associated stellar population: many stars likely do have $25>$A$_{V,fg}>17$ because they are at different depths into the Galactic Center and Galactic Plane, but are still behind primarily low-density ISM dust.
The $\sim3-5$ magnitudes of uncertainty in the amount of foreground material has a small ($<0.1$ dex) effect on the inferred \referee{hydrogen} column density.


\subsection{Extinction Curve}
\label{sec:extinctioncurve}
\begin{figure*}[!h]
    \centering
    \includegraphics[width=0.32\linewidth]{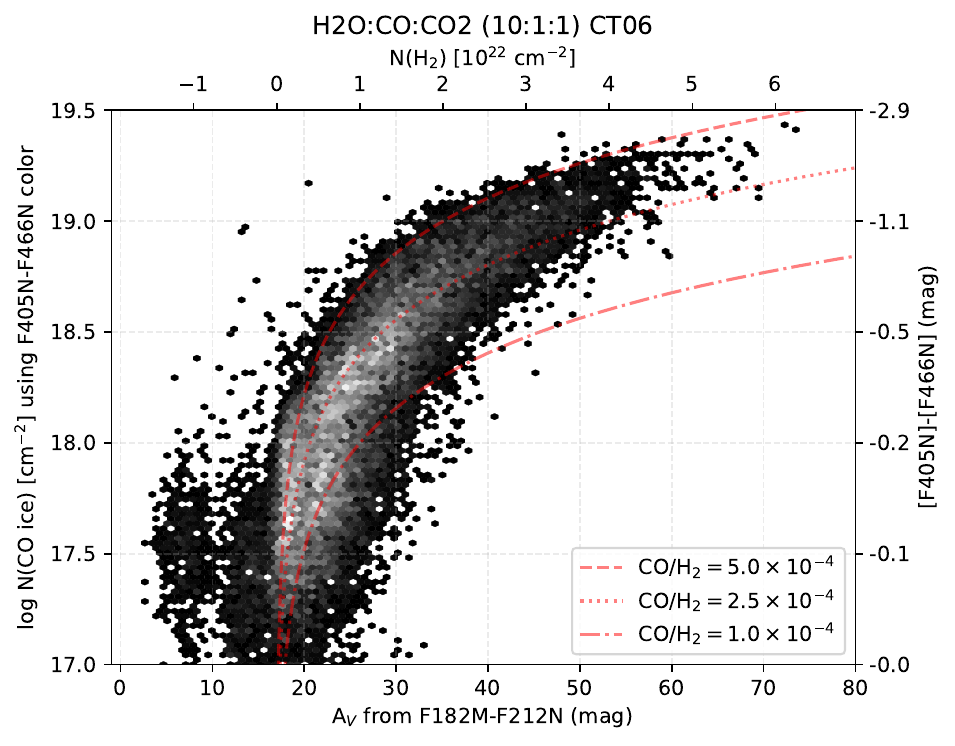}
    \includegraphics[width=0.32\linewidth]{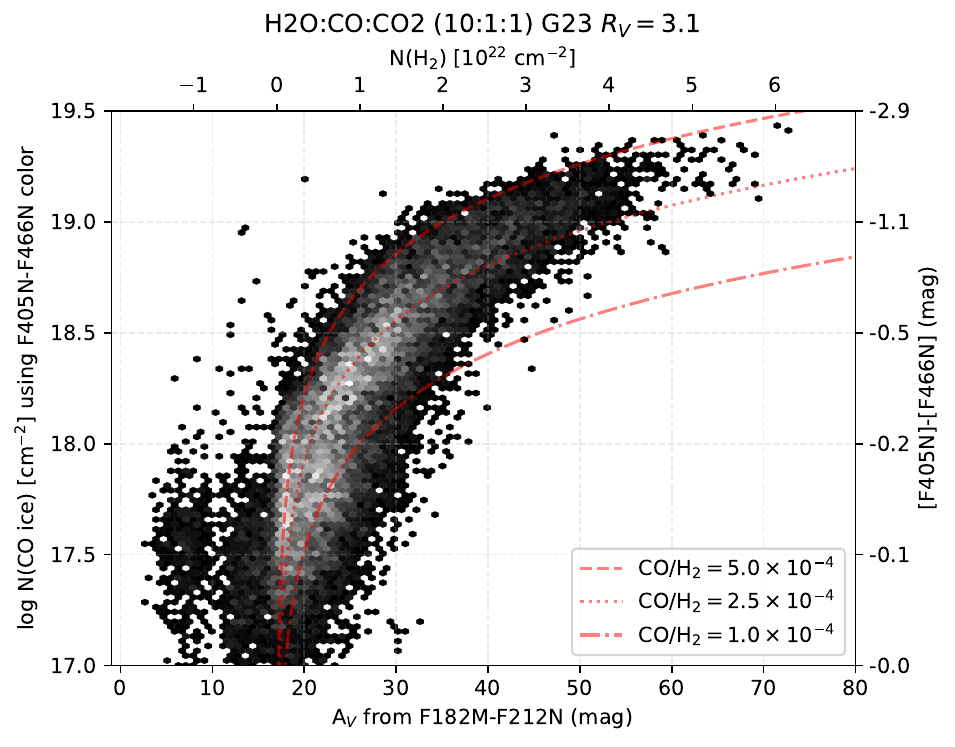}
    \includegraphics[width=0.32\linewidth]{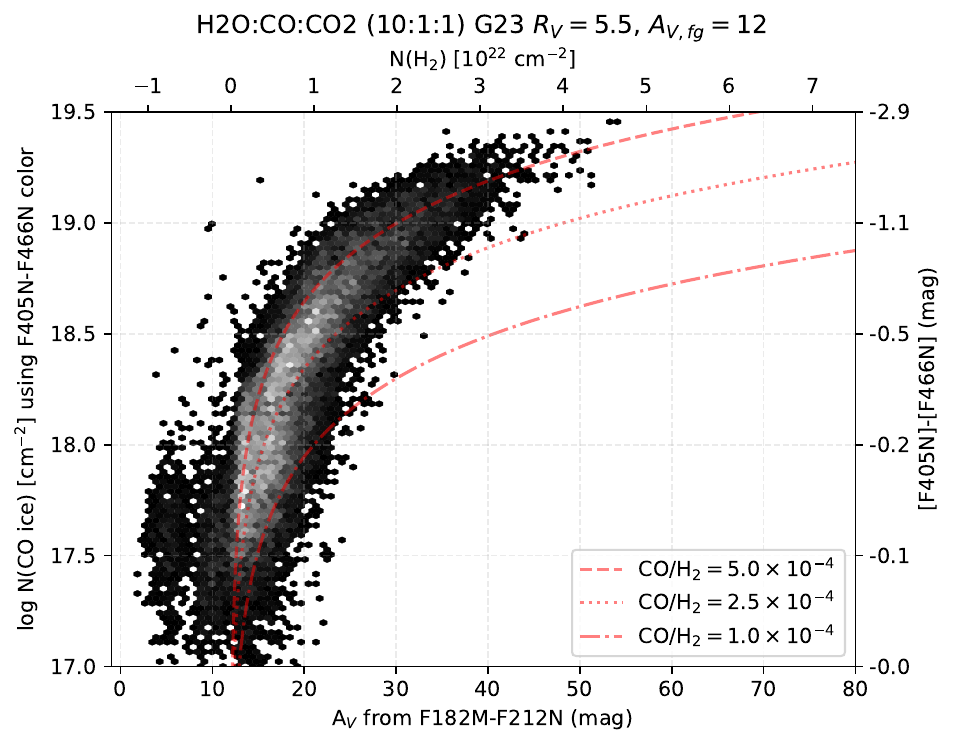}
    \caption{Comparison of different possible adopted extinction curves.
    \referee{See \S \ref{sec:extinctioncurve}.}
    }
    \label{fig:curve-uncertainty}
\end{figure*}

\begin{figure}[!h]
    \centering
    \includegraphics[width=0.6\linewidth]{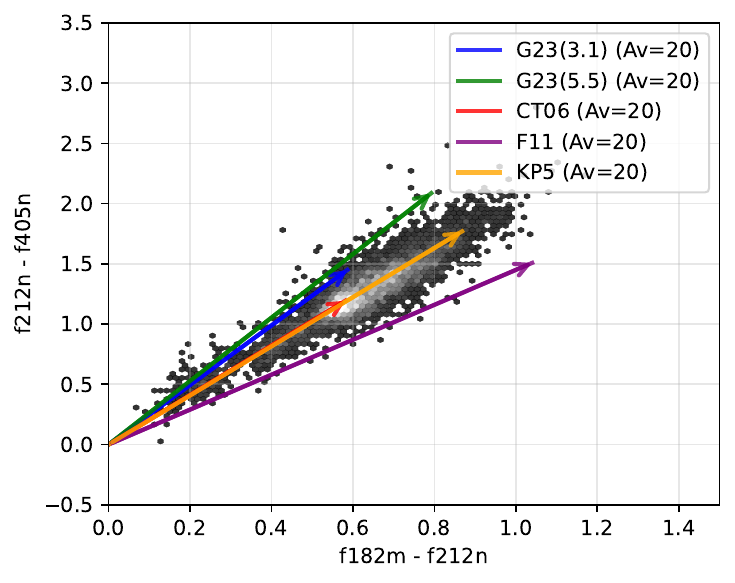}
    \caption{Comparison of \referee{five} plausible extinction vectors describing the foreground medium.
    The data shown are the ice-free data selected from the Brick data set.
    \referee{See \S \ref{sec:extinctioncurve}.}
    }
    \label{fig:extinctionvectors}
\end{figure}

The choice of extinction curve can have a large effect on inferred total extinction.
Figure \ref{fig:curve-uncertainty} compares the \citet{Chiar2006} curve to the \citet{Gordon2023} curves with R$_V$=3.1 (middle) and 5.5 (right).
The former two are effectively identical.
Dust with higher R$_V$ produces a substantially different A$_V$ and therefore N(H$_2$) by about 30\% (0.5 dex), however, this R$_V$ is disfavored for foreground dust.
Figure \ref{fig:extinctionvectors} shows that the higher $R_V$ curve is disfavored, while the other two are equally acceptable; the data in that figure are all photometric measurements from the Brick data set with [F405N]-[F466N] $> -0.1$, i.e., those with no clear ice absorption signal.
While this 30\% difference might be an upper limit, we regard it as unlikely and adopt a lower $\sim10\%$ uncertainty from extinction curve uncertainty.

\clearpage
\subsection{Ice Mixture and total CO abundance}
\label{appendix:icemixes}
\begin{figure*}[!h]
 \includegraphics[width=0.32\linewidth]{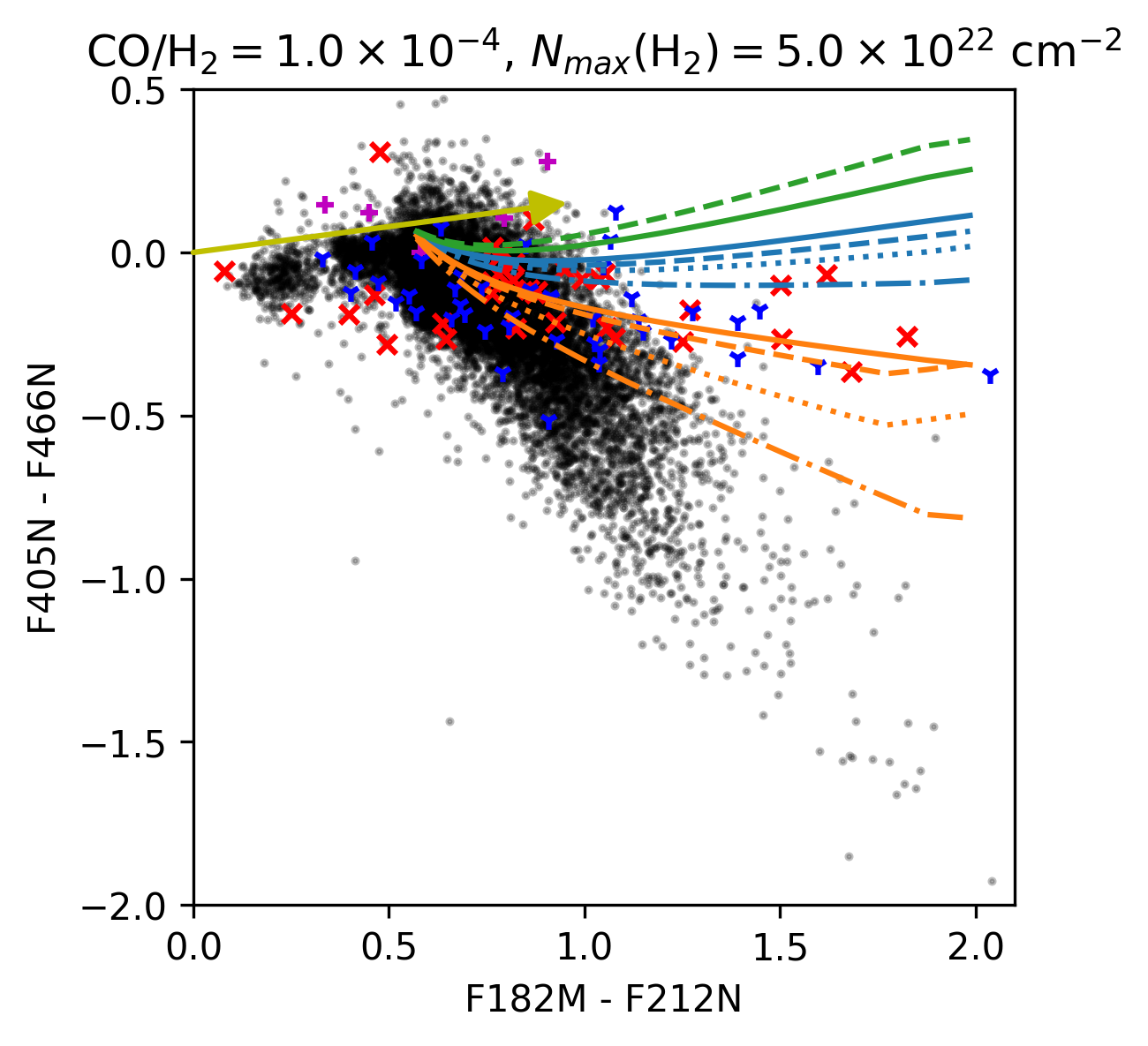}
 \includegraphics[width=0.32\linewidth]{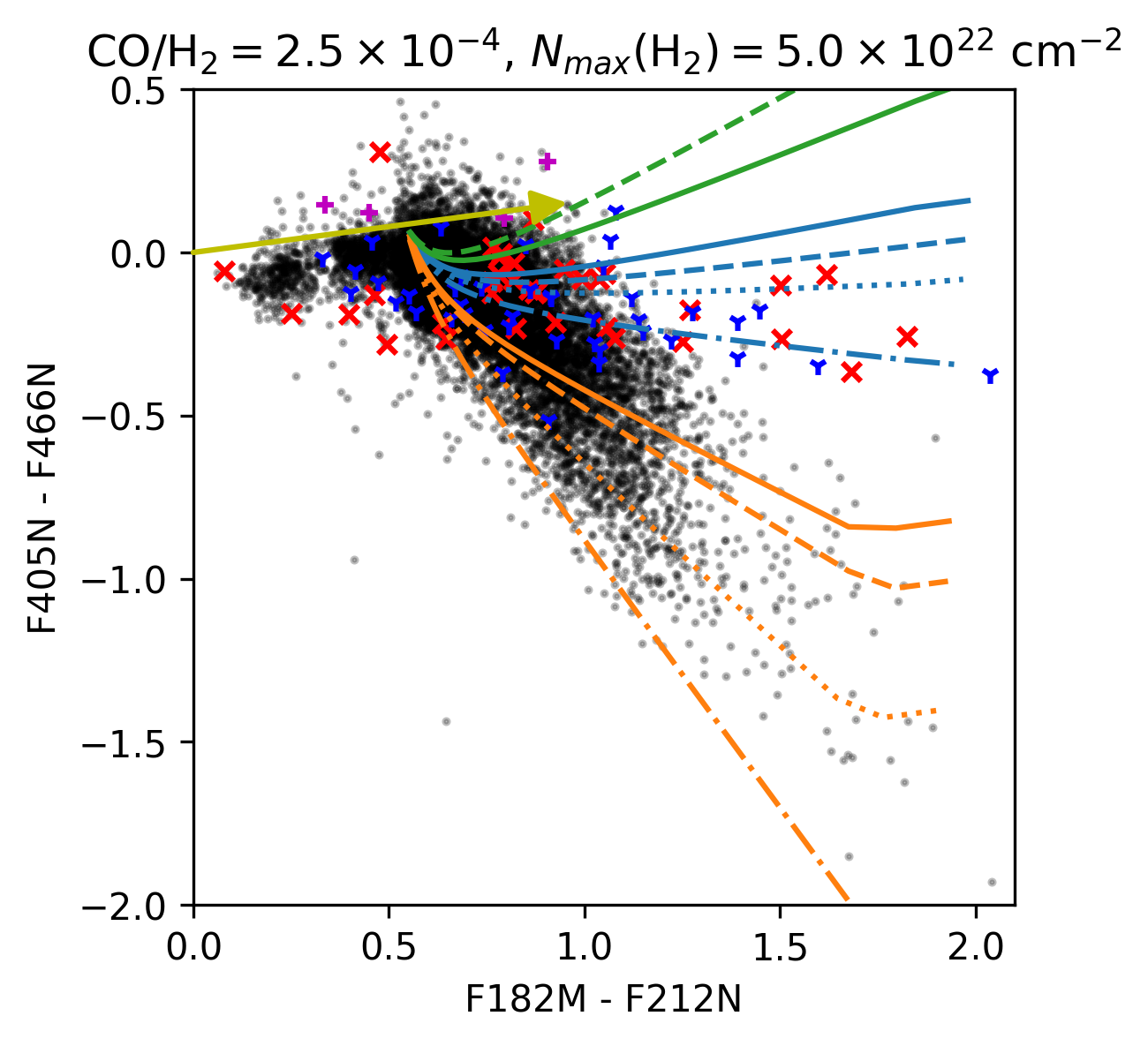}
 \includegraphics[width=0.32\linewidth]{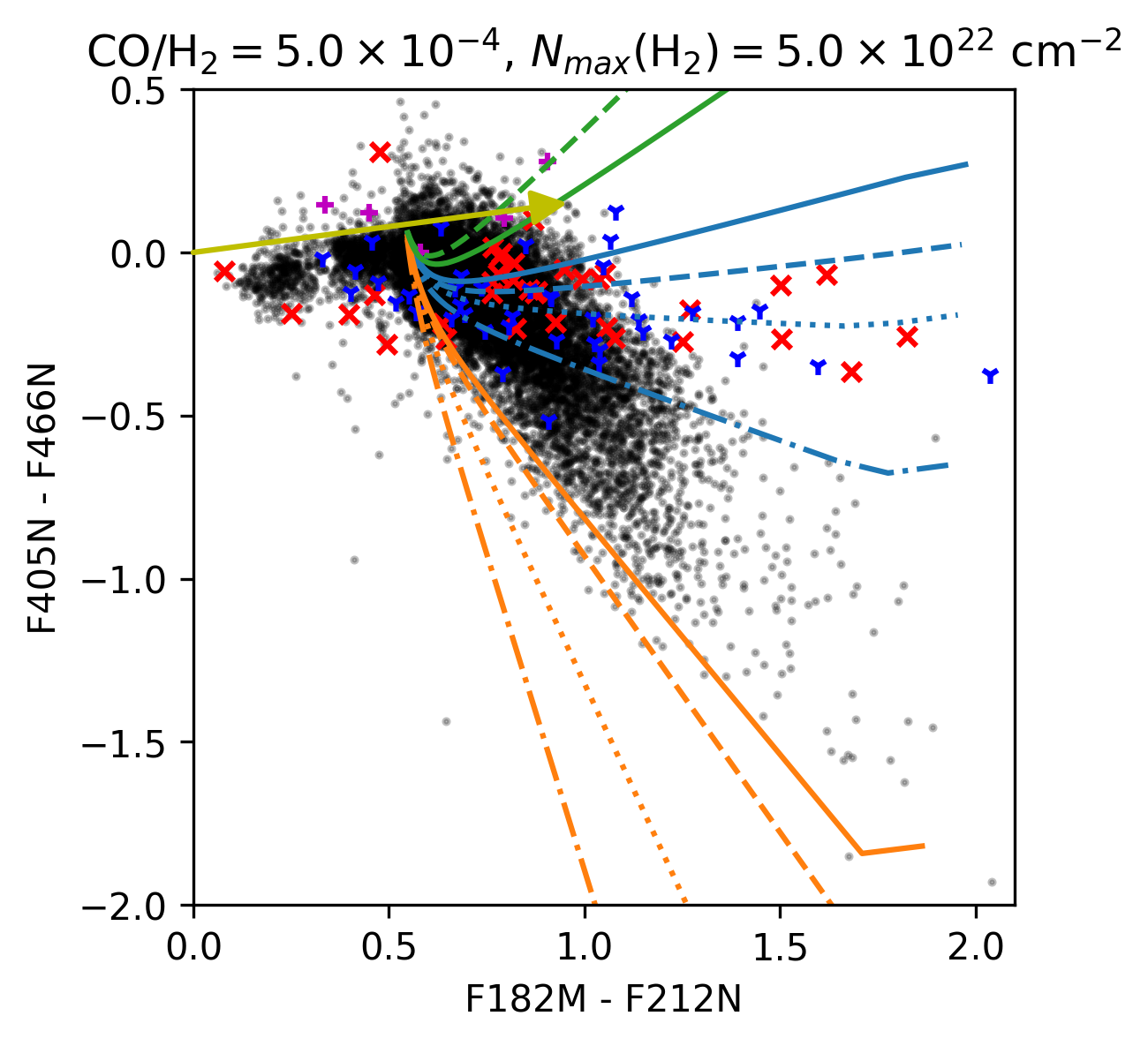}
    \caption{Comparison of different assumed CO/H$_2$ ratios in color-color space used to evaluate the valid range of ice ratios H$_2$O:CO:CO$_2$ as shown in Figure \ref{fig:colorcolor}.
    \referee{See \ref{sec:icemixes}.}
    }
    \label{fig:icemix-uncertainty-colorcolor}
\end{figure*}

The choice of ice mixture is likely the dominant uncertainty in our measurements of ice abundances, and it is degenerate with the total CO abundance.
The possible range of total CO abundance (CO/H$_2$) is from $1\times10^{-4}$, the solar neighborhood value, to $2\times10^{-3}$, which is the upper limit assuming all carbon is in CO and the Galactic center carbon abundance is C/H=$10^{-3}$ ([C/H-12] = 9.0).
The latter extreme is not very plausible, however, as $\gtrsim25-75\%$ of carbon is expect to be in other forms, e.g., carbonaceous grains, CO$_2$, or PAHs, \referee{leaving $\lesssim75-25\%$ of carbon available to form CO \citep{Jenkins2009}}.
Figure \ref{fig:icemix-uncertainty-colorcolor} shows the effect of different CO abundance measurements on the predicted colors of ice mixtures.
The assumed abundance changes the shape of the model curves because we assume that, starting at the cloud edge, the column of total ice is proportional to the column of H$_2$ and 100\% of the CO is in ice.
While that latter assumption is bad (we know there is gas-phase \referee{CO} at least at the cloud edge), it is useful for evaluating the extremes, as 100\% freezeout is expected at the highest densities, as the extreme values of the model curves must accommodate the most extinguished data.

The endpoint of these curves is relevant, as the highest column density of H$_2$ we probe with F466N is about N(H$_2)\sim5\times10^{22}$.
The lowest-abundance plot, with CO/H$_2$ = 10$^{-4}$, cannot produce the deepest absorption observed with any of the mixtures.
One can extrapolate that a much higher \water:CO abundance ratio would match the data, but such extreme O/C ratios seem unlikely.
The two color-color diagrams with higher CO abundance have a range of curves that cover the full observed data, with higher CO abundance corresponding to lower \water:CO ratio, though none of the plausible CO abundances are consistent with solar-neighborhood water ratios.

\begin{figure*}[!h]
    \centering
    \includegraphics[width=0.32\linewidth]{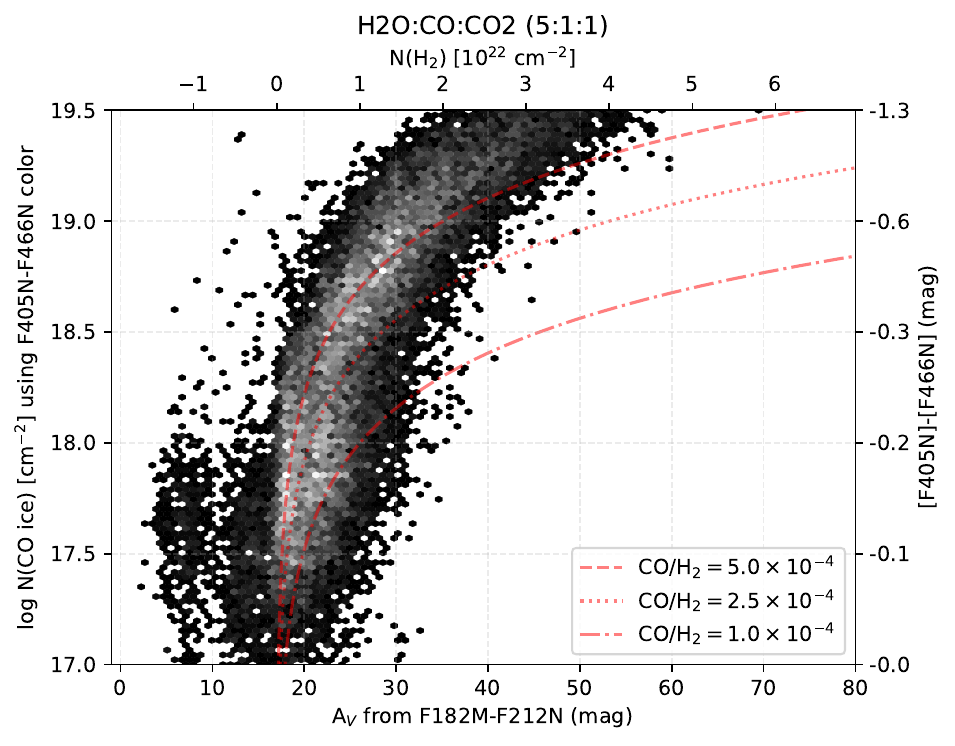}
    \includegraphics[width=0.32\linewidth]{figures/NCO_H2O_CO_CO2_10_1_1_vs_AV_F405N-F466N_F182M-F212N_hexbin_with1182_log.pdf}
    \includegraphics[width=0.32\linewidth]{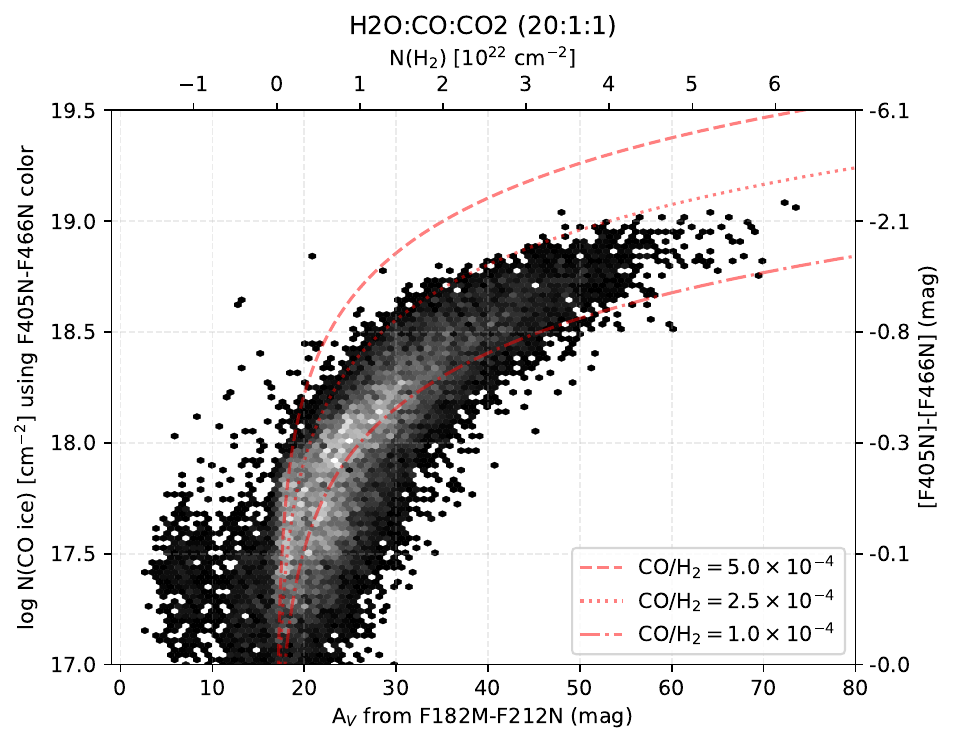}
    \caption{Comparison of different assumed ice ratios H$_2$O:CO:CO$_2$.
    \referee{See \ref{sec:icemixes}.}
    }
    \label{fig:icemix-uncertainty}
\end{figure*}

To retrieve N(ice) from a set of measurements, we compute the absorption as a function of total ice column, but that total ice is comprised of a fixed mixture of molecular ices.
The effective opacity in a given line per unit mass is diluted by the presence of other ices.
Figure \ref{fig:icemix-uncertainty} shows the effect of different ice mixtures, specifically on the relative amount of \water, on the inferred CO ice column.
The column density of CO is approximately linearly anticorrelated with the assumed amount of \water.

Summarizing, \water:CO:CO$_2 \gtrsim 15$ is inconsistent with essentially all of the data, as is \water:CO:CO$_2 \lesssim 8$.
Within these constraints, we then have $\sim2\times$ uncertainty, or $\sim0.3$ dex.


\clearpage
\section{CO abundance growth}
\label{app:coabundance}
\begin{figure*}[!h]
    \centering
    \includegraphics[width=0.49\linewidth]{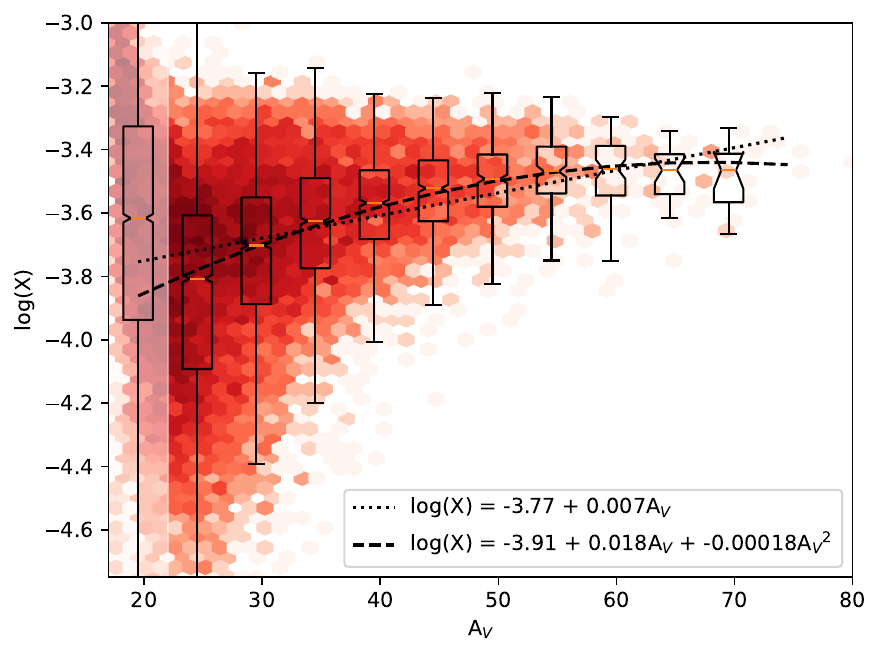}
    \includegraphics[width=0.49\linewidth]{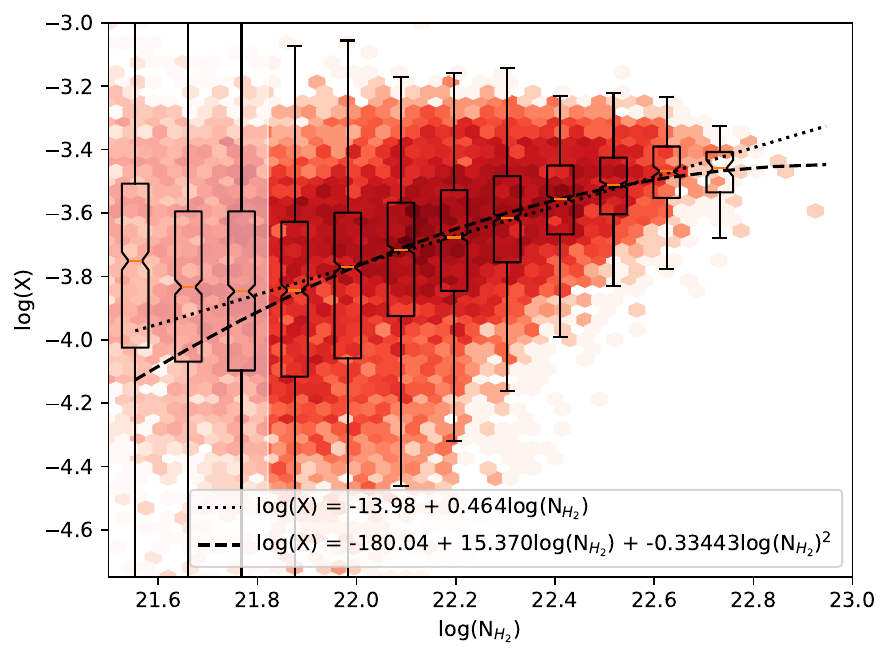}
    \caption{CO ice abundance vs A$_V$ and column density in the Brick.
    The data are the same as those shown in Figure \ref{fig:fig9colvsav}, but with N(CO) divided by N(H$_2$) as the Y-axis.
    The box-and-whisker plots are binned in bins of width A$_V=5$.
    \referee{The region at $A_V<22$ and $\log\left(N(H_2)\right)<21.8$ contains the low signal-to-noise data for which we have likely not detected CO ice.
    The dotted and dashed lines show fitted relations as labeled in the legend.
    }
    \referee{See \S \ref{app:coabundance}.}
    }
    \label{fig:coabundancevsavcolumn}
\end{figure*}

The CO abundance appears to consistently rise with column density.
We plot CO abundance vs A$_V$ and column density in Figure \ref{fig:coabundancevsavcolumn}.
The hexbin plot shows the Brick data with box-and-whisker plots in bins of width A$_V=5$ or $\log \mathrm{N}(\mathrm{H}_2) = 0.1$ overlaid.
We fit both linear and second-order polynomial functions to the medians of the bins above A$_V>22$, i.e., starting one bin past the foreground.
In the N(H$_2$) plot, the column density starts at A$_V=17$, i.e., we assume ice growth begins at the edge of the cloud, ignoring foreground A$_V$.
The power-law fit gives $X = 10^{-11} \mathrm{N}(\mathrm{H}_2)^{0.33}$, though the second-order fit better matches the data, indicating that the abundance increase slows down.
Theoretically, the CO abundance may turn around as CO is converted into other species, but our data do not show this definitively.
The rise from zero CO ice in the atomic ISM to the first solid measurement at X$_{\mathrm{CO}} = 10^{-3.8}$ is well-probed within the data sample, but is dominated by systematic uncertainty about the foreground extinction level, which may be surmountable with spectroscopic observations.

The amount of scatter in the CO column also decreases with increasing column density.
This effect comes from a combination of spatial selection, in which the higher column densities only exist closer to the center of the Brick cloud and therefore select for a more homogeneous subsample, and line-of-sight selection.
The higher scatter at low column densities is produced by an effect that disappears at higher columns: we sample stars at different depths into the Galactic center, and even behind the Galactic center, that have different total foreground dust extinction at a constant amount of CO.
Given that the total extinction to the Galactic Center is A$_V\sim17$, even stars on the other side of the galaxy with double the diffuse ISM path length would be limited to A$_V<40$, so above this threshold stars can only be extinguished by Galactic Center clouds.

\clearpage
\section{Comparison to KP5}
\label{app:kp5}
\begin{figure*}[!h]
    \includegraphics[width=0.49\linewidth]{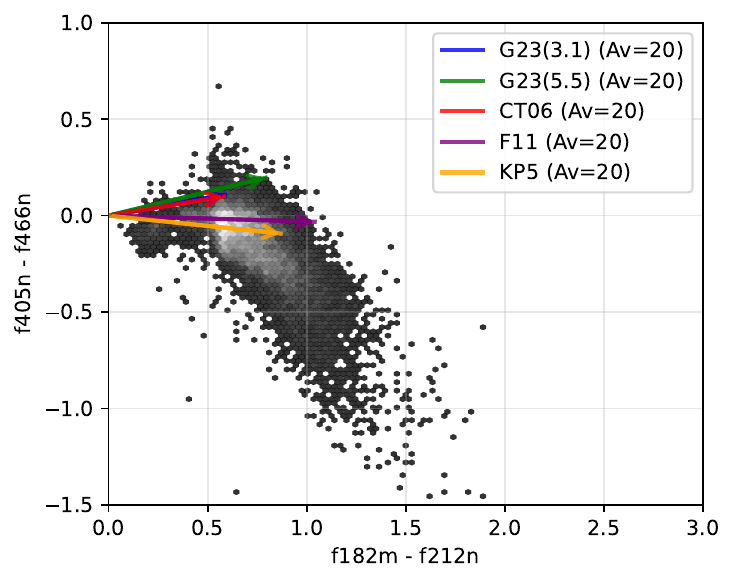}
    \includegraphics[width=0.49\linewidth]{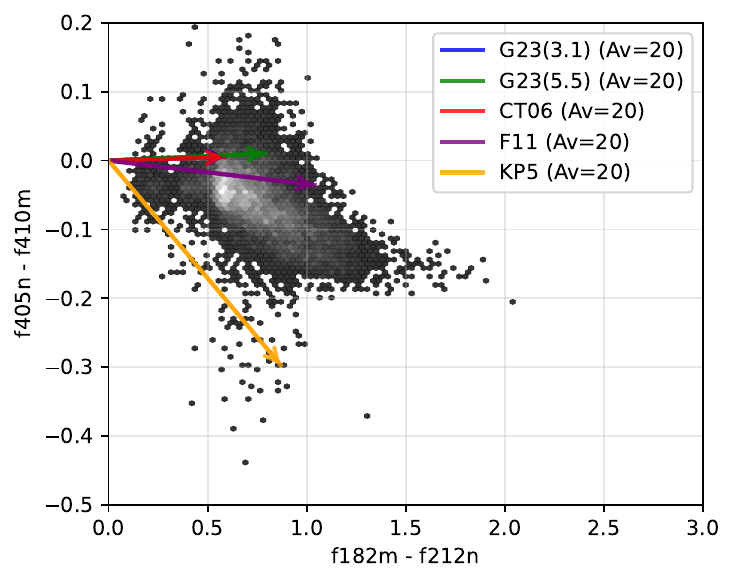}
    \caption{Color-color diagrams showing the extinction-tracing F182M-F212N on the X axis vs F405N-F466N tracing CO (left) and F405N-F410M tracing CO$_2$ (right).
    While the KP5 model produces qualitative agreement with the trend to the blue in F405N-F466N, which other extinction models (also plotted) do not, its slope does not match that of the Galactic Center data.
    }
    \label{fig:kp5}
\end{figure*}

\citet{Pontoppidan2024} presented a constrained dust plus ice model similar to those we have created based on a fit to Spitzer c2d data as presented in \citet{Chapman2009}. This icy dust opacity model has been used successfully to model extinction of gas-phase CO \citep{Rubinstein2024} and H$_2$ \citep{Okoda2025} lines in NIRSpec and MIRI spectra of nearby protostars, where ice features can be very deep \citep{Yang2022, Federman2024}.
In our modeling approach, we have separated the extinction measurement and subsequent dereddening from the measurement of excess absorption, but the end result is the same: the amount of ice absorption scales linearly with the total column density and therefore the total extinction.
To make the KP5 model work in our framework, we treat it as a pure ice, no-dust model and let the ice column set the dust extinction.
We calculate the number fraction ice in the KP5 model is H$_2$O:CO$_2$:CO = 72:25:2.7 in the NIRCam band (KP5 uses different ice mixtures at different wavelengths, so it is not formally self-consistent), which results in a CO/H $=2.7\times10^{-6}$, much lower than adopted elsewhere in this work.
The KP5 model is not a good fit to our data, confirming the conclusion that Galactic Center ice is substantially different from the ice observed in Galactic disk protostars.
In particular, it under-predicts the F466N absorption from CO ice and over-predicts F410M absorption from CO$_2$.
Figure \ref{fig:kp5} shows this comparison.

\clearpage
\section{Where are the icy stars}
\label{app:spatialplotstwo}
\begin{figure*}[!h]
    \includegraphics[width=\linewidth]{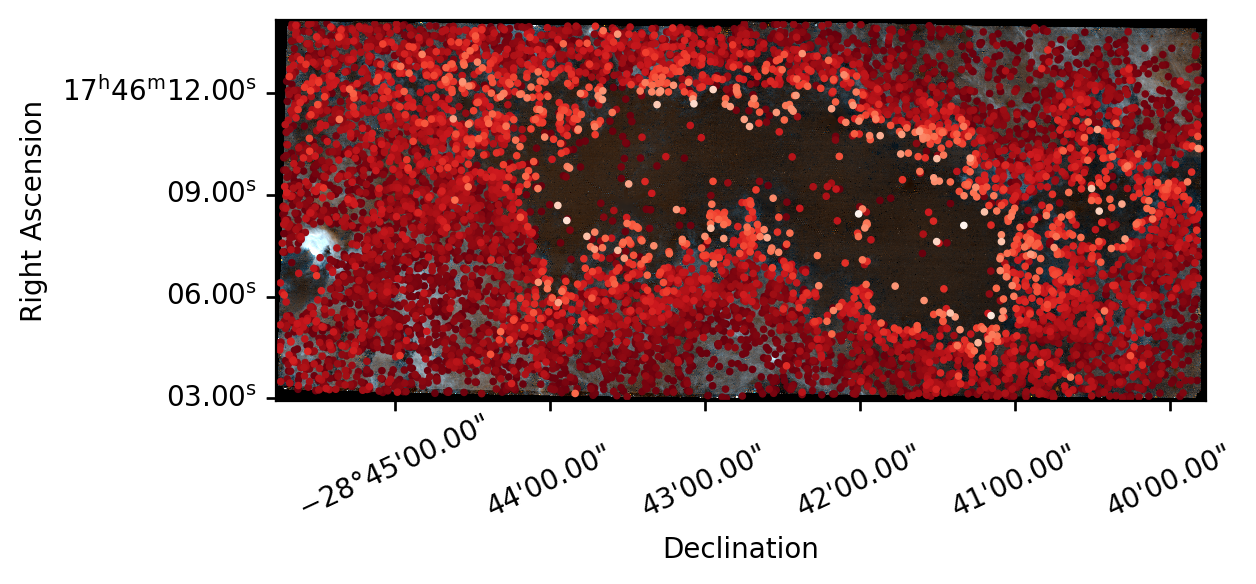}
    \includegraphics[width=\linewidth]{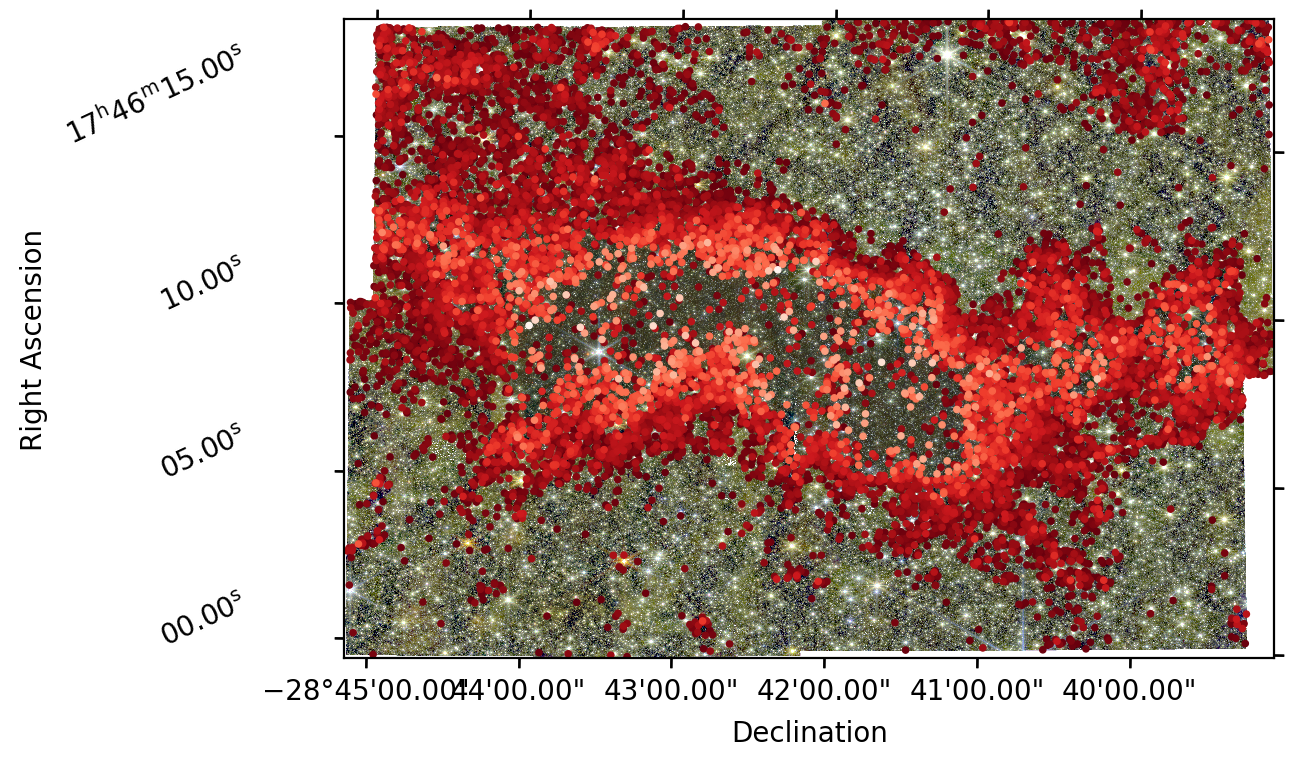}
    \caption{We show where the most ice-absorbed stars occur.
    The first figure shows stars with dereddened F405N-F466N $<-0.3$ and A$_V>17$, with extinction measured from F182M-F212N color.
    The backdrop is the star-subtracted F405N+F466N image from \citet{Ginsburg2023}.
    The second shows stars with F356W-F444W $>0.3$ and A$_V>17$, with extinction measured from F200W-F400W color.
    The backdrop is an RGB image composed of F444W (R), F356W (G), and F200W (B) from program 1182.
    The colormap is the dereddened color, going from the threshold value in dark red to white at the most extreme.
    }
    \label{fig:spatialmap}
\end{figure*}

Figure \ref{fig:spatialmap} shows where the icy stars reside \referee{overlaid on JWST images, in contrast with Figure \ref{fig:spatialdistribution}, which showed the same stars on ACES millimeter emission maps.
In particular, Figure \ref{fig:spatialmap}b shows the extent of the wide-band images from program 1182, emphasizing that there are no F356W-excess sources outside of the Brick cloud despite the large extent of the JWST images.
}
Stars with substantial color excess in the F466N filter, tracing CO, and the F356W filter, hypothesized to trace \methanol, are shown overlaid on background images made from the narrowband project 2221 filters and the wideband project 1182 filters, respectively.
The figures highlight that stars with excess absorption are not detected toward the center of the cloud because extinction in that region is too high.
They also highlight that the CO-absorbed stars are more widespread than the F356W-absorbed stars, contributing to the hypothesis that chemical processing at high density is required to produce that excess and hinting that a more complex molecule, like \methanol, is responsible.
The F356W-excess stars are seen slightly further into the center of the cloud because the wideband filters provide greater sensitivity.

\clearpage
\section{A deeper check on NIRCam filters}
\label{sec:1e19table}
We repeat Table \ref{tab:nircam_ice_absorption} but with a higher column density in Table \ref{tab:nircam_ice_absorption_19}.
In this table, we adopt a column density of each ice molecule of 10$^{19}$ \persc, which is expected for some molecules (CO, \water) and unlikely (but not impossible) to be observed anywhere for some others (e.g., OCS, HCN, CH$_3$OH).
If a molecule does not appear in this table and it has been measured in lab, it is unlikely to produce substantial, detectable absorption in any NIRCam filter.

\begin{table*}[ht]
\centering
\caption{Ice molecules that significantly absorb NIRCam filters [N(ice)=10$^{19}$ cm$^{-2}$]}
\label{tab:nircam_ice_absorption_19}
\begin{tabular}{|l|p{10cm}|p{6cm}|}
\hline
\textbf{Filter} & \textbf{Ice Molecules} & \textbf{$\Delta$mag Values} \\
\hline
F250M & CO$_2$ & 0.33 \\
F277W & CO$_2$, H$_2$O, CH$_3$OH, NH$_3$ & 0.71, 0.62, 0.34, 0.14 \\
F300M & H$_2$O, CH$_3$OH, CO$_2$, NH$_3$, HCN & 3.6, 1.5, 0.88, 0.41, 0.17 \\
F323N & H$_2$O, CH$_3$OH, HCN, CO$_2$, NH$_3$ & 3.0, 1.9, 1.9, 0.83, 0.11 \\
F322W2 & CO$_2$, CH$_3$OH, H$_2$O, HCN & 0.73, 0.55, 0.39, 0.21 \\
F335M & CH$_3$OH, CO$_2$, H$_2$O, HCN, CH$_4$ & 1.9, 0.77, 0.65, 0.63, 0.13 \\
F356W & CH$_3$OH, CO$_2$, HCN, H$_2$O & 0.81, 0.72, 0.44, 0.34 \\
F360M & CO$_2$, CH$_3$OH, HCN, H$_2$O & 0.69, 0.55, 0.31, 0.11 \\
F405N & CO$_2$, H$_2$O, HCN & 0.8, 0.26, 0.17 \\
F410M & CO$_2$, H$_2$O, HCN, CH$_3$OH & 0.99, 0.29, 0.17, 0.11 \\
F430M & CO$_2$, H$_2$O, HCN & 2.0, 0.38, 0.11 \\
F444W & CO$_2$, H$_2$O, OCS, HCN & 0.7, 0.25, 0.17, 0.14 \\
F460M & CO$_2$, H$_2$O, CO & 0.43, 0.24, 0.23 \\
F466N & CO, CO$_2$, H$_2$O & 0.64, 0.43, 0.22 \\
F470N & CO$_2$, CO, H$_2$O & 0.42, 0.35, 0.18 \\
F480M & OCS, CO$_2$, CO, HCN, H$_2$O & 1.1, 0.41, 0.26, 0.22, 0.15 \\
\hline
\end{tabular}
\par Molecules that absorb NIRCam filters by at least 0.1 mag when their column density is 10$^{19}$ cm$^{-2}$.  This table is not comprehensive, since some molecules are potentially much more abundant (e.g., \water), and the more complex molecules are likely to be rarer.  NIRCam filters excluded from this table do not have significant ($>0.1$ mag) ice absorption at N(ice)=10$^{19}$ cm$^{-2}$.  
\end{table*}

\subsection{MIRI filters}
For completeness, we computed the same tables as in Appendix \ref{sec:1e19table} for MIRI.
We have not examined the MIRI filters closely, but these measurements provide a first-look idea of what MIRI might detect.
Tables \ref{tab:miri_ice_absorption} and \ref{tab:miri_ice_absorption_19} mirror tables \ref{tab:nircam_ice_absorption} and \ref{tab:nircam_ice_absorption_19}, but for MIRI rather than NIRCam.

\begin{table*}[ht]
\centering
\caption{Ice molecules that significantly absorb MIRI filters [N(ice)=10$^{18}$ cm$^{-2}$]}
\label{tab:miri_ice_absorption}
\begin{tabular}{|l|p{10cm}|p{6cm}|}
\hline
\textbf{Filter} & \textbf{Ice Molecules} & \textbf{$\Delta$mag Values} \\
\hline
F1130W & H$_2$O & 0.15 \\
F1140C & H$_2$O & 0.15 \\
F1280W & H$_2$O & 0.14 \\
F1500W & CO$_2$ & 0.18 \\
F1550C & CO$_2$ & 0.25 \\
\hline
\end{tabular}
\par Molecules that absorb MIRI filters by at least 0.1 mag when their column density is 10$^{18}$ cm$^{-2}$.  This table is not comprehensive, since some molecules are potentially much more abundant (e.g., \water), and the more complex molecules are likely to be rarer.  MIRI filters excluded from this table do not have significant ($>0.1$ mag) ice absorption at N(ice)=10$^{18}$ cm$^{-2}$.  Several molecules in the ice database are excluded because they have not been reported in the ISM, including NH$_4$CN, N$_2$H$_4$, and HC$_3$N.  
\end{table*}

\begin{table*}[ht]
\centering
\caption{Ice molecules that significantly absorb MIRI filters [N(ice)=10$^{19}$ cm$^{-2}$]}
\label{tab:miri_ice_absorption_19}
\begin{tabular}{|l|p{10cm}|p{6cm}|}
\hline
\textbf{Filter} & \textbf{Ice Molecules} & \textbf{$\Delta$mag Values} \\
\hline
F560W & CH$_{3}$COCH$_{3}$, CO$_2$, H$_2$O, HCN, OCS & 0.31, 0.3, 0.26, 0.22, 0.13 \\
F770W & CH$_{3}$COCH$_{3}$, CH$_3$OH, H$_2$O, CO$_2$, HCN & 0.36, 0.27, 0.26, 0.24, 0.12 \\
F1000W & NH$_3$, CH$_3$OH, H$_2$O, CO$_2$, CH$_{3}$COCH$_{3}$ & 0.39, 0.34, 0.31, 0.29, 0.15 \\
F1065C & H$_2$O, CO$_2$ & 0.74, 0.29 \\
F1130W & H$_2$O, CO$_2$, CH$_3$OH, OCS, HCN, CH$_{3}$COCH$_{3}$ & 1.4, 0.28, 0.22, 0.14, 0.11, 0.11 \\
F1140C & H$_2$O, CO$_2$, CH$_3$OH, OCS, CH$_{3}$COCH$_{3}$ & 1.5, 0.28, 0.22, 0.12, 0.11 \\
F1280W & H$_2$O, CH$_3$OH, HCN, CO$_2$, OCS & 1.3, 0.83, 0.4, 0.34, 0.1 \\
F1500W & H$_2$O, CO$_2$, CH$_3$OH & 0.91, 0.6, 0.54 \\
F1550C & H$_2$O, CO$_2$, CH$_3$OH & 0.81, 0.68, 0.46 \\
F1800W & H$_2$O, CO$_2$, NH$_3$ & 0.49, 0.27, 0.14 \\
F2100W & NH$_3$, H$_2$O, CO$_2$, C$_{4}$N$_{2}$, CH$_{3}$C$_{3}$N & 0.34, 0.26, 0.12, 0.1, 0.1 \\
F2300C & NH$_3$, H$_2$O, C$_{4}$N$_{2}$ & 0.23, 0.15, 0.15 \\
F2550W & CH$_{3}$CH$_{2}$CN & 0.1 \\
\hline
\end{tabular}
\par Molecules that absorb MIRI filters by at least 0.1 mag when their column density is 10$^{19}$ cm$^{-2}$.  This table is not comprehensive, since some molecules are potentially much more abundant (e.g., \water), and the more complex molecules are likely to be rarer.  MIRI filters excluded from this table do not have significant ($>0.1$ mag) ice absorption at N(ice)=10$^{19}$ cm$^{-2}$.  
\end{table*}


\clearpage
\section{ISO spectra of GC sources}
\label{sec:isospectra}
Since no Galactic Center JWST spectra are public at the time of writing, and none have been acquired toward any of our target sources, we refer to ISO spectra  \referee{in the Galactic Center} as references for chemically similar environments.
\citet{Gerakines1999} published spectra of GCS 3 and 4, pointings toward the Quintuplet cluster, focusing on the CO$_2$ feature.
\citet{Moneti2001} measured \water, CO, and XCN toward GCS 3.
These spectra are still rising at 4 \um, indicating the dominance of warm dust emission in the large $14\times20$\arcsec\ ISO SWS aperture.
The XCN feature at 4.6 \um peaks at the same depth as CO, hinting that this feature may be substantial in the Brick spectra, taken from only a few arcminutes away.
However, the derived synthetic colors from this spectrum do not match the NIRCam data, and the ISO spectra have relatively low total optical depth.
Both of these lines of evidence mean these spectra are not a perfect analog of the stars we measure.

\begin{figure*}[!h]
    \includegraphics[width=0.45\linewidth]{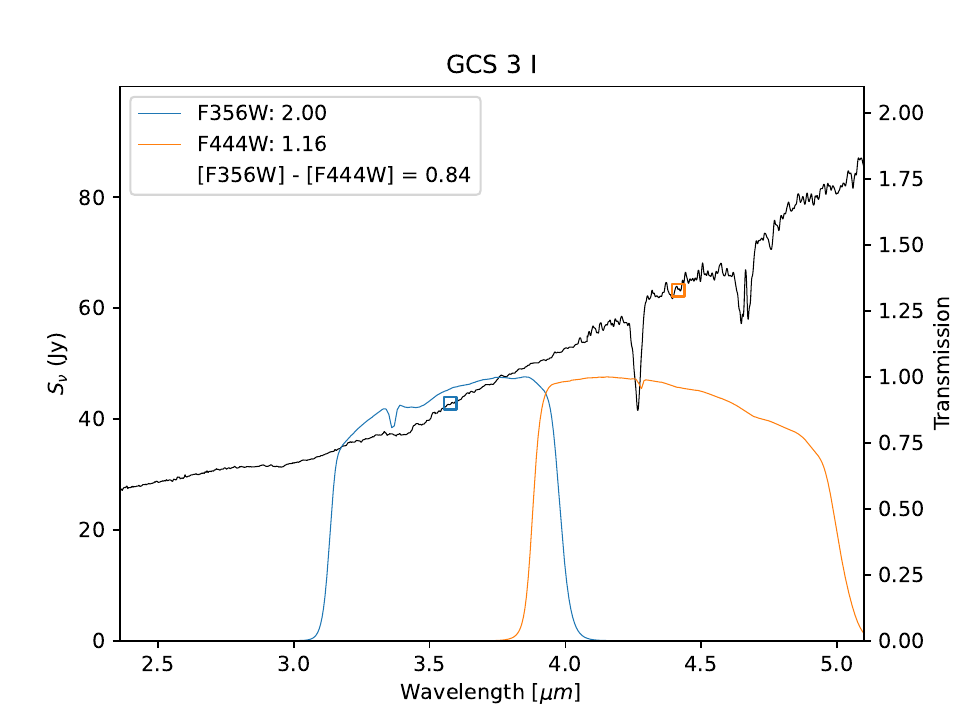}
    \hfill
    \includegraphics[width=0.45\linewidth]{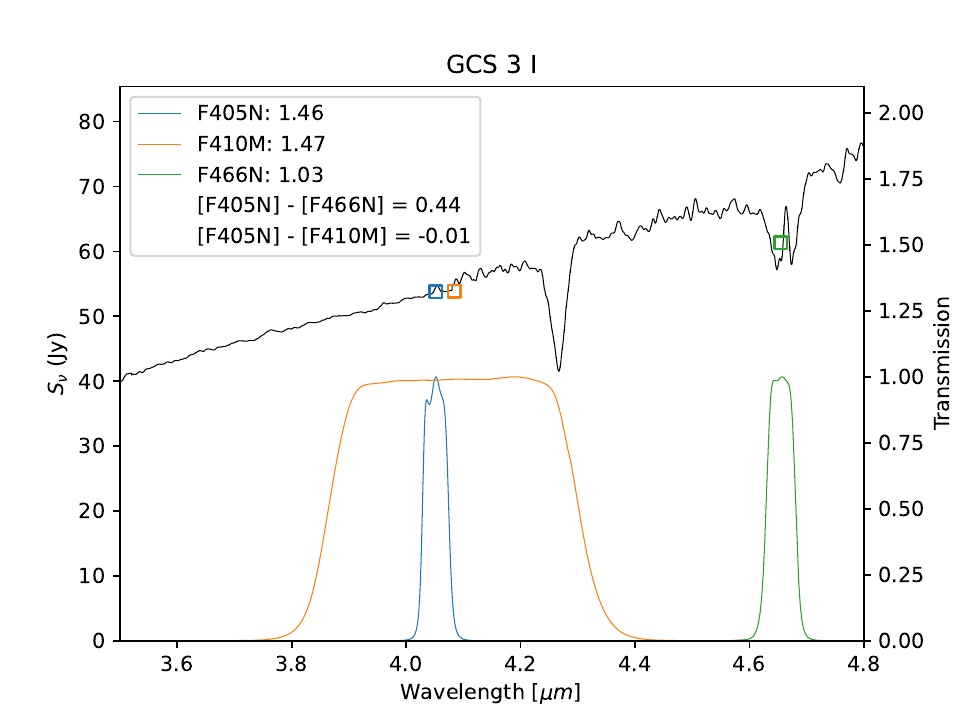}
    
    \includegraphics[width=0.45\linewidth]{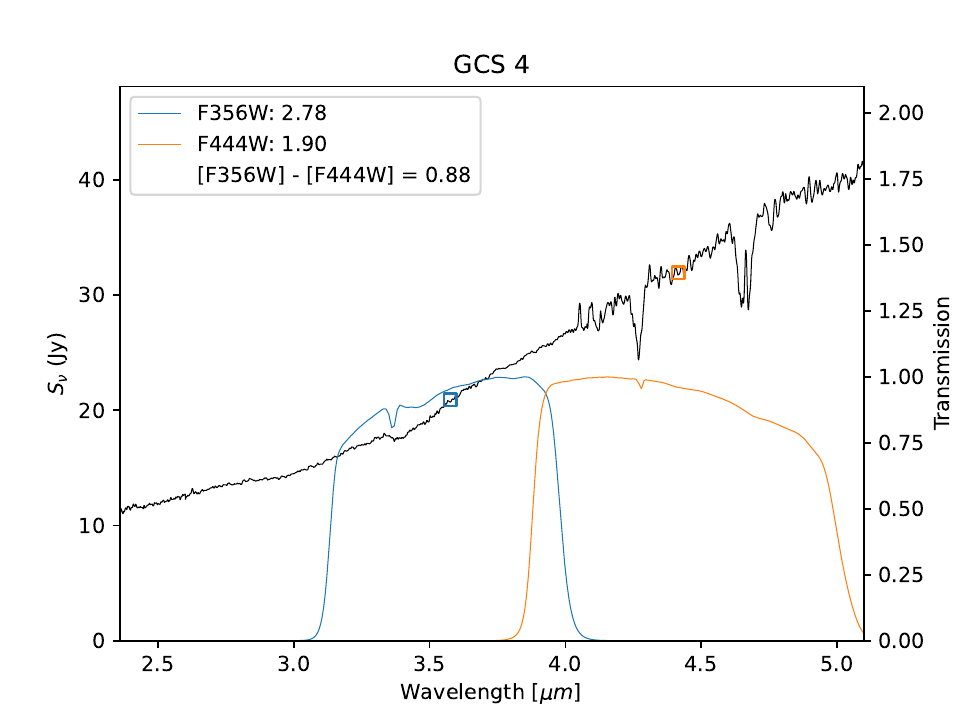}
    \hfill
    \includegraphics[width=0.45\linewidth]{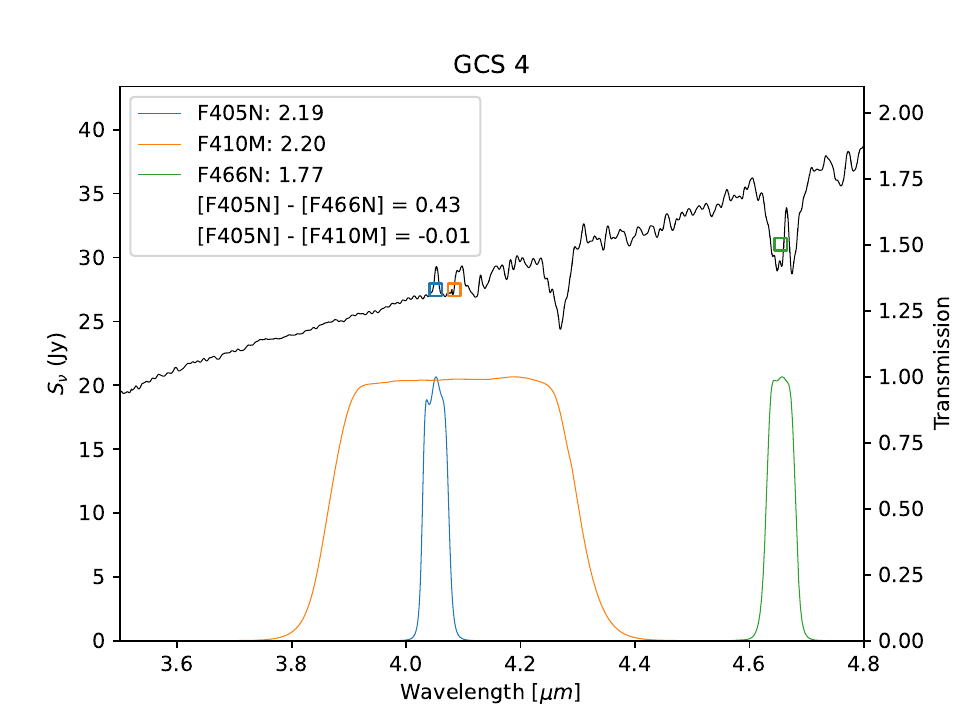}
    \caption{ISO spectra of GCS 3 (top) and GCS 4 (bottom), pointings toward the Quintuplet cluster.
    \referee{See \S \ref{sec:isospectra}.}
    }
    \label{fig:isospectra}
\end{figure*}

\clearpage
\section{Effects of mixed ice}
\label{sec:icemixturesredux}

\begin{figure}[!h]
    \centering
    \includegraphics[width=0.6\linewidth]{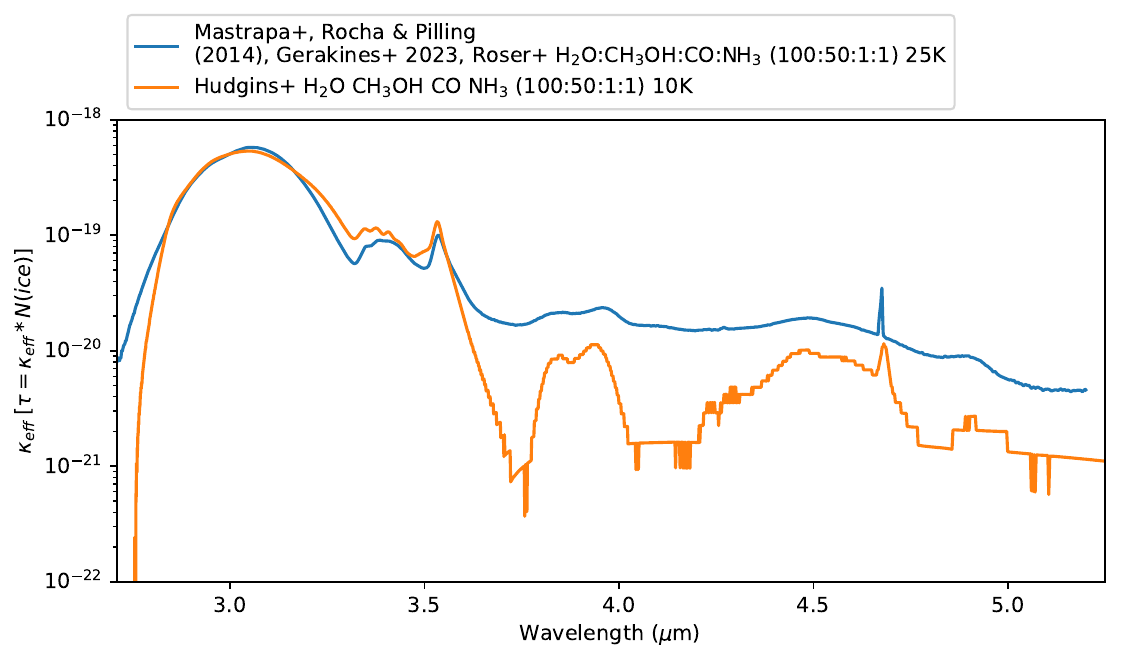}
    \caption{Comparison of our linear combination of ices to a real lab mixture.
    The orange curve shows a laboratory measurement from \citet{Hudgins1993} of a \water:\methanol:CO:\ammonia mixture of 100:50:1:1.
    The blue curve shows our linear combination of four independent laboratory measurements of pure ices of these molecules from \citet{Mastrapa2009}, Fraser (via LIDA; no reference is provided on their site), \citet{Gerakines2023}, and \citet{Roser2021}.
    There are very significant differences between these opacity curves, which we discuss in \S \ref{sec:icemixturesredux}.
    }
    \label{fig:icemixcomparison}
\end{figure}

Figure \ref{fig:icemixcomparison} shows a comparison between a linear combination of pure ices and a direct mixture of these ices.
The curves are \referee{broadly similar}.
\referee{There are structures in the lab-measured ice, dips between the peaks, that are not seen in the linear combination.
The missing dips are because they are not present in the \methanol spectrum from \citet{Rocha2014}.
}

\begin{figure*}[!h]
    \centering
    \includegraphics[width=0.7\linewidth]{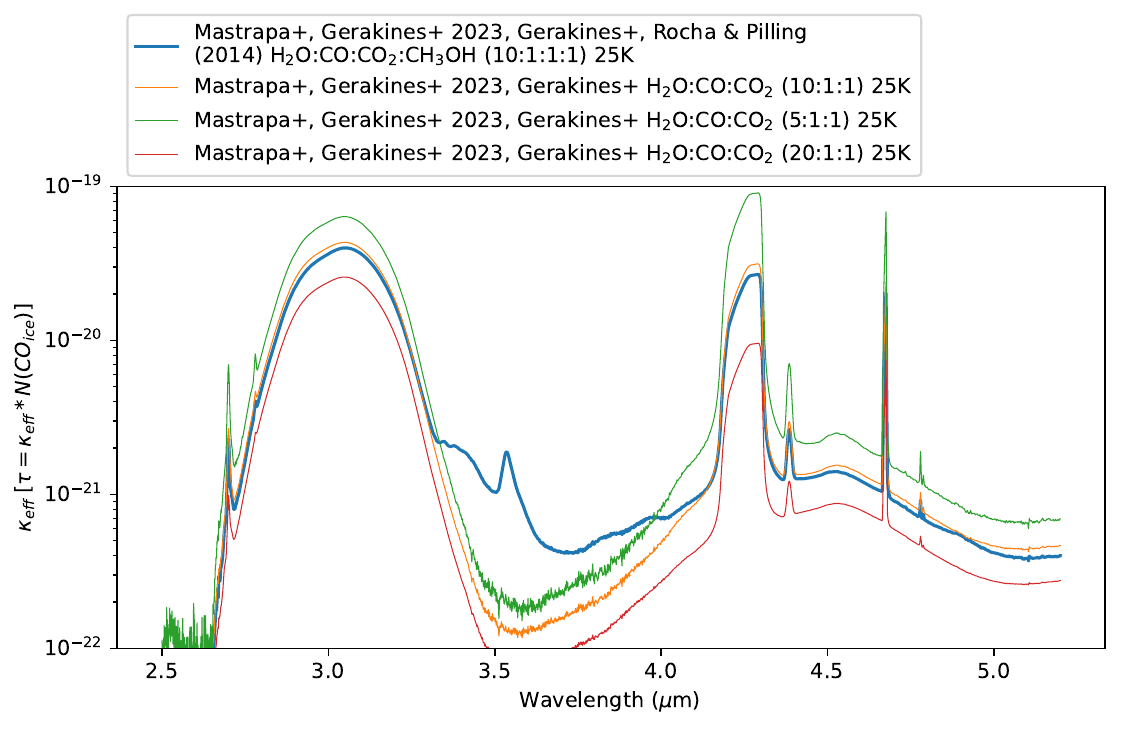}
    \includegraphics[width=0.7\linewidth]{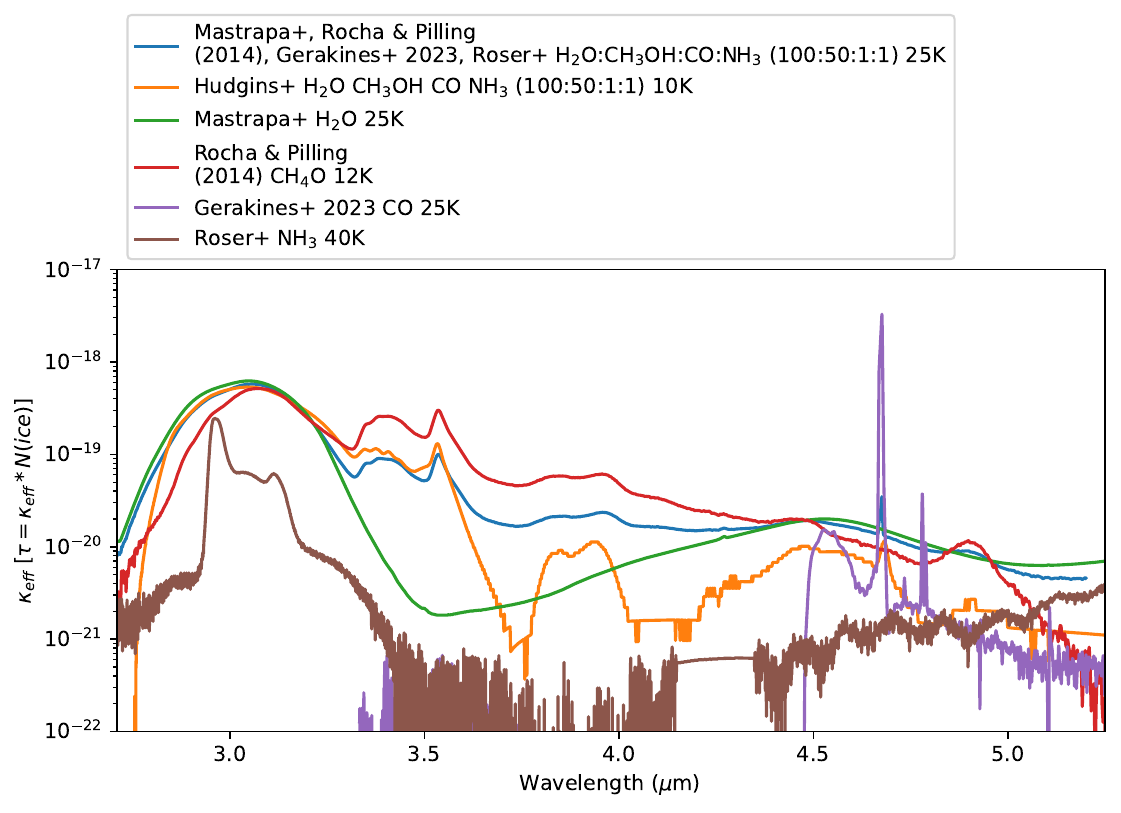}
    \caption{Comparison of the effective opacities of the \water:CO:CO$_2$ ice mixtures considered.
    These are all normalized to the CO column density such that the opacity plotted is the total opacity of the ice per CO molecule.
    The second figure shows the same, but now with the individual pure ice components also shown and the opacity normalized to the column of the specific ice.
    \referee{See \S \ref{sec:icemixturesredux}.}
    }
    \label{fig:icemix_overlays}
\end{figure*}

\clearpage
\section{Additional absorbance plots}
\label{sec:moreabsorptionplots}
\begin{figure}[!h]
    \centering
    \includegraphics[width=0.6\linewidth]{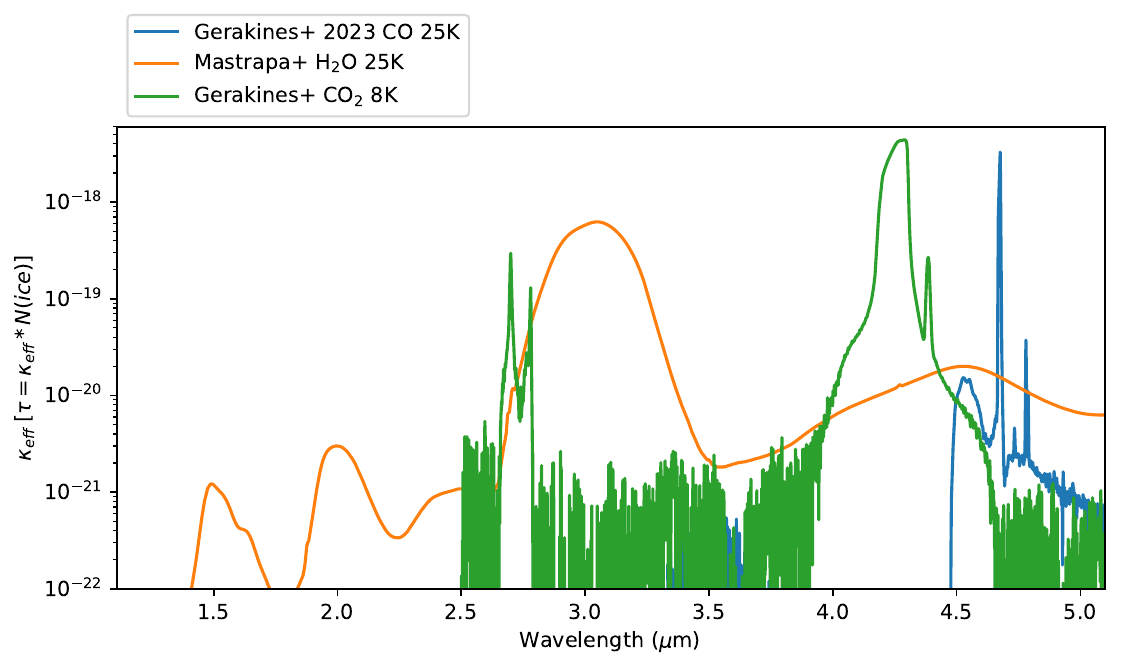}
    \caption{Opacities used shown overplotted on the full NIRCam band.  This plot can be used to guide filter selection or, at a glance, infer which ice species may affect a given photometric measurement.
    Curves are from \citet{Gerakines2020}, \citet{Gerakines2023}, and \citet{Mastrapa2009}.
    \referee{The NIRCAM filters are intentionally excluded for visual clarity; Figures \ref{fig:opacitiesf277etc}-\ref{fig:filter_overlays_continued} show the NIRCam filter transmission profiles overlaid on model opacity curves.}
    }
    \label{fig:opacitiesonfull}
\end{figure}

\begin{figure}[!h]
    \centering
    \includegraphics[width=0.6\linewidth]{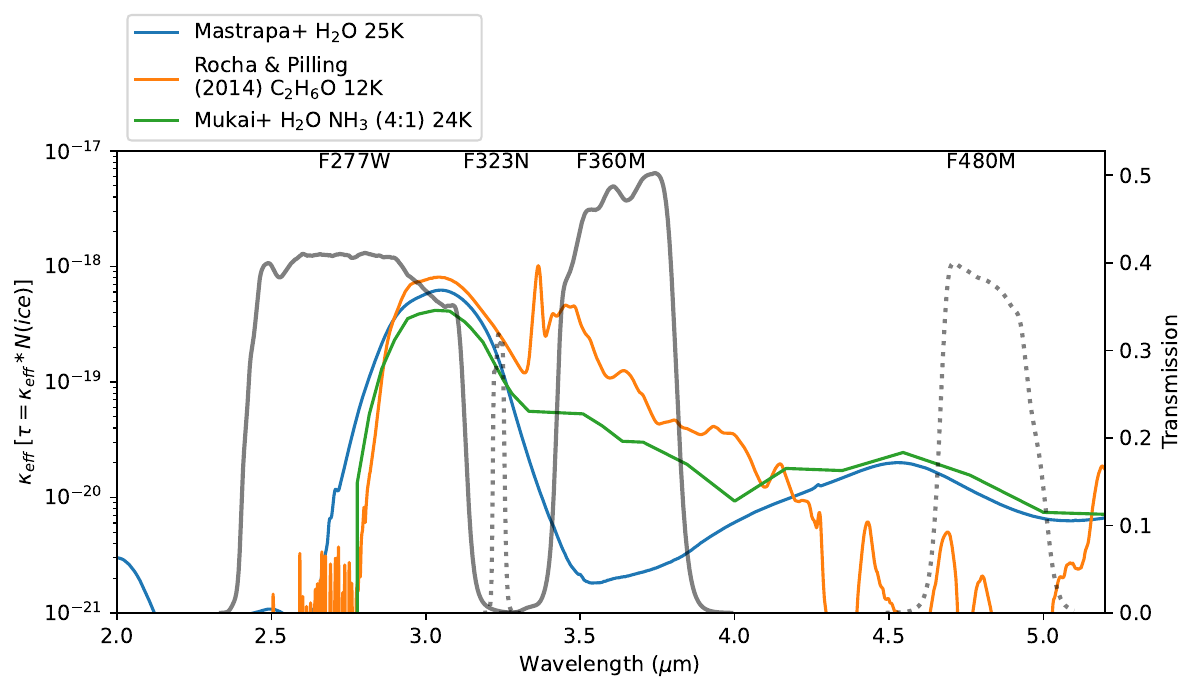}
    \caption{Opacities overplotted on the F277W, F323N, F360M, and F480M bands.  Similar to the other opacity plots (Fig. \ref{fig:F466Nplusopacities}, \ref{fig:opacitiesonfull}, and \ref{fig:opacities_on_widebands}, the main aim is to show which ices are likely to affect a given photometric band.
    Curves come from  \citet{Rocha2014}, \citet{Mastrapa2009}, and \citet{Mukai1986}.
    \referee{The right axis displays the filter transmission curve of the selected filters (gray).}
    }

    \label{fig:opacitiesf277etc}
\end{figure}

\begin{figure*}[!h]
\centering
    \includegraphics[width=0.75\linewidth]{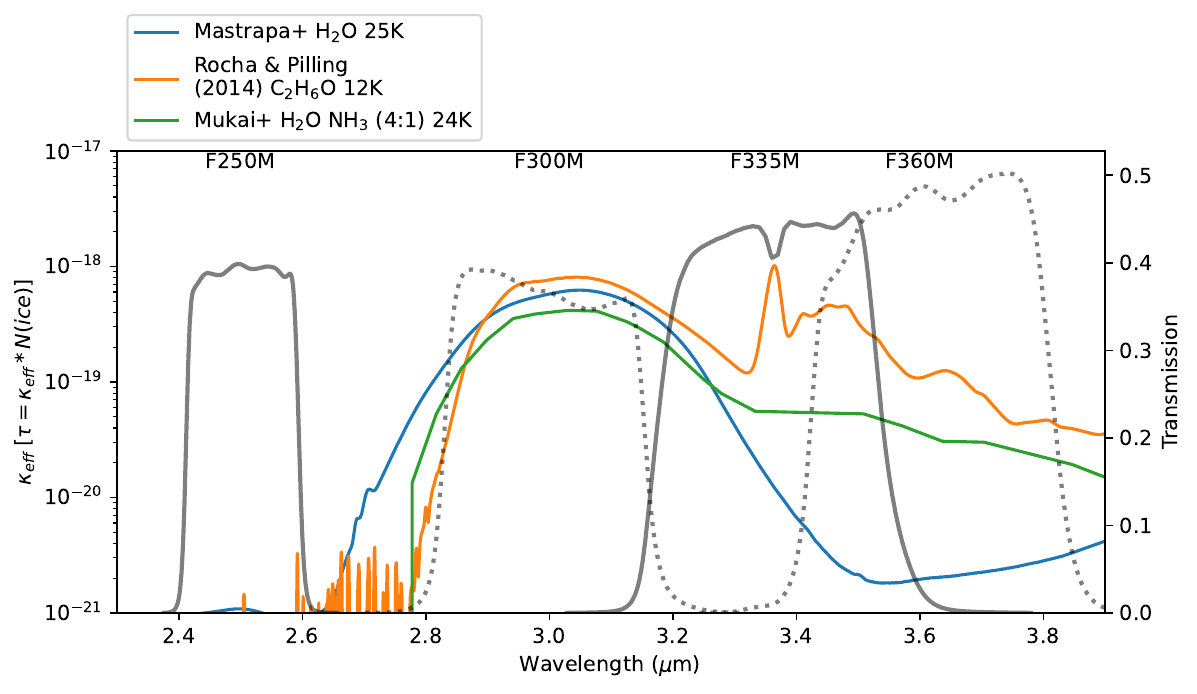} \\
    \includegraphics[width=0.75\linewidth]{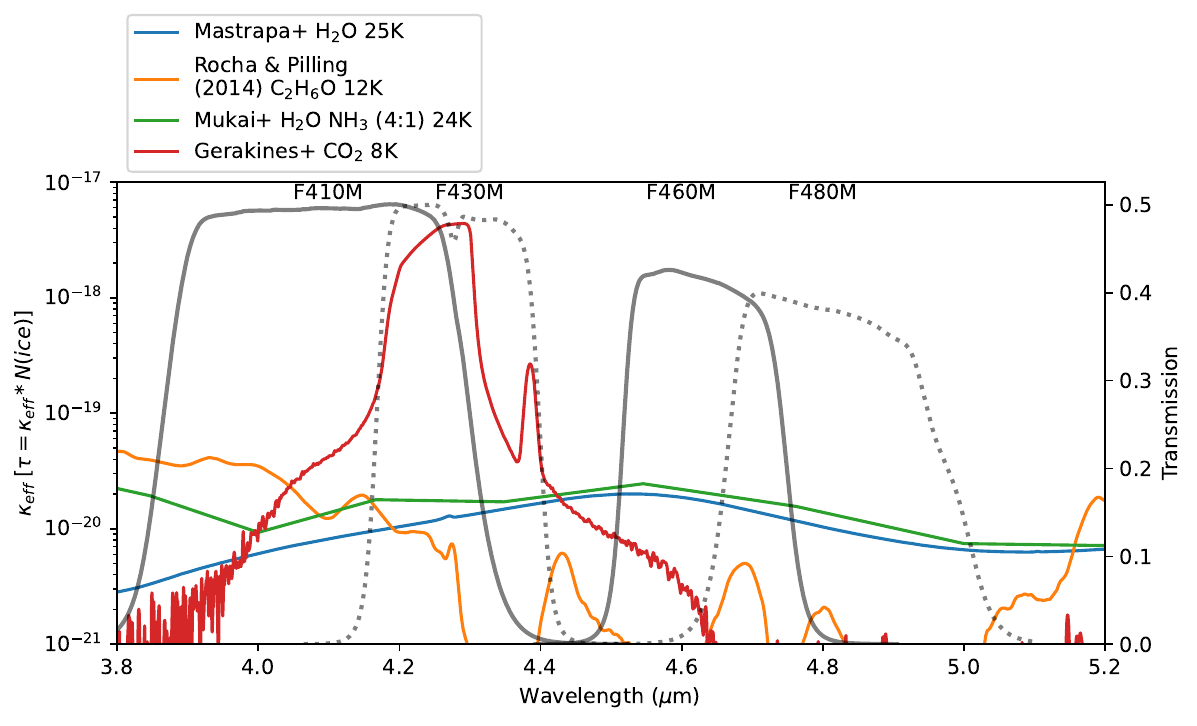} \\
    \includegraphics[width=0.75\linewidth]{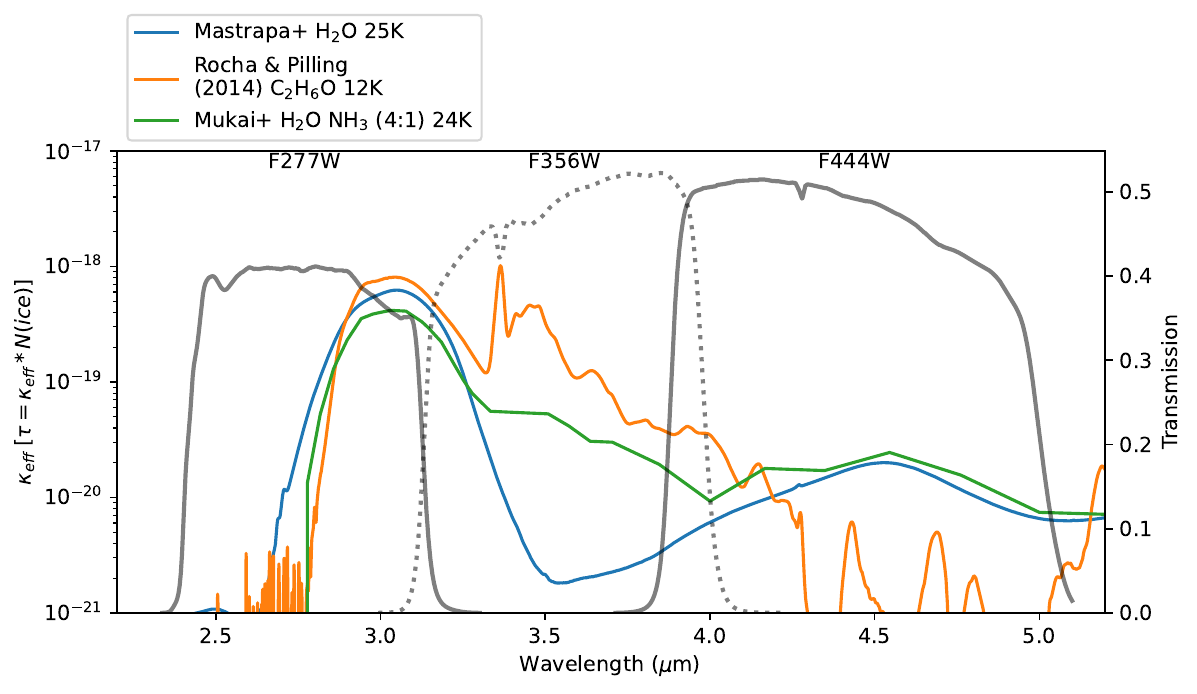}
    \caption{\referee{Reference figures showing opacities of water, ethanol, and water-ammonia ices on medium- and broad-band filters.  In the 4\um medium-band filter figure, CO$_2$ ice is also shown.}
    \referee{The right axis displays the filter transmission curve of the selected filters (gray).}
    }
    \label{fig:filter_overlays_continued}
\end{figure*}

Figures \ref{fig:opacitiesonfull} and \ref{fig:opacitiesf277etc} show ice opacities from selected ice species overlaid on the transmission curve of NIRCam filters.
To first order, any peak above $\kappa_{eff}>10^{-18}$ cm$^2$ will have a strong effect even at low column densities, and any absorption peaks above $\kappa_{eff}>10^{-19}$ cm$^2$ cannot be ignored.
Some molecules apparently have substantial opacity, such as NH$_3$, but they are not expected to freeze out at the more modest column densities (i.e., N(H$_2)<10^{23}$ \persc) that allow background stars to be detected \citep{Caselli2022}.
\referee{
In Figure \ref{fig:filter_overlays_continued}, we show the same curves overlaid on the remaining NIRCam filters for completeness.
}

\section{Stellar SED plots with and without extinction and ice}
\label{sec:stellarsedplots}
\begin{figure}
    \centering
    \includegraphics[width=1\linewidth]{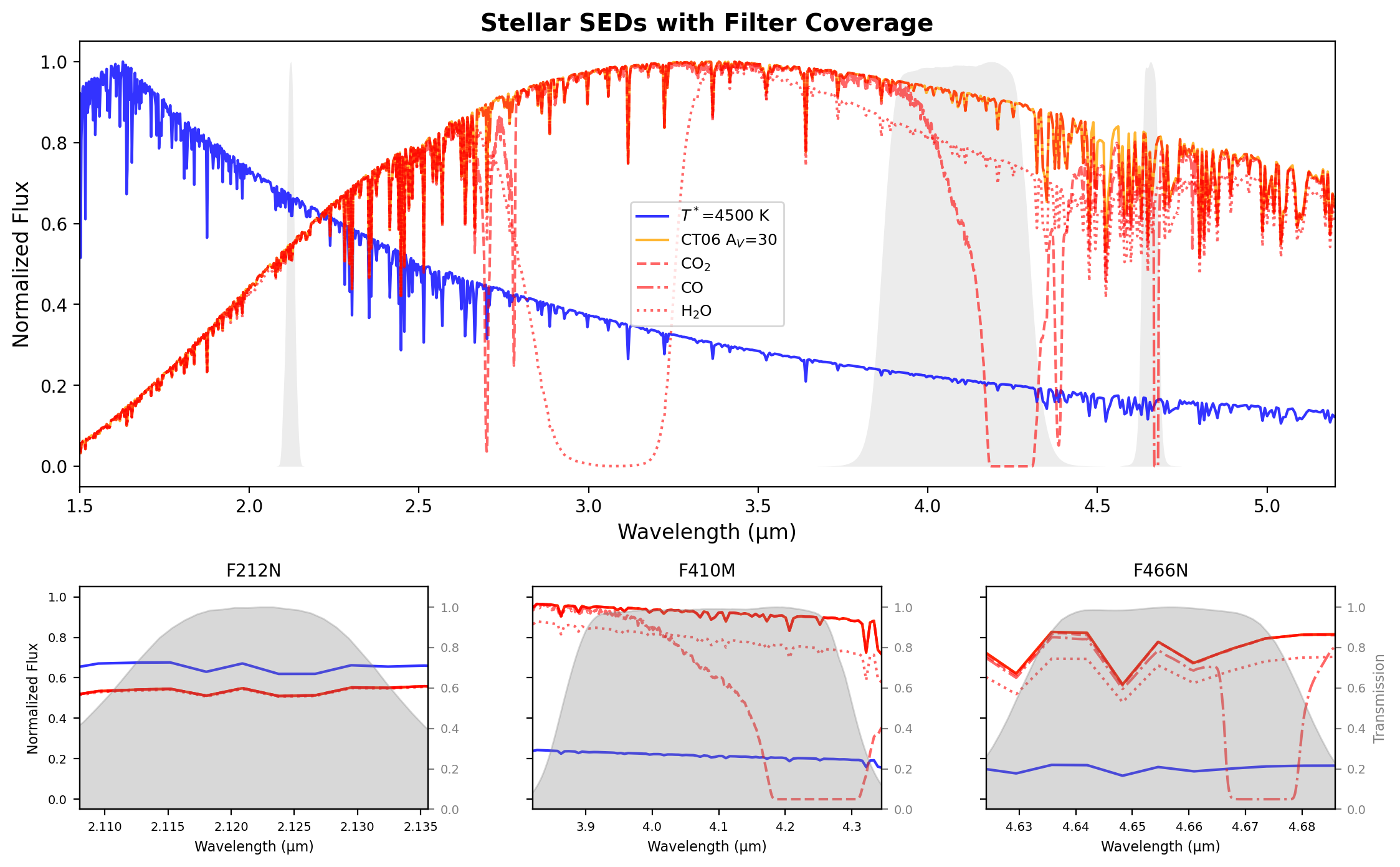}
    \caption{\referee{Example stellar spectrum from the PHOENIX solar metallicity model grid with T=4500 K and log(g)=2.5 (blue), dust-extincted stellar spectrum with $A_V=30$ (orange), and dust-extincted plus ice-absorbed stellar spectrum with $A_V=30$ and N(CO)=10$^{19}$ \persc\ (red dash-dotted), N(H$_2$O)=10$^{19}$ \persc\ (red dotted), and N(CO$_2$)=10$^{19}$ \persc\ (red dashed).  These high column densities represent typical values among the most highly extinguished stars at $A_V\gtrsim30$ within the cloud.  The insets show three filters: F212N, F410M, and F466N, highlighting the effect of ice absorption on the latter two and the lack of ice absorption on the former.   The other filters are excluded, as they show redundant information (F182M and F187N have no ice features, F405N rests within F410M).}}
    \label{fig:stellarSEDexample}
\end{figure}

Figure \ref{fig:stellarSEDexample} shows a 4500 K, g=2.5 PHOENIX \citep{Husser2013} model stellar SED.
It is shown in three colors: in blue unadulterated, in orange attenuated by dust only, and in red, in three linestyles, absorbed by CO$_2$, H$_2$O, and CO ice.
The NIRCam filters F212N, F410M, and F466N are overlaid and shown as zoomed-in inset plots.
This figure shows the features that go into the photometric modeling throughout the paper.

\section{Color vs Column Density}
\label{sec:colorvscolumn}
\begin{figure}
    \centering
    \includegraphics[width=1.0\linewidth]{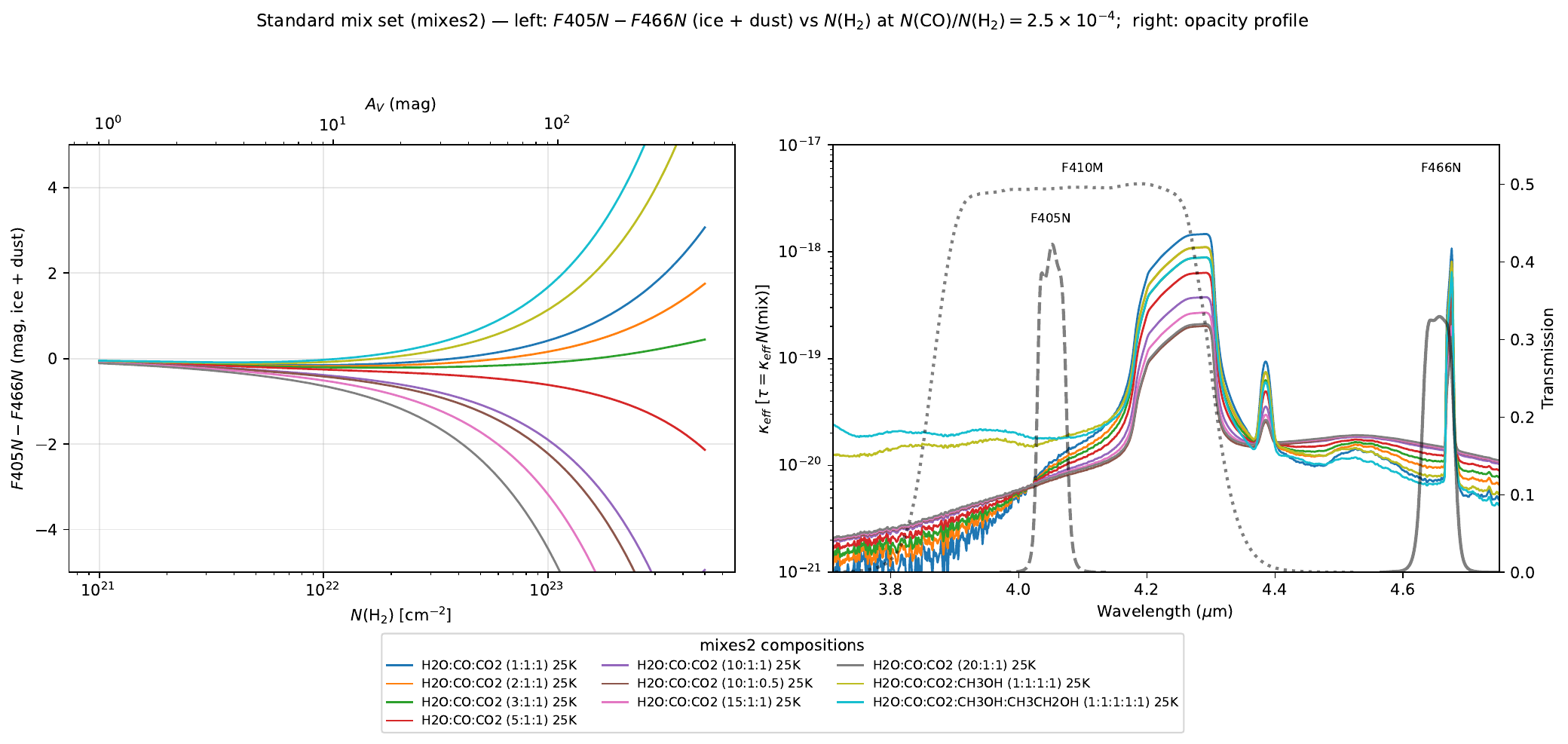}
    \includegraphics[width=1.0\linewidth]{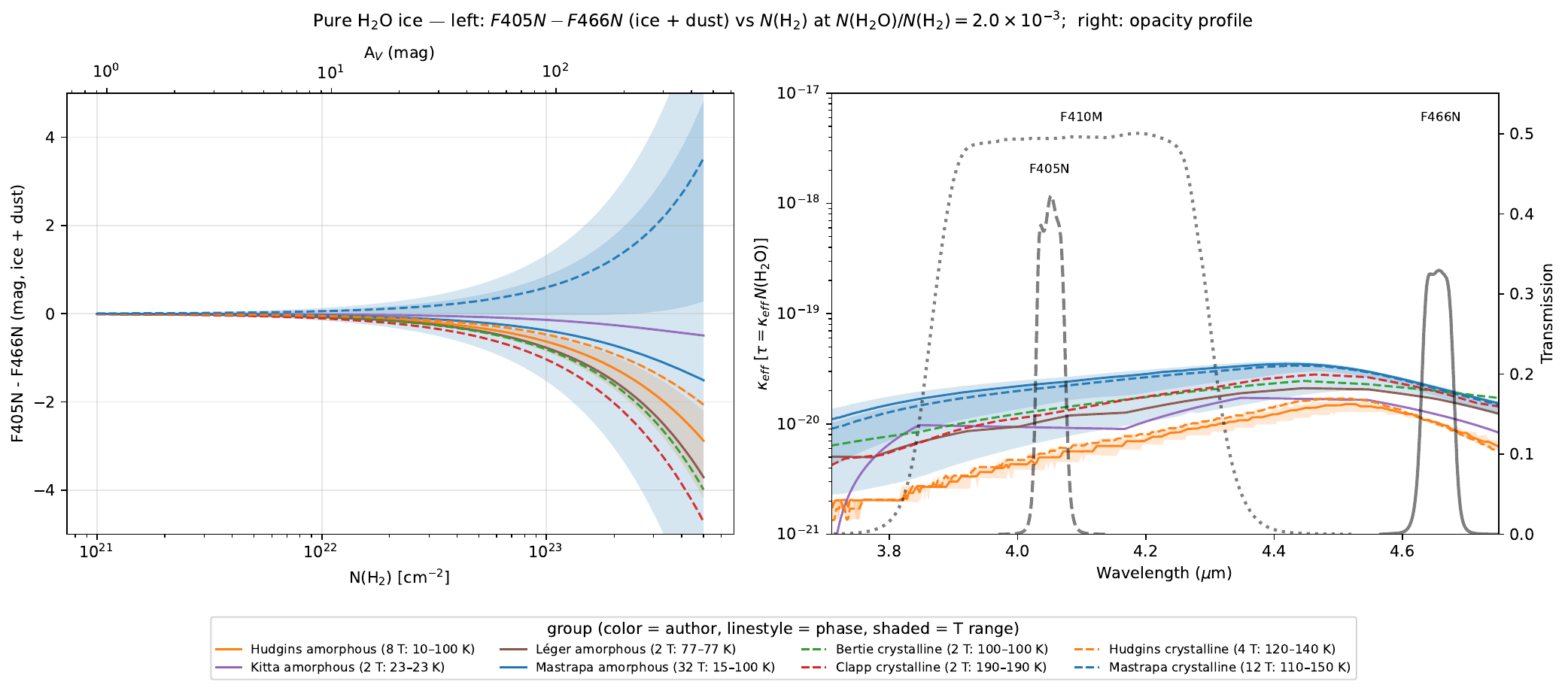}
    \caption{Color vs. opacity plots (left) and opacity curves (right) for a selection of ice mixtures.  See \S \ref{sec:colorvscolumn}.}
    \label{fig:color_vs_av}
\end{figure}

\referee{
To clarify the relations shown in the color-color diagrams in Figures  \ref{fig:colorcolor}, \ref{fig:colorcolorwithcloudc}, \ref{fig:ccd_f405nmf410m}, \ref{fig:colorcolorwide},  we show color as a function of column density and extinction side-by-side with the opacity curves for various ice species and mixtures in Figure \ref{fig:color_vs_av}.
The top panel shows the ice mixtures we consider in those previous color-color diagrams; these are labeled \texttt{mixes2} in the source code.
The bottom panel shows the range of opacities for a single ice species: pure water, in both amorphous and crystalline form, at a range of temperatures.
The water opacities illustrate a few key points: first, that depending on the adopted opacity curve, water ice on dust can produce either red or blue colors in F405N-F466N, and second, that even in the most extreme opacities, water ice alone cannot produce the blue colors observed at the column densities observed (F405N-F466N $\sim-1.5$ at $A_V\sim 50$), while the ice mixtures in the top panel can and do.
To demonstrate this point, we show the color vs column density for water assuming that 100\% of oxygen is in water and that the oxygen abundance is 10$^{-3}$ with respect to hydrogen, both of which are extreme assumptions (we expect perhaps $\sim10-50\%$ of oxygen to be in water ice \cite{Jenkins2009,Whittet2010} and oxygen abundance in the GC is a little less than 10$^{-3}$ \cite{Mendez-Delgado2022}).
Adopting lower abundance of water results in only reddening: the dust dominates the weak blueing effect of water, while CO-driven blueing remains significant.
However, both of these plots point to the hydrogen column being overestimated by the local H/A$_V$ relation, since both violate total or specific metal budgets.
}

\clearpage
\section{The OJA}

\end{document}